\documentclass[11pt,draftcls,onecolumn]{IEEEtran}

\usepackage{cite}
\usepackage{amsmath} 
\usepackage{amssymb} 
\usepackage{dsfont}
\usepackage[mathcal]{euscript}
\usepackage{empheq}
\usepackage{graphicx}
\usepackage{graphics}
\usepackage{bm}
\usepackage{psfrag}
\usepackage{mathrsfs}
\usepackage{bbm}
\usepackage{color}
\usepackage{cancel}
\input{epsf}
\newtheorem{theorem}{Theorem}
\newtheorem{lemma}{Lemma}
\newtheorem{corollary}{Corollary}
\newtheorem{proposition}{Proposition}

\renewcommand{\P}{\mathbb{P}}
\newcommand{\E}{\mathbb{E}}
\newcommand{\V}{\mathbb{V}}
\newcommand{\beq}{\begin{equation}}
\newcommand{\eeq}{\end{equation}}
\newcommand{\beqa}{\begin{eqnarray}}
\newcommand{\eeqa}{\end{eqnarray}}
\newcommand{\dfz}{\triangleq}

\begin{document}
%
\title{Consistent Tomography under Partial Observations over Adaptive Networks}

\author{Vincenzo~Matta and Ali~H.~Sayed
\thanks{V.~Matta is with DIEM, University of Salerno,
via Giovanni Paolo II, I-84084, Fisciano (SA), Italy (e-mail: vmatta@unisa.it).

A.~H.~Sayed is with the \'Ecole Polytechnique F\'ed\'erale de Lausanne EPFL, School of Engineering, CH-1015 Lausanne, Switzerland (e-mail: ali.sayed@epfl.ch).
His work was supported in part  by NSF grants CCF-1524250 and ECCS-1407712.
}
}

\maketitle

\begin{abstract}
This work studies the problem of inferring whether an agent is directly influenced by another agent over an adaptive diffusion network. 
Agent $i$ influences agent $j$  if they are connected (according to the network topology), {\em and} if agent $j$ uses the data from agent $i$ to update its online statistic. 
The solution of this inference task is challenging for two main reasons. 
First, only the output of the diffusion learning algorithm is available to the external observer that must perform the inference based on these indirect measurements.
Second, only output measurements from a fraction of the network agents is available, with the total number of agents itself being also unknown.
The main focus of this article is ascertaining under these demanding conditions whether {\em consistent tomography is possible}, namely, whether it is possible to reconstruct the interaction profile of the observable portion of the network, with negligible error as the network size increases.
We establish a critical achievability result, namely, that for symmetric combination policies and for any given fraction of observable agents, the interacting and non-interacting agent pairs split into two separate clusters as the network size increases.
This remarkable property then enables the application of clustering algorithms to identify the interacting agents influencing the observations. We provide a set of numerical experiments that verify the results for finite network sizes and time horizons. 
The numerical experiments show that the results hold for asymmetric combination policies as well, which is particularly relevant in the context of {\em causation}.
\end{abstract}

\begin{IEEEkeywords}
Diffusion networks, network tomography, causation, combination policy, Erd\"os-Renyi model.
\end{IEEEkeywords}

\section{Introduction}
\IEEEPARstart{A} \,significant number of complex real-world systems are modeled well by {\em networks}. 
At an abstract level, a network is an ensemble of interconnected agents.
The interactions among neighboring agents enable the flow of information across the graph, and give rise to interesting and complex patterns of coordinated behavior.

One problem of immense value in network science is the inverse modeling problem. 
In this problem, the network structure (topology) is assumed to be largely unknown and one is interested in inferring relationships between agents based on some dataset arising from the agents' operations.  
Formulations of this type are of great interest in several application domains, such as communications, computer science, cyber-security, control, physics, biology, economics, and social sciences.

Inverse problems over networks are challenging because, in the vast majority of applications, direct access to the data exchanged between agents is often unavailable, and the inference of inter-agent relations must be based on {\em indirect} observations. 
Another source of difficulty in such inverse problems is that the access to observations is generally {\em limited to a subset of the network agents}. 
The process of discovering inter-agent interactions from indirect measurements is broadly referred to as {\em network tomography}.
Some useful applications of network tomography include, among other possibilities: 
tracing the routes of clandestine information flows across communications networks~\cite{Venk-He-Tong-IT, He-Tong-Forens,He-Tong-IT,He-Agaskar-Tong-SP, Kim-Tong-SP2012, Marano-Matta-He-Tong-IT}; revealing agent interactions over social networks, where disclosing commonalities within groups of agents might be useful for commercial as well as security purposes~\cite{KrimTSIPN, Earnest}; inferential problems related to group testing and identification of defective items~\cite{Saligrama}; brain networks, where interactions among neurons might be of paramount importance to the understanding of particular diseases~\cite{NatureComm, Kiyavash0}; anomaly detection in communications networks, where one tries to reveal the activities of intermediate nodes through destination-only measurements~\cite{Giannakis1}.

One useful type of networks is the class of adaptive networks~\cite{SayedFoundTrends,SayedProcIEEE}.
These networks consist of spatially dispersed agents that continually exchange information through diffusion mechanisms, and which are specifically designed to enable simultaneous {\em adaptation} and 
{\em learning} from streaming data (such as tracking targets moving in formation from streaming spatial observations; modeling the prey-predator behavior of animal groups on the move; allocating frequency resources over cognitive communication networks)~\cite{SayedFoundTrends,SayedProcIEEE}.
Using a powerful form of agent cooperation, adaptive diffusion networks are able to solve rather sophisticated inferential tasks, including: estimation tasks~\cite{SayedProcIEEE}, detection tasks~\cite{AdaptiveDetectionIT,AdaptiveDetectionTSIPN}, optimization and online learning tasks~\cite{ChenSayedIT2015Part1,ChenSayedIT2015Part2}.
This work is focused on {\em tomography over partially observed diffusion networks}.

\subsection{Related Work}
From a merely theoretical perspective, the aforementioned problems fall under the umbrella of signal processing over graphs~\cite{SayedFoundTrends,SayedProcIEEE,OrtegaSPmag,Ortegaetal2016,TsitsiveroBarbarossaDiLorenzo2016}. 
They deal with the objective of retrieving a graph topology (a connection topology, or an ``effective'' topology corresponding to the exchange of information) from a set of indirect measurements taken at some accessible network locations. 
There are several works addressing a similar construction, albeit with different specific goals. 
With no pretence of exhaustiveness, we give a brief summary of the works that we believe are most related to the present article. 

In~\cite{MaterassiSalapakaTAC2012}, an unknown network topology is reconstructed by taking advantage of the locality properties of the  Wiener filter. Exact reconstruction results are provided for self-kin networks, while reconstruction of the smallest self-kin network embracing the true network topology is guaranteed for general networks.
In~\cite{Kiyavash1}, the authors introduce the concept of directed information graphs, which is used to capture the dependencies in networks of interacting processes linked by causal dynamics.
The setting is further enriched in~\cite{Kiyavash2}, where a new metric to learn causal relationships by means of {\em functional dependencies}, in a possibly nonlinear dynamical network, is proposed.

More closely related to the network model considered in our work is the problem of estimating a graph when the relationships are encoded into an autoregressive model --- see, e.g.,~\cite{Monetaetal}, where several methods to address causal inference in such a context are reviewed.
Causal graph processes are also exploited in~\cite{MeiMoura}, where a computationally tractable algorithm for graph structure recovery is proposed, along with a detailed convergence analysis.
In~\cite{Geigeretal}, the authors examine a causal inference problem that is modeled through a vector autoregressive process. 
The objective is that of reconstructing the important parts of the transition matrix through observation of a subset of the random process. Special technical conditions for exact reconstruction are provided. 
In~\cite{KiyavashPolytrees}, new methods are proposed to learning directed information polytrees where samples are available from only a subset of processes.

Some recent works exploit spectral graph properties, in conjunction with sparsity constraints.
In~\cite{RabbatGraphs, RabbatGraphsAllerton}, the problem of inferring the graph topology from observations of a random signal diffusing over the network is addressed. It is shown that the space of feasible matrices is a convex polytope, and two inferential strategies are proposed to select a point in this space as the topology estimate. 
In~\cite{MateosGraphsAsilomar}, the authors 
exploit convex optimization and sparsity in order to reconstruct an unknown graph from observable indirect relationships generated by diffusive signals defined on the graph nodes. The main idea is identifying a graph shift operator given only its eigenvectors, with the corresponding spectral templates being obtained from the sample covariance of independent graph signals diffused over the network. 

\subsection{Main Result}
This work complements the previous efforts: it establishes an important identifiability condition and clarifies the asymptotic behavior of the recovery error as a function of the network size. An identification procedure is also developed to carry out the tomography calculations using a standard clustering technique. We summarize the main contributions of this work as follows. 
We consider a network of agents running a diffusion strategy to solve some inference task of interest (such as a distributed detection problem).
The overall network size is unknown, and the outputs of the diffusion algorithm are available from only a {\em limited} fraction of agents. 
The goal is determining the relationships among these agents, namely, establishing whether the datum of an individual agent is influencing another individual agent.  
We show that, under some regularity conditions, the interaction relationships existing within the observable portion of the network can be recovered, with negligible error for sufficiently large network sizes.

More specifically, we consider the class of Erd\"os-R\'enyi random graphs, where the probability of two agents being connected follows a Bernoulli distribution. For these graphs we discover that, for any fraction of observable agents, the group of interacting agent pairs and the group of non-interacting agent pairs split into two well-defined clusters. These clusters emerge as clearly separate for sufficiently large network sizes.
This result is established for diffusion strategies employing {\em symmetric} combination matrices, a feature that enables analytical tractability of the problem. 
However, from a ``physical'' viewpoint, separability of the clusters does not appear to be limited to the symmetry assumption. 
Accordingly, we examine also the relevant case of asymmetric matrices, which is of interest especially in the context of {\em causal} networks. With reference to a typical form of asymmetric combination matrices, numerical simulations show that separability of the clusters can be preserved.

One distinctive feature of our work is that we establish {\em theoretical achievability} results for {\em consistent} tomography from {\em partially observed} networks. In answering these particular questions, in addition to what has been answered before, our work complements well existing results and recent progresses in the field of network tomography.


{\em Notation.} We use boldface letters to denote random variables, and normal font letters for their realizations. 
Capital letters refer to matrices, small letters to both vectors and scalars. 
Sometimes we violate the latter convention, for instance, we denote the total number of network agents by $N$. 

For $i,j=1,2,\dots,N$, the $(i,j)$-th entry of an $N\times N$ matrix $Z$ will be denoted by $z_{ij}$, or alternatively by $[Z]_{ij}$.
Moreover, the sub-matrix that lies in the rows of $Z$ indexed by the set $\mathcal{S}_1\subseteq\{1,2,\dots, N\}$ and in the columns indexed by the set $\mathcal{S}_2\subseteq\{1,2,\dots, N\}$, will be denoted by $Z_{\mathcal{S}_1 \mathcal{S}_2}$, or alternatively by $[Z]_{\mathcal{S}_1 \mathcal{S}_2}$.
If $\mathcal{S}_1=\mathcal{S}_2=\mathcal{S}$, the sub-matrix $Z_{\mathcal{S}_1 \mathcal{S}_2}$ will be abbreviated as $Z_{\mathcal{S}}$, or as $[Z]_{\mathcal{S}}$. 

The symbols $\P$, $\E$, and $\V$ are used to denote the probability, expectation, and variance operators, respectively. 
The notation $\stackrel{\textnormal{p}}{\longrightarrow}$ denotes convergence in probability as $N\rightarrow\infty$.

\section{The Problem}

\subsection{The Adaptive Diffusion Network}
A network of $N$ agents collects streaming data from the environment. 
The datum collected by the $i$-th agent at time $n$ is denoted by $\bm{x}_i(n)$, and the global sequence of data 
is assumed to be formed by spatially (i.e., w.r.t. index $i$) and temporally (i.e., w.r.t. index $n$) independent and identically distributed random variables. Without loss of generality, we assume that the variables have zero mean and unit variance. 

In order to track drifts in the phenomenon that the network is monitoring, the agents implement a distributed adaptive strategy, where each agent relies on sharing information with local neighbors. In this work we focus on the Combine-Then-Adapt (CTA) diffusion mechanism, whose useful properties in terms of estimation and online learning performance have been already studied in some revealing detail in~\cite{SayedFoundTrends,SayedProcIEEE,ChenSayedIT2015Part1,ChenSayedIT2015Part2}. 
The CTA algorithm can be described as follows.

First, during the {\em combination} step, agent $i$ mixes the data received from its neighbors by using a sequence of convex (i.e., nonnegative and adding up to $1$) combination weights $w_{i\ell}$, for $\ell=1,2,\dots,N$, giving rise to the following intermediate variable:
\beq
\bm{v}_i(n-1)=\sum_{\ell=1}^N w_{i\ell}\,\bm{y}_{\ell}(n-1).
\label{eq:combine}
\eeq
Then, during the {\em adaptation} step, agent $i$ updates its output variable by comparison with the incoming streaming data $\bm{x}_i(n)$, employing a (typically small) step-size $\mu\in(0,1)$:  
\beq
\bm{y}_i(n)=\bm{v}_i(n-1) + \mu[\bm{x}_i(n) - \bm{v}_i(n-1)].
\label{eq:adapt}
\eeq
Equations~(\ref{eq:combine}) and~(\ref{eq:adapt}) can be merged into a single step as:   
\beq
\bm{y}_i(n)=(1-\mu)\sum_{\ell=1}^N w_{i\ell}\, \bm{y}_{\ell}(n-1) + \mu\,\bm{x}_i(n).
\label{eq:CTA1}
\eeq
For later use, it is convenient to introduce the {\em scaled} combination matrix $A$, whose entries are defined as:
\beq
a_{ij}\dfz(1-\mu)w_{ij},
\label{eq:aweightsmatdef}
\eeq
which, in the sequel, will be simply referred to as the combination matrix. We remark that, since we use a sequence of {\em convex} combination weights, the matrix $A/(1-\mu)$ is a {\em right-stochastic} matrix. 

If we now stack the observations across the network at time $n$ into the $N\times 1$ vector $\bm{x}_n$, and the state variables at time $n$ into the $N\times 1$ vector $\bm{y}_n$, Eq.~(\ref{eq:CTA1}) can be compactly rewritten as: 
\beq
\bm{y}_n=A\,\bm{y}_{n-1} + \mu \,\bm{x}_n,
\label{eq:origdef}
\eeq
which, by iteration, allows us to express the diffusion output vector as a function of the streaming data: 
\beq
\boxed{
\bm{y}_n=\mu \sum_{m=1}^n A^{n-m} \bm{x}_m, \quad n\geq 1
}
\label{eq:origdef2}
\eeq
starting from state $\bm{y}_0=0$, i.e., neglecting the {\em transient} term.

\begin{figure}[t]
\centerline{\includegraphics[width=.45\textheight]{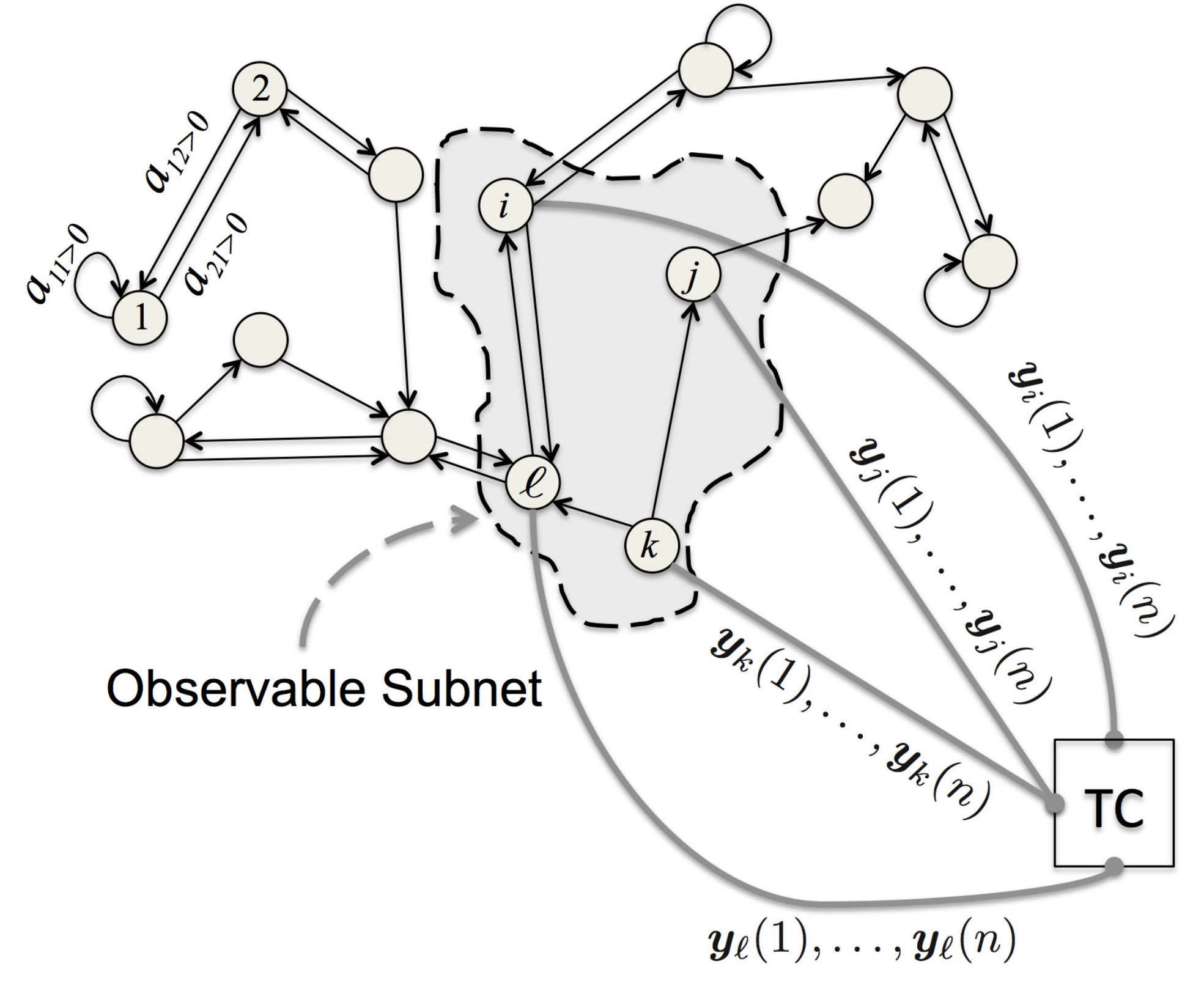}}
\caption{Conceptual sketch of the tomography problem addressed in this article.}
\label{fig:concept}
\end{figure}

\subsection{Network Tomography}
\label{subsec:NT}
An illustration of the setting considered in this work is given in Fig.~\ref{fig:concept}. 
An entity external to the network (hereafter named Tomography Center, TC) is interested in reconstructing the interaction profile of the network, namely, it is interested in ascertaining which agent is influencing which other agent. 

The TC is assumed to have access to a {\em subset} of the network agents, and is able to collect the streams of outputs exchanged by such agents during their communication activity. 
Letting $\Omega\subset \{1,2,\dots,N\}$ be the observable subnet, the data available to the TC at time $n$ are $\{\bm{y}_i(1),\bm{y}_i(2),\dots,\bm{y}_i(n)\}_{i\in\Omega}$.
We shall focus on the asymptotic regime of large networks ($N\rightarrow\infty$), for the meaningful case where the fraction of observed agents does not vanish. Letting $K=|\Omega|$, such regime is defined in terms of the following condition: 
\beq
\boxed{
\frac{K}{N}\rightarrow \xi
}
\label{eq:csidef}
\eeq
where $\xi\in(0,1)$ takes on the meaning of the asymptotic fraction of observed agents.

In our setting, the TC does not know the overall number of agents in the network. Accordingly, the main goal of the TC is producing an estimate of the interaction profile for the {\em observed} agents.
This problem is challenging for the following reasons.
Let us ignore for a moment the fact that the network is partially observed, and assume that the TC is able to collect all output sequences from all agents at all times.
Using such a dataset, there exist several well-established strategies to make inference about the influence that one agent has on another agent. 
The most intuitive is an estimate of the correlation between the outputs of two agents, which, however, is problematic for {\em directed} flows of information (where agent $i$ can influence agent $j$ but not vice versa). 
Such asymmetry would be reflected into the $(i,j)$-th and $(j,i)$-th entries of the combination matrix, yielding $a_{ij}=0$ while $a_{ji}>0$.
The case of asymmetric influence is well studied within the framework of {\em causal} inference, or {\em causation}. Many solutions exist for disclosing causal relationships~\cite{Kiyavash1, Kiyavash2, Monetaetal, MeiMoura, Geigeretal, KiyavashPolytrees}. 

There is, however, a challenging problem that is peculiar to the network setting of this work due to the streaming nature of the data. 
In general, when the TC starts collecting data, the network would have been in operation since some time already.
Therefore, the output signals at the agents would have benefited from sufficient exchanges of information.
While this exchange of information is beneficial for solving inference tasks, it nevertheless can become detrimental for reconstructing the network tomography. This is because, over a strongly connected network and after a ``transient'' phase, all agents would have become ``correlated!''

In order to overcome this difficulty, we exploit knowledge of the diffusion mechanism.
To this aim, let us introduce the correlation matrix of the diffusion output vector:
\beq
R_0(n)\dfz\E[\bm{y}_n \bm{y}_{n}^T],
\eeq
which, from~(\ref{eq:origdef2}), admits the following closed-form representation:
\beq
R_0(n)=\mu^2 \sum_{i=0}^{n-1} A^i (A^i)^T 
\stackrel{n\rightarrow\infty}{\longrightarrow} R_0=\mu^2 \sum_{i=0}^\infty A^i (A^i)^T,
\label{eq:Lyapunov}
\eeq 
where the latter series is guaranteed to converge since all eigenvalues of $A$ are strictly inside the unit disc.
The limiting correlation matrix, $R_0$, can be interpreted as the (unique) solution to the discrete-time Lyapunov equation~\cite{SayedFoundTrends}:
\beq
R_0 - A R_0 A^T = \mu^2 I_N,
\eeq
where $I_N$ denotes the $N\times N$ identity matrix.
We note that $R_0(n)$ is positive definite for each $n$, and so is $R_0$, due to the stability of $A$~\cite{KailathSayed}. We also introduce the one-lag correlation matrix, which, in view of~(\ref{eq:origdef}), takes on the form:
\beq
R_1(n)\dfz\E[\bm{y}_n \bm{y}_{n-1}^T]=AR_0(n-1) \stackrel{n\rightarrow\infty}{\longrightarrow} R_1=A R_0.
\label{eq:R1nR0n}
\eeq
Therefore, we obtain the following relationship:
\beq
A=R_1(n)(R_0(n-1))^{-1}, \quad n\geq 2,
\label{eq:fundamentalestimate}
\eeq
and, at steady-state,
\beq
\boxed{
A=R_1R_0^{-1}
}
\label{eq:fundamentalestimate_asy}
\eeq
In principle, since there exist many ways to estimate $R_0$ and $R_1$ consistently as $n\rightarrow\infty$, expression~(\ref{eq:fundamentalestimate_asy}) reveals one possible strategy to estimate $A$ from the output of the diffusion process. 
However, the approach described so far suffers from a critical problem: the network is only {\em partially} observed and, hence, not all entries in the matrices $R_0$ and $R_1$ can be estimated. This in turn means that the evaluation of~(\ref{eq:fundamentalestimate_asy}) is in general problematic. 

In order to estimate the combination (sub-)matrix corresponding to the observable agents, i.e.,
\beq
A^{\textnormal{(obs)}}\dfz A_\Omega,
\eeq
one might be tempted to simply replace the matrices involved in~(\ref{eq:fundamentalestimate_asy}) by their {\em observable} counterparts, $R_0^{\textnormal{(obs)}}\dfz[R_0]_{\Omega}$, and $R_1^{\textnormal{(obs)}}\dfz[R_1]_{\Omega}$, obtaining the rough estimate:\footnote{Actually, the matrices computed by the TC will generally be renumbered versions of $[R_0]_\Omega$ and $[R_1]_\Omega$, since the labeling used by the TC need not correspond to the ordering in $[R_0]_\Omega$ and $[R_1]_\Omega$. 
This is because the TC does not know the original labeling of the agents in the network, nor the network size. 
However, our focus is on evaluating the interaction between {\em pairs} of agents, irrespective of their labels, and, hence, we can safely keep the notation used in~(\ref{eq:A11_first}).} 
\beq
\boxed{
\hat A^{\textnormal{(obs)}}=R_1^{\textnormal{(obs)}}(R_0^{\textnormal{(obs)}})^{-1}
}
\label{eq:A11_first}
\eeq 
Needless to say, calculation on the right-hand side of~(\ref{eq:A11_first}) does not lead to the true $A^{\textnormal{(obs)}}$, except in some special cases. For this reason, we are denoting the result of~(\ref{eq:A11_first}) by using the hat notation. 
If we could recover $A^{\textnormal{(obs)}}$ exactly, then we could deduce the desired influence relations. However, given that we only have the estimate $\hat A^{\textnormal{(obs)}}$, it is not clear at all whether the mutual influence relationships existing between the observed nodes can be consistently retrieved from $\hat A^{\textnormal{(obs)}}$. Answering this nontrivial question in the affirmative is one of the main contributions of this work.

In order to highlight the key ideas without added complexity, we focus in this article on the case of symmetric combinations matrices. 
The symmetry assumption is made because, if $A$ is symmetric, the autocorrelation matrix of the diffusion output takes on the following convenient form --- see ~(\ref{eq:Lyapunov}):
\beq
R_0(n)=\mu^2 \sum_{i=0}^{n-1} A^{2 i}\Rightarrow
R_0=\mu^2 (I_N-A^2)^{-1},
\label{eq:origdef3}
\eeq
which is exploited in the proof of Theorem~$1$.
The extension of the results to the asymmetric case requires adjustments in the arguments used in the proof of Theorem~$1$. 
For example, this can be pursued by appealing instead to the Kronecker representation of the solution of discrete-time Lyapunov equations~\cite{Sayed2008adaptive}. 
We note in passing that, since the matrix $A/(1-\mu)$ is always right-stochastic, the symmetry assumption makes $A/(1-\mu)$ {\em doubly}-stochastic. 

\section{Error Caused by Partial Observations}
Using the estimate~(\ref{eq:A11_first}), we can write:
\beq
\hat A^{\textnormal{(obs)}}=A^{\textnormal{(obs)}} + E,
\label{eq:errmatdef}
\eeq
where $E$ denotes the {\em error} matrix.
In terms of the individual entries, we can write, for $i,j=1,2,\dots, K$:\footnote{A pair $(i,j)$ refers to agents $\omega_i$ and $\omega_j$, where $\omega_1<\omega_2<\dots<\omega_K$ span the {\em observable} subnet $\Omega$.}
\beq
\hat a^{\textnormal{(obs)}}_{ij}=a^{\textnormal{(obs)}}_{ij} + e_{ij}.
\label{eq:a11eqhata11}
\eeq 
In this work we are interested in establishing whether the estimated values $\hat a^{\textnormal{(obs)}}_{ij}$ allow us to identify the condition $a^{\textnormal{(obs)}}_{ij}>0$ or $a^{\textnormal{(obs)}}_{ij}=0$, which would reveal whether the agents $i$ and $j$ influence each other. In order to enable this determination, we start by characterizing the behavior of the error terms $e_{ij}$ in~(\ref{eq:a11eqhata11}).

\begin{theorem}[Concentration of errors]
For a symmetric combination matrix, the entries of the error matrix defined in~(\ref{eq:errmatdef}) are nonnegative, and satisfy for all $i=1,2,\dots, K$:
\beq
\boxed{
\sum_{j=1}^K e_{ij} 
\leq 1-\mu
}
\label{eq:mainclaim}
\eeq
\end{theorem}

\begin{IEEEproof}
See Appendix~\ref{app:Theo1}.
\end{IEEEproof}
We have the following corollary, which is particularly tailored to our application. 

\begin{corollary}[Number of errors exceeding a threshold]
For a symmetric combination matrix, we have, for all $i=1,2,\dots,K$, and for any $\epsilon>0$:
\beq
\boxed{
\sum_{j=1}^K \mathbb{I}\{e_{i j} > \epsilon\}
\leq 
\frac{1-\mu}{\epsilon}
}
\label{eq:corollclaim}
\eeq
where $\mathbb{I}\{\cdot\}$ is the indicator function (which is equal to one when its argument is true and zero otherwise). 
\end{corollary}

\begin{IEEEproof}
Suppose that~(\ref{eq:corollclaim}) is false. Then, we would have (recall that $e_{i j}\geq 0$ for all $i,j=1,2,\dots,K$):
\beqa
\sum_{j=1}^K e_{ij}&=&\sum_{j: e_{ij} > \epsilon} e_{ij} + \sum_{j: e_{ij} \leq \epsilon} e_{ij}\geq\sum_{j: e_{ij} > \epsilon} e_{ij}\nonumber\\
&>&
\epsilon\sum_{j=1}^K \mathbb{I}\{e_{i j} > \epsilon\}
>\frac{1-\mu}{\epsilon}\times\epsilon=1-\mu,\nonumber\\
\label{eq:Markovlike}
\eeqa
which contradicts~(\ref{eq:mainclaim}).
\end{IEEEproof}

Theorem~$1$ and its corollary provide useful information about the concentration of the entries in the error matrix. 
In particular, Eq.~(\ref{eq:mainclaim}) reveals that the row-sum of the entries in the error matrix cannot exceed $1-\mu$, whereas Eq.~(\ref{eq:corollclaim}) places an upper bound on the number of entries that exceed any positive threshold.

Consider now a small threshold $\epsilon$, and the fraction of off-diagonal (because we are interested in inter-agent interactions) entries that exceed $\epsilon$:
\beq
\frac{1}{K(K-1)}\,\sum_{i=1}^K\sum_{j\neq i} \mathbb{I}\{e_{i j} > \epsilon\}.
\eeq
In the regime dictated by~(\ref{eq:csidef}), where $K\rightarrow\infty$ as $N\rightarrow\infty$, we see from~(\ref{eq:corollclaim}) that such fraction vanishes, namely, that most entries of the error matrix will be small in the asymptotic regime of large networks.
Therefore, in view of definition~(\ref{eq:a11eqhata11}), it will hold for large networks that, for $i\neq j$,
\beq
\hat{a}^{\textnormal{(obs)}}_{ij}=
\left\{
\begin{array}{l}
a^{\textnormal{(obs)}}_{ij} + \textnormal{\footnotesize{small quantity}},~~~~~\textnormal{ if } a^{\textnormal{(obs)}}_{ij}>0,\\
\\
\textnormal{\footnotesize{small quantity}}, ~~~~~~~~~~~~~~\textnormal{ if } a^{\textnormal{(obs)}}_{ij}=0.
\end{array}
\right.
\label{eq:aobsarray}
\eeq
This useful dichotomy suggests that the nonzero entries of $A^{\textnormal{(obs)}}$ will make the estimated entries $\hat{a}^{\textnormal{(obs)}}_{ij}$ stand out above the error floor as $N$ increases.
As a result, we should be able to decide whether $a^{\textnormal{(obs)}}_{ij}=0$ or $a^{\textnormal{(obs)}}_{ij}>0$ by comparing the estimated value, $\hat{a}^{\textnormal{(obs)}}_{ij}$, against some threshold. 

However, and unfortunately, the behavior of the error matrix alone is not sufficient to conclude that this inferential procedure is feasible, for one crucial reason. 
For most typical combination matrices, the nonzero entries $a_{ij}^{\textnormal{(obs)}}$ decrease with $N$ as well, so that the nonzero entries appearing in the first line of~(\ref{eq:aobsarray}) vanish for large network sizes, along with the errors. 
This means that the estimated entries, $\hat{a}^{\textnormal{(obs)}}_{ij}$, would vanish even when agents $i$ and $j$ are interacting (i.e., even when $a^{\textnormal{(obs)}}_{ij}>0$). 
For this reason, a closer examination of the behavior of the error quantities is necessary before we can conclude that this inference procedure is viable. Specifically, it is necessary to assess how fast the error signals $e_{ij}$ decay to zero in relation to the desired entries $a_{ij}^{\textnormal{(obs)}}$.
In order to carry out this analysis, we need to make some assumptions about the network structure. We will select some typical and popular random models, as explained next.

\section{Behavior of Interacting Agents $(a_{ij}>0)$}
\label{eq:subsecNIM}
The interaction profile can be conveniently described in terms of a symmetric {\em interaction} matrix $G$, defined by the following conditions:
\beq
g_{ij}=
\left\{
\begin{array}{l}
1, \textnormal{ if } a_{ij}>0,\\
0, \textnormal{ if } a_{ij}=0.
\end{array}
\right.
\label{eq:GGGdef}
\eeq
We shall set $g_{ii}=1$ for all $i=1,2,\dots,N$, which corresponds to the assumption that  an agent does always use its own state variable in the combination step.

In our analysis, the combination matrix, $A$, will be constructed through the following two-step procedure. 
First, an interaction matrix $G$ is generated according to a {\em random graph} model~\cite{ErdosRenyi,BollobasBook}.
Then, $A$ is determined by a {\em combination policy}, $\gamma(G)$, which sets the values of the combination weights corresponding to the nonzero entries of $G$. 
Formally:
\beq
\boxed
{
A=\gamma(G)
}
\eeq  
Note that $\gamma(G)$ must always assign {\em positive} weights at the locations corresponding to {\em nonzero} entries of $G$, otherwise the interaction matrix related to $A$ would be different from $G$.
Moreover, since in this article we focus on symmetric combination matrices, we must have that $[\gamma(G)]_{ij}=[\gamma(G)]_{ji}$ (a condition that would not be directly implied by the symmetry of $G$).

\subsection{Random Model for the Interaction Profile}
As we have stated, in this work we examine the useful case where the interaction profile of the network is generated according to a {\em random graph} model~\cite{ErdosRenyi,BollobasBook}.
In particular, we consider the classic Erd\"os-R\'enyi model.
This model, which we denote by $\mathscr{G}(N,p_N)$, is an undirected (i.e., symmetric) graph, where the presence/absence of the $N(N-1)/2$ edges is determined by a sequence of $N(N-1)/2$ independent Bernoulli random variables with success (i.e., interaction) probability $p_N$.
Accordingly, the variables $\bm{g}_{ij}$, for $i=1,2,\dots, N$ and $j>i$, are independent Bernoulli random variables with $\P[\bm{g}_{ij}=1]=p_N$, and the matrix $G$ is symmetric. 

One meaningful regime to examine the random graph properties is the regime where the probability $p_N$ decreases as $N$ increases~\cite{BollobasBook}. 
Examining the asymptotic regime where $p_N$ is kept constant while $N$ increases would generally produce networks that are too much connected with respect to what happens in typical applications. 
For instance, in the regime of constant $p_N$, the network diameter (i.e., the maximum of the shortest distance between any pair of nodes) is equal to $2$ with high probability~\cite{binomialgraphs}. 
On the other hand, $p_N$ could not vanish in an arbitrary fast way, otherwise the number of neighbors of an agent would be too small, and the network would become very scarcely connected.
A well-known result that holds true for the Erd\"os-R\'enyi graph is that the following scaling law:
\beq
\boxed{
p_N=\frac{\ln N + c_N}{N}
}
\label{eq:pconn}
\eeq
with $c_N\rightarrow\infty$ (in an arbitrarily slow fashion, i.e., even with $p_N\rightarrow 0$), ensures that the graph remains connected with probability tending to $1$ as $N$ diverges~\cite{BollobasBook}.
In the following, we shall focus on the regime of connected Erd\"os-R\'enyi graph with vanishing $p_N$, namely, on the regime where~(\ref{eq:pconn}) is satisfied with $c_N\rightarrow\infty$ and $p_N\rightarrow 0$.  
Such a regime will be denoted by the symbol $\mathscr{G}^\star(N,p_N)$.

\subsection{Statistical Properties of Fundamental Graph Descriptors}
Some useful descriptors of the network graph can be conveniently represented in terms of the matrix $G$ in~(\ref{eq:GGGdef}). 
In particular, the (interaction) neighborhood of agent $i$ (which includes $i$ itself) is defined as:
\beq
\mathcal{N}_i=\{\ell\in\{1,2,\dots,N\}: g_{i\ell}=1\},
\eeq
while the {\em degree} of agent $i$ is: 
\beq
d_i=|\mathcal{N}_i|=\sum_{\ell=1}^N g_{i\ell}=1 + \sum_{\ell\neq i} g_{i\ell}=1 + \sum_{\ell\neq i} g_{\ell i}.
\label{eq:degdef}
\eeq
We define the network maximal degree as:
\beq
d_{\textnormal{max}}\dfz \displaystyle{\max_{i=1,2,\dots N} d_i}.
\eeq 
Let us now highlight some useful statistical properties of the degree and maximal degree variables for the Erd\"os-Renyi model.

We observe from~(\ref{eq:degdef}) that, for an Erd\"os-Renyi graph, the random variable $\bm{d_i} -1$ is a binomial random variable with parameters $N-1$ and $p_N$, which shall be denoted by $\beta(N-1,p_N)$. 
Note also that $\bm{d}_i$ and $\bm{d}_j$, for $i\neq j$, are {\em not} independent, because of the implied graph symmetry. 
Now, for a binomial random variable $\beta(N,p_N)$, we have: 
\beq
\E\left[\frac{\bm{\beta}(N,p_N)}{N p_N}\right]=1,
~\V\left[\frac{\bm{\beta}(N,p_N)}{N p_N}\right] =\frac{N p_N (1-p_N)}{(N p_N)^2}.
\label{eq:muvarbin}
\eeq
Under the $\mathscr{G}^\star(N,p_N)$ model, we see from~(\ref{eq:pconn}) that the product $N p_N$ diverges as $N\rightarrow\infty$, and, hence, the variance in~(\ref{eq:muvarbin}) vanishes as $N\rightarrow\infty$, implying in particular the following convergence in probability~\cite{shao}:
\beq
\boxed{
\frac{\bm{\beta}(N,p_N)}{N p_N}\stackrel{\textnormal{p}}{\longrightarrow} 1
}
\label{eq:binoconv}
\eeq
which reveals that the degrees of the nodes scale as $N p_N$.

It is also of interest to characterize the asymptotic behavior of $\bm{d}_{\textnormal{max}}$, which, by being the maximum of $N$ degrees, is expected to grow faster than $N p_N$. However, and interestingly, the following lemma shows that it cannot grow {\em much} faster. 
 
\begin{lemma}[Behavior of maximal degree]
Under the $\mathscr{G}^\star(N,p_N)$ model we have, for all $i,j=1,2,\dots,N$, with $i\neq j$: 
\beq
\boxed{
\P[\bm{d}_{\textnormal{max}} \geq N p_N e\, | \, \bm{g}_{ij}=1]
\leq 
\left(e + \frac{2 e^2}{N}\right) e^{-c_N}\stackrel{N\rightarrow\infty}{\longrightarrow} 0
}
\label{eq:theo1claim}
\eeq
where $e$ is Euler's number. 
\end{lemma}
\begin{IEEEproof}
See Appendix~\ref{app:Theo2}. 
\end{IEEEproof}
According to~(\ref{eq:theo1claim}), the maximal degree exceeds the level $N p_N e$ with negligible probability, i.e., it cannot grow  substantially faster than $N p_N$.
We note that the probability in~(\ref{eq:theo1claim}) is computed conditionally on the event that two agents interact. This choice is made because, in the following, we need to know the behavior of the maximal degree in relation to {\em interacting} agent pairs.

\subsection{Stable Combination Policies}
\label{subsec:cp}
We can now use~(\ref{eq:theo1claim}) to characterize the asymptotic behavior of the (off-diagonal) nonzero entries in the combination matrix.
In order to illustrate the main idea, we start by examining a popular combination policy, known as the Laplacian rule, which is given by~\cite{SayedFoundTrends}:
\beq
a_{ij}=
\left\{
\begin{array}{l}
g_{ij}\,(1-\mu)\,\lambda/d_{\textnormal{max}},\quad\textnormal{for } i\neq j,
\\
\\
(1-\mu) - \sum_{\ell\neq i} a_{i\ell}\quad\textnormal{for } i=j.
\end{array}
\right.
\eeq
for some $\lambda$, with $0<\lambda\leq 1$. Therefore, from Lemma~$1$ we can write, for all $i\neq j$:
\beqa
\lefteqn{\P[N p_N \bm{a}_{ij} > (1-\mu) \lambda/e \, | \, \bm{g}_{ij}=1]}\nonumber\\
&=&
\P[\bm{d}_{\textnormal{max}} < N p_N e \, | \, \bm{g}_{ij}=1]\nonumber\\
&\geq&
1 - \left(e + \frac{2 e^2}{N}\right) e^{-c_N}\stackrel{N\rightarrow\infty}{\longrightarrow} 1.
\label{eq:lowbounonzero}
\eeqa
Equation~(\ref{eq:lowbounonzero}) has the following important implication: for large enough $N$, any nonzero entry of the Laplacian combination matrix, scaled by the factor $N p_N$, stays ``almost always'' above a certain threshold (namely, the value $(1-\mu)\lambda/e$). 
Therefore, multiplying $a_{ij}$ by the scaling factor $N p_N$ would keep the nonzero entries stable (in the sense that they will not vanish) as $N$ diverges.
The same scaling factor is relevant for other combination policies. 
Therefore, it makes sense to introduce the following general class of combination policies.

{\em Combination-policy class $\mathscr{C}_\tau$.} 
A combination policy belongs to class $\mathscr{C}_\tau$ if there exists $\tau>0$ such that, for all $i,j=1,2,\dots,N$, with $i\neq j$:
\beq
\boxed
{
\P[N p_N \bm{a}_{ij} > \tau \, | \, \bm{g}_{ij}=1]
\geq
1 - \epsilon_N
}
\label{eq:AssumptionA1}
\eeq
where $\epsilon_N$ goes to zero as $N\rightarrow\infty$, and where the probability is evaluated under the $\mathscr{G}^\star(N,p_N)$ model.
$\hfill\square$

Let us now examine the physical meaning of~(\ref{eq:AssumptionA1}) in connection to network tomography applications. 
For a policy belonging to class $\mathscr{C}_\tau$, we can rephrase~(\ref{eq:aobsarray}) as: 
\beq
N p_N \hat{a}^{\textnormal{(obs)}}_{ij}=
\left\{
\begin{array}{l}
\underbrace{N p_N a^{\textnormal{(obs)}}_{ij}}_{\textnormal{not vanishing}}
+
\underbrace{N p_N e_{ij}}_{\textnormal{small quantity?}}, ~~~~\textnormal{ if } a^{\textnormal{(obs)}}_{ij}>0,\\
\\
\underbrace{N p_N e_{ij}}_{\textnormal{small quantity?}}, ~~~~~~~~~~~~~~~~~~~\textnormal{ if } a^{\textnormal{(obs)}}_{ij}=0,
\end{array}
\right.
\label{eq:aobsarray2}
\eeq
where the qualification of being ``not vanishing'' is a consequence of~(\ref{eq:AssumptionA1}). 
In light of~(\ref{eq:aobsarray2}), if we will able to show that $N p_N e_{ij}$ is {\em still} a small quantity, then $\hat{a}^{\textnormal{(obs)}}_{ij}$ would be effectively useful for tomography purposes, because the nonzero entry $Np_N a_{ij}^{\textnormal{(obs)}}$ would stand out from the error floor as $N$ gets large.
Actually, this heuristic argument will be made rigorous in the proof of Theorem~$2$, which is presented in the next section.

Before ending this section, we would like to introduce another useful combination policy that belongs to class $\mathscr{C}_\tau$, namely, the Metropolis rule, which is given by~\cite{SayedFoundTrends}:
\beq
a_{ij}=
\left\{
\begin{array}{l}
g_{ij}\,(1-\mu)/\max(d_i,d_j),\quad\textnormal{for } i\neq j,
\\
\\
(1-\mu) - \sum_{\ell\neq i} a_{i\ell},~~~~~~~~~\textnormal{for } i=j.
\end{array}
\right.
\label{eq:MetMatdef}
\eeq
Since $\max(d_i,d_j)\leq d_{\textnormal{max}}$, for the Metropolis rule we have the following implication, for all $i\neq j$:
\beqa
\lefteqn{
\{\bm{d}_{\textnormal{max}}< N p_N e \, | \, \bm{g}_{ij}=1\}
}\nonumber\\
&\Rightarrow&
\{\max(\bm{d}_i,\bm{d}_j)< N p_N e \, | \, \bm{g}_{ij}=1\}
\nonumber\\
&=&\{N p_N \bm{a}_{ij} > (1-\mu)/e \, | \, \bm{g}_{ij}=1\},
\label{eq:implicev}
\eeqa
where the latter equality comes from~(\ref{eq:MetMatdef}). 
Now, since for two events $\mathcal{E}_1$ and $\mathcal{E}_2$, the condition $\mathcal{E}_1\Rightarrow \mathcal{E}_2$ implies that $\P[\mathcal{E}_2]\geq \P[\mathcal{E}_1]$, from~(\ref{eq:implicev}) we can write: 
\beqa
\lefteqn{\P[N p_N \bm{a}_{ij} > (1-\mu)/e \, | \, \bm{g}_{ij}=1]}\nonumber\\
&\geq&
\P[\bm{d}_{\textnormal{max}} < N p_N e \, | \, \bm{g}_{ij}=1]\nonumber\\
&\geq&
1 - \left(e + \frac{2 e^2}{N}\right) e^{-c_N}\stackrel{N\rightarrow\infty}{\longrightarrow} 1,
\label{eq:lowbounonzeroMet}
\eeqa
where the latter inequality follows by Lemma~$1$.
Equation~(\ref{eq:lowbounonzeroMet}) reveals that the Metropolis rule satisfies~(\ref{eq:AssumptionA1}) with the choice $\tau=(1-\mu)/e$.

\section{Consistent Tomography}
In order to establish whether the estimated matrix, $\hat{A}^{\textnormal{(obs)}}$, can be used to infer the interaction pattern contained in $A^{\textnormal{(obs)}}$, we still need to provide a statistical characterization for the entries of $\hat{A}^{\textnormal{(obs)}}$.
We pursue this goal by characterizing the asymptotic behavior of two conditional distributions: the empirical distribution given that two agents do not interact, and the empirical distribution given that two agents interact.
In particular, owing to the $N p_N$ scaling factor that we have obtained in the previous section, we shall focus on the {\em scaled} matrix $N p_N \hat{A}^{\textnormal{(obs)}}$.

Let us preliminarily introduce the number of non-interacting ($\bm{N}_0$) and the number of interacting ($\bm{N}_1$) agent pairs over the observed set, defined respectively as:
\beq
\bm{N}_0\dfz\sum_{i=1}^K\sum_{j\neq i} (1-\bm{g}_{ij}^{\textnormal{(obs)}}),
\qquad
\bm{N}_1\dfz\sum_{i=1}^K\sum_{j\neq i} \bm{g}_{ij}^{\textnormal{(obs)}},
\label{eq:N0N1def}
\eeq
where $\bm{g}_{ij}^{\textnormal{(obs)}}=\mathbb{I}\{\bm{a}_{ij}^{\textnormal{(obs)}}>0\}$ is the $(i,j)$-the entry of the interaction sub-matrix corresponding to the observable agents.
Next we introduce the number of entries in $N p_N \hat{A}^{\textnormal{(obs)}}$ that stay below some positive level $\alpha$, for the case of non-interacting and interacting agent pairs, respectively:
\beqa
\bm{N}_0(\alpha)&\dfz&\sum_{i=1}^K\sum_{j\neq i} \mathbb{I}\{N p_N \bm{\hat{a}}_{i j}^{\textnormal{(obs)}}\leq\alpha, \,
\bm{g}_{i j}^{\textnormal{(obs)}}=0\},
\label{eq:N0obsemp}
\\
\bm{N}_1(\alpha)&\dfz&\sum_{i=1}^K\sum_{j\neq  i} \mathbb{I}\{N p_N \bm{\hat{a}}_{i j}^{\textnormal{(obs)}}\leq\alpha, \,
\bm{g}_{i j}^{\textnormal{(obs)}}=1\}.
\label{eq:N1obsemp}
\eeqa
Using~(\ref{eq:N0N1def}),~(\ref{eq:N0obsemp}) and~(\ref{eq:N1obsemp}), we can compute the conditional empirical distributions given that the agents are not interacting and given that they are interacting, defined respectively as:
\beq
\bm{\mathcal{F}}_0(\alpha)=\frac
{\bm{N}_0(\alpha)}
{\bm{N}_0},\qquad
\bm{\mathcal{F}}_1(\alpha)=\frac
{\bm{N}_1(\alpha)}
{\bm{N}_1},
\label{eq:E01indx}
\eeq
where $\bm{\mathcal{F}}_0(\alpha)$ (resp., $\bm{\mathcal{F}}_1(\alpha)$) are conventionally set to $1/2$ when $\bm{N}_0=0$ (resp., $\bm{N}_1=0$).
It is also convenient to introduce the complementary distribution $\bar{\bm{\mathcal{F}}}_1(\alpha)=1 - \bm{\mathcal{F}}_1(\alpha)$. 
Accordingly, the quantity $\bm{\mathcal{F}}_0(\alpha)$ represents the fraction of entries in $N p_N \hat{A}^{\textnormal{(obs)}}$ that correspond to {\em non-interacting} agent pairs and that stay {\em below} $\alpha$.
In contrast, the quantity $\bar{\bm{\mathcal{F}}}_1(\alpha)$ represents the fraction of entries in $N p_N \hat{A}^{\textnormal{(obs)}}$ that correspond to {\em interacting} agent-pairs and that stay {\em above} $\alpha$. 

The next theorem establishes achievability of consistent tomography through the asymptotic characterization of the aforementioned empirical distributions. 
\begin{theorem}[Achievability of consistent tomography]
Let the network interaction profile obey a $\mathscr{G}^\star(N,p_N)$ model, and let the combination policy belong to class $\mathscr{C}_\tau$. Then, for any nonzero fraction of observable agents satisfying~(\ref{eq:csidef}), we have in the limit as the network size increases ($N\rightarrow \infty$):
\beq
\boxed{
\bm{\mathcal{F}}_0(\epsilon)\stackrel{\textnormal{p}}{\longrightarrow} 1~~\forall \epsilon>0,\qquad
\bar{\bm{\mathcal{F}}}_1(\tau)\stackrel{\textnormal{p}}{\longrightarrow} 1
}
\label{eq:maintheo33claim}
\eeq
\end{theorem}
\begin{IEEEproof}
See Appendix~\ref{app:Theo3}.
\end{IEEEproof}

\begin{figure}[t]
\centerline{\includegraphics[width=.45\textheight]{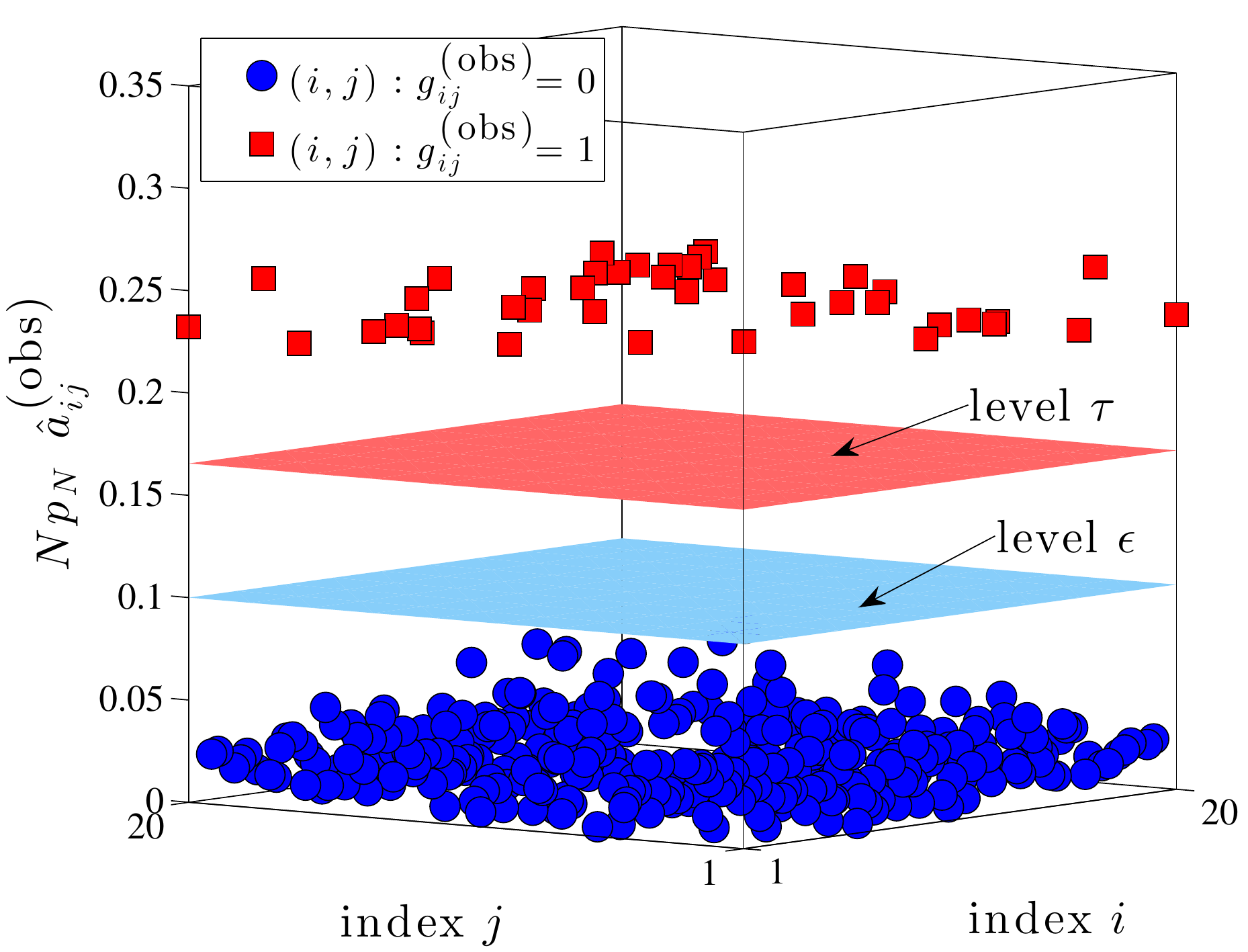}}
\caption{A pictorial illustration of Theorem~$2$, which establishes that the scaled estimated entries $\hat{a}_{ij}^{\textnormal{(obs)}}$ cluster in two groups depending on whether agents $i$ and $j$ are interacting or not.}
\label{fig:figpict}
\end{figure}

Theorem~$2$ reveals that, from the knowledge of the estimated combination matrix, $\hat{A}^{\textnormal{(obs)}}$, it is possible to separate the zero/nonzero entries of the true combination matrix, $A^{\textnormal{(obs)}}$. 
In fact, we see from Theorem~$2$ that the following dichotomous behavior is observed, asymptotically with $N$: 
$i)$ when agents $i$ and $j$ are {\em not} interacting, most of the (scaled) estimated matrix entries, $N p_N \hat{a}^{\textnormal{(obs)}}_{ij}$, stay below {\em any} level $\epsilon$ (and, hence, also below an arbitrarily small value $\epsilon<\tau$); $ii)$ when agents $i$ and $j$ are interacting, most of the (scaled) estimated matrix entries, $N p_N \hat{a}^{\textnormal{(obs)}}_{ij}$, stay above a positive value $\tau$.
Therefore, a separation between the two classes of non-interacting and interacting agent pairs arises, which translates into the emergence of two separate clusters, one corresponding to the region $N p_N \hat{a}^{\textnormal{(obs)}}_{ij}\leq \epsilon$, and the other one corresponding to the region $N p_N \hat{a}^{\textnormal{(obs)}}_{ij}> \tau$.
This situation is illustrated in Fig.~\ref{fig:figpict}. 

Theorem~$2$ establishes the separability of the two classes based upon knowledge of $\hat{A}^{\textnormal{(obs)}}$, whose computation requires knowledge of the {\em exact} correlation sub-matrices, $[R_0]_\Omega$ and $[R_1]_\Omega$ --- see~(\ref{eq:A11_first}). 
Since, in principle, such matrices can be estimated with arbitrarily large precision as the steady-state is approached (i.e., as the number of collected diffusion output samples increases) the result in Theorem~$2$ is an {\em achievability} result.
In addition, for many of the methods available to estimate correlation matrices, the behavior of the estimation error is known, at least for $n$ sufficiently large. 
A useful extension of the present treatment would be examining the interplay between the two sources of error, namely, the error caused by partial observations, and the error caused by estimation of the correlation sub-matrices.

When some prior knowledge about the value of $\tau$ and the typical number of neighbors, $N p_N$, is available, Theorem~$2$ provides a constructive recipe to perform the reconstruction of the interaction profile across the observed network. 
In fact, the separation between the two classes of interacting and non-interacting agent pairs can be performed by comparing each entry $\hat{a}^{\textnormal{(obs)}}_{ij}$ to an intermediate threshold lying between $0$ and $\tau/(N p_N)$. 

On the other hand, in many contexts it is unrealistic to assume precise knowledge of the parameters $\tau$ and $N p_N$. 
When such knowledge is not available, the aforementioned reconstruction strategy is not applicable, but the achievability result in Theorem~$2$ still has a fundamental implication in that it guarantees the existence of the clustering structure! 
The existence of two separate thresholds, $\epsilon<\tau$, and the related (asymptotic) separation of the scaled estimated entries in two separate clusters ($N p_N \hat{a}^{\textnormal{(obs)}}_{ij}\leq\epsilon$ and $N p_N \hat{a}^{\textnormal{(obs)}}_{ij}>\tau$), opens up the possibility of employing universal and non-parametric pattern recognition strategies to perform cluster separation.
In particular, in our numerical experiments, we shall verify the validity of this argument by employing the $k$-means clustering algorithm.

\section{Higher Order Asymptotic Analysis}
As illustrated in the previous section, and as we can infer from Fig.~\ref{fig:figpict}, it is undesirable to have $N p_N \hat{a}^{\textnormal{(obs)}}_{ij} > \epsilon$ when $\bm{g}_{ij}=0$.
For the sake of brevity, let us refer to the agent pair for which such event occurs as being identified as ``mistakenly-interacting''.
Examining~(\ref{eq:N0N1def}),~(\ref{eq:N0obsemp}) and~(\ref{eq:E01indx}), we deduce that the number of mistakenly-interacting pairs is given by:
\beq
\bm{N}_0 - \bm{N}_0(\epsilon)=\bm{N}_0 [1 - \bm{\mathcal{F}}_0(\epsilon)]
\approx N^2 \underbrace{[1 - \bm{\mathcal{F}}_0(\epsilon)]}_{\rightarrow 0,\textnormal{ see~(\ref{eq:maintheo33claim})}},
\label{eq:falseN1}
\eeq
where we used the fact that, under the $\mathscr{G}^\star(N,p_N)$ model, $\bm{N}_0$ scales as $N(N-1)(1-p_N)$, and $p_N$ vanishes as $N\rightarrow\infty$.

On the other hand, for the number of {\em truly-interacting} agent pairs we have that:
\beq
\bm{N}_1\approx N^2 p_N.
\label{eq:trueN1}
\eeq 
The fact that $p_N$ vanishes as $N$ increases causes the following issue.
According to~(\ref{eq:falseN1})--(\ref{eq:trueN1}) and also to~(\ref{eq:oneminusF0})--(\ref{eq:E0bound}), without any information about the speed of convergence of $1 - \bm{\mathcal{F}}_0(\epsilon)$ (in comparison to $p_N$), we cannot exclude that the number of {\em mistakenly-interacting} pairs is larger than the number of {\em truly-interacting} pairs.
In order to ward off the presence of such unpleasant feature, we should prove that the quantity $1 - \bm{\mathcal{F}}_0(\epsilon)$ goes to zero faster than $p_N$, formally:
\beq
\frac{1 - \bm{\mathcal{F}}_0(\epsilon)}{p_N}\stackrel{\textnormal{p}}{\longrightarrow} 0,
\eeq 
a condition that will be abbreviated as $1 - \bm{\mathcal{F}}_0(\epsilon)=o_p(p_N)$.
Verification of such desired requirement is addressed in the forthcoming Proposition~$1$. 
In order to prove Proposition~$1$, we require that the combination policy possesses two additional regularity properties, which are again automatically satisfied by the Laplacian and Metropolis rules. 
 
{\em Property P$1$.} There exists $\kappa>0$ such that, for all $i,j=1,2,\dots,N$, with $i\neq j$: 
\beq
\boxed
{
\bm{a}_{ij}
\leq
\frac{\kappa}{\bm{d}_i}\,\bm{g}_{ij}
}
\label{eq:AssumptionA1bis}
\eeq
with the inequality being trivially an equality for $\bm{g}_{ij}=0$.
$\hfill\square$

It is useful to make the following comparison between~(\ref{eq:AssumptionA1}) and~(\ref{eq:AssumptionA1bis}).
Equation~(\ref{eq:AssumptionA1}) means that the nonzero entries in the combination matrix, scaled by $N p_N$, do not collapse to zero, i.e., they are stable {\em from below}.
On the other hand, in view of~(\ref{eq:binoconv}), the upper bound in~(\ref{eq:AssumptionA1bis}) implies that the nonzero entries in the combination matrix, scaled by $N p_N$, are asymptotically stable {\em from above}. This is because $\bm{d}_i$ approaches $N p_N$ asymptotically as $N\rightarrow \infty$ so that~(\ref{eq:AssumptionA1bis}) translates into the asymptotic condition that $N p_N \bm{a}_{ij} \leq \kappa$.

Moreover, it is straightforward to verify that~(\ref{eq:AssumptionA1bis}) holds for the Laplacian rule, with $\kappa=(1-\mu)\lambda$, as well as for the Metropolis rule, with $\kappa=1-\mu$.

{\em Property P$2$.} 
Consider a certain permutation of the agents, i.e., a renumbering of the rows and columns of the interaction matrix $G$, which leads to the interaction matrix $\tilde{G}$. Such operation can be represented by means of a permutation matrix $P$ (see Appendix~\ref{app:permprop}), as $\tilde{G}=P G P^T$.
Property P$2$ holds if:
\beq
\gamma(P G P^T) = P \,\gamma(G)\, P^T,
\label{eq:propertyP2def}
\eeq
namely, if applying the combination policy $\gamma(\cdot)$ to the {\em renumbered} interaction matrix is equivalent to renumbering (through the same $P$) the initial combination matrix $A=\gamma(G)$.~$\hfill\square$

\begin{figure}[t]
\centerline{\includegraphics[width=.45\textheight]{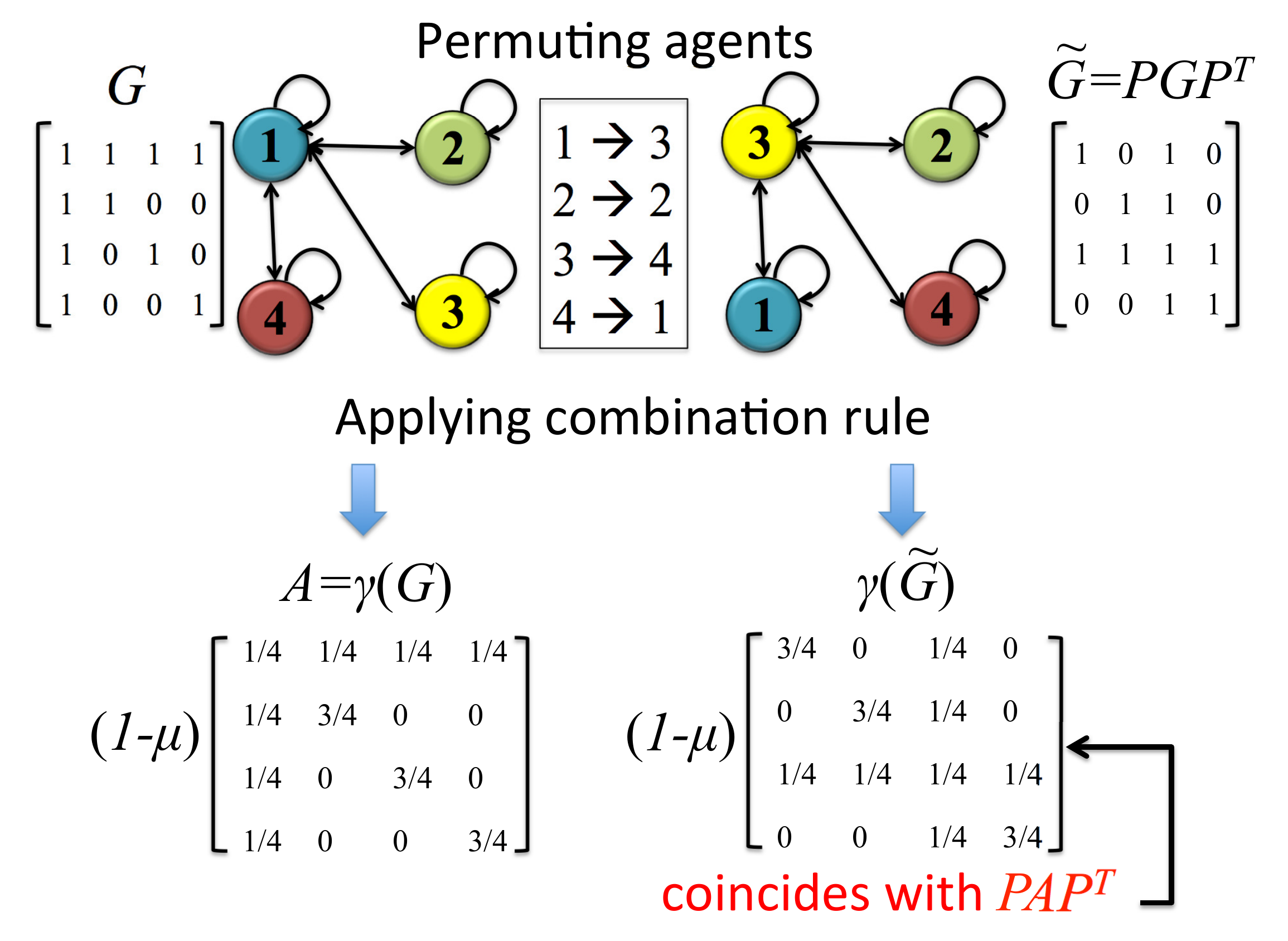}}
\caption{Illustration of property P$2$: permuting the agents, leads to a permuted version of the combination policy.}
\label{fig:figperm}
\end{figure}

The meaning of property P$2$ is illustrated in Fig.~\ref{fig:figperm}.
In the leftmost panel, we represent a network graph, along with the corresponding combination matrix obtained by applying the Metropolis rule. 
In the rightmost panel, the agents are exchanged as detailed in the figure, using the permutation matrix:
\beq
P=\left[\begin{array}{cccc}0 & 0 & 0 & 1 \\0 & 1 & 0 & 0 \\1 & 0 & 0 & 0 \\0 & 0 & 1 & 0\end{array}\right]
\label{eq:permatinfigure}
\eeq
Then, the Metropolis rule is applied to the new (i.e., renumbered) interaction matrix. 
It is seen that the resulting combination matrix corresponds to renumbering the original combination matrix using the permutation~(\ref{eq:permatinfigure}). 

Property P$2$ is particularly relevant for the following reasons.
First, under the Erd\"os-Renyi model, the statistical distribution of the interaction matrix is invariant to agents' permutations. 
Owing to~(\ref{eq:propertyP2def}), such invariance is automatically inherited by the combination matrix.   
Moreover, property P$2$ is satisfied by typical combination policies, such as the Laplacian rule and the Metropolis rule. 
It is also useful to provide a counterexample that shows why property P$2$ is not always verified. 
Consider a network with three agents, and with the following interaction matrix:
\beq
G=\left[\begin{array}{ccc}1 & 1 & 0 \\1 & 1 & 1 \\0 & 1 & 1\end{array}\right]
\label{eq:intmat1}
\eeq
The combination policy of our counterexample works as follows. 
First, a Metropolis rule is applied. 
Second, the resulting self-combination weight assigned to agent $1$ is slightly increased by adding a small extra-weight. 
The extra-weight assigned to agent $1$ is then subtracted, in equal parts, from the other nonzero weights of agent $1$, in order to meet the requirement $\sum_{\ell\neq 1} a_{1\ell}=1-\mu$. 
Finally, the other entries of the combination matrix are adjusted so as to make $A/(1-\mu)$ symmetric and right-stochastic. 
The final result is (with $\epsilon\ll 1$):
\beq
A=(1-\mu)\left[\begin{array}{ccc} 2/3+\epsilon & 1/3-\epsilon & 0 \\1/3-\epsilon & 1/3+\epsilon & 1/3 \\0 & 1/3 & 2/3\end{array}\right]
\label{eq:Amat1}
\eeq
Assume now that agents $1$ and $2$ are exchanged, which would yield the following interaction matrix:
\beq
G=\left[\begin{array}{ccc}1 & 1 & 1 \\1 & 1 & 0 \\1 & 0 & 1\end{array}\right]
\label{eq:intmat2}
\eeq
Applying the combination policy described before, we would end up with the following combination matrix:
\beq
A=(1-\mu)\left[\begin{array}{ccc} 1/3+\epsilon & 1/3-\epsilon/2 & 1/3-\epsilon/2 \\
1/3-\epsilon/2 & 2/3+\epsilon/2 & 0 \\
1/3-\epsilon/2 & 0 & 2/3+\epsilon/2\end{array}\right]
\label{eq:Amat2}
\eeq 
We see that: $i)$ the interaction matrix in~(\ref{eq:intmat2}) is a renumbered version of the interaction matrix in~(\ref{eq:intmat1}), which corresponds to exchanging agents $1$ and $2$, while $ii)$ the combination matrix in~(\ref{eq:Amat2}) is {\em not obtained by applying the same renumbering} to the combination matrix in~(\ref{eq:Amat1}).  
Therefore, property P$2$ is violated.  
The presented counterexample shows that, while in practice constructing a combination policy that violates property P$2$ might look rather artificial, from a purely theoretical viewpoint it must be stated that property P$2$ is not always verified.


A combination policy satisfying the additional properties P$1$ and P$2$ describes the following class of combination policies.

{\em Combination-policy class $\mathscr{C}_\tau^\prime$.}
A combination policy belongs to class $\mathscr{C}_\tau^\prime$ if it belongs to class $\mathscr{C}_\tau$ {\em and} possesses properties P$1$ and P$2$.
$\hfill\square$

Proposition~$1$, which is stated below, will be proved by resorting to an approximation that will be referred to as the {\em independence approximation}. 
More specifically, the proof of the proposition will rely on characterizing the variance of the entries in the error matrix (see Appendix~\ref{app:prop1}). 
Unfortunately, the exact evaluation of the error variance is generally a formidable task. 
In order to gain insight into the asymptotic behavior of the variance, in the proof of Proposition~$1$ we make a simplified evaluation by treating the entries of the various matrices involved in the calculations as {\em independent} random variables. 
The accuracy of the results arising from this approximation is tested by means of numerical experiments --- see Fig.~\ref{fig:varver} in Appendix~\ref{app:prop1}. 
A more rigorous analysis would require estimating the order of the error introduced by the independence approximation.

\begin{proposition}[Rate of convergence of $1 - \bm{\mathcal{F}}_0(\epsilon)$]
Let the network interaction profile obey a $\mathscr{G}^\star(N,p_N)$ model, and let the combination policy belong to class $\mathscr{C}_\tau^\prime$. 
Then, the result in~(\ref{eq:maintheo33claim}) holds true because $\mathscr{C}_\tau^\prime\subset \mathscr{C}_\tau$.
In addition, for any nonzero fraction of observable agents satisfying~(\ref{eq:csidef}), and for all $\epsilon>0$, we have:
\beq
\boxed{
1 - \bm{\mathcal{F}}_0(\epsilon)\approx o_p(p_N)
}
\label{eq:propclaim}
\eeq
where the approximation in~(\ref{eq:propclaim}) arises from the independence approximation used in the proof.
\end{proposition}

\begin{IEEEproof}
See Appendix~\ref{app:prop1}.
\end{IEEEproof}

\section{Illustrative Examples}
We now examine the practical significance of the asymptotic results derived in the previous sections, with reference to three combination matrices that are rather popular in the literature of adaptive networks.
The first two strategies lead to symmetric combination matrices, which therefore match the hypotheses of our theorems. The third strategy corresponds to an asymmetric combination matrix. Even if the asymmetric case is not contemplated by our theorems, it is relevant in practice and, as we shall see from the forthcoming experiments, consistent tomography works also for such a relevant case. The presentation of the examples is organized through the following steps. 
\begin{itemize}
\item[S1)]
We consider first the case that the projections of the correlation matrices, $R_0$ and $R_1$, onto the observable part of the network, are available without error. For this case, we compute the inversion of the observable part, which leads to the matrix $\hat A^{\textnormal{(obs)}}$, through:
\beq
\boxed{
\hat A^{\textnormal{(obs)}}=R_1^{\textnormal{(obs)}}(R_0^{\textnormal{(obs)}})^{-1}
}
\label{eq:perfknow}
\eeq 
\item[S2)]
We use the off-diagonal entries of $\hat A^{\textnormal{(obs)}}$, and apply a $k$-means clustering algorithm in order to split these entries into two clusters. The cluster with smallest arithmetic average is labeled as ``non-interacting'', while the other cluster is labeled as ``interacting''. We remark that such {\em classification strategy is implemented in a universal, fully non-parametric way}.
\item[S3)]
Then, we enlarge the setting to the case that the projections of the correlation matrices are estimated from the diffusion outputs. 
In the simulations, the observations fed into the diffusion algorithm, $\{\bm{x}_i(n)\}_{i,n}$, follow a standard normal distribution.
As an estimator for $R_0^{\textnormal{(obs)}}\dfz[R_0]_{\Omega}$, we use the empirical correlation, namely:
\beq
\hat R_0^{\textnormal{(obs)}}=\frac 1 n Y Y^T,
\label{eq:Rest1}
\eeq
where, for $i=1,2,\dots,K$, the $i$-th row of the $K\times n$ matrix $Y$ is given by: 
\beq
\bm{y}_{\omega_i}(1),\bm{y}_{\omega_i}(2),\dots,\bm{y}_{\omega_i}(n),
\label{eq:Rest2}
\eeq 
with the indices $\omega_1<\omega_2<\dots<\omega_K$ spanning the observable set $\Omega$.
The estimate $\hat{R}_1^{\textnormal{(obs)}}$ is computed similarly. 
We remark that in this work we do not focus on optimizing the performance of the correlation matrix estimators, since our focus is on ascertaining the fundamental limits of tomography. 
There are already considerable works in the literature on perfecting correlation estimation from ensemble data. One challenge that arises in the network case is the interplay between the matrix size and the number of samples used for the estimation. 

Now, using~(\ref{eq:perfknow}) with the true correlation matrices replaced by their estimated counterparts, we get the following estimate:
\beq
\boxed{
\hat A^{\textnormal{(obs)}}=\hat{R}_1^{\textnormal{(obs)}}(\hat{R}_0^{\textnormal{(obs)}})^{-1}
}
\label{eq:imperfknow}
\eeq 
\item[S4)]
We run the $k$-means clustering algorithm over the entries of $\hat{A}^{\textnormal{(obs)}}$ in~(\ref{eq:imperfknow}).   
\end{itemize}

\subsection{Laplacian Combination Rule}
\label{sec:Laplacian}
Under the Laplacian combination rule, the off-diagonal combination weights are zero when the agents are not interacting, and are otherwise equal to a constant across the network. 
Therefore, we see that the weights reflect perfectly the structure of the underlying network graph. 
In fact, several important properties of the graph are encoded in the properties of the Laplacian matrix~\cite{TsitsiveroBarbarossaDiLorenzo2016}. 

\begin{figure*}
{\centering{\bf ~~~~~~Laplacian combination rule}\par\medskip}
\begin{minipage}{.33\linewidth}
\centering
{
\includegraphics[scale=.28]{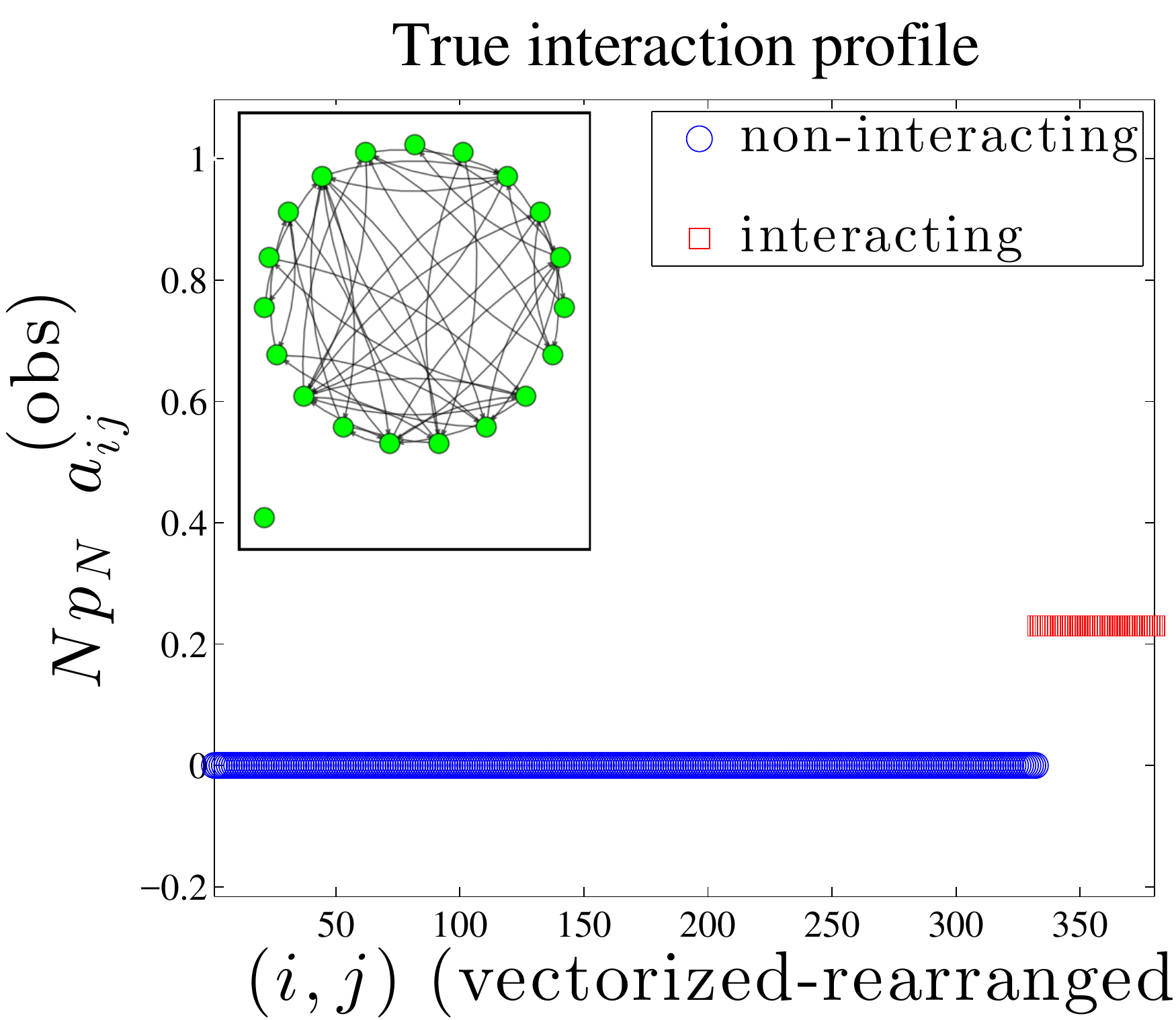}}
\end{minipage}
\begin{minipage}{.33\linewidth}
\centering
{
\includegraphics[scale=.28]{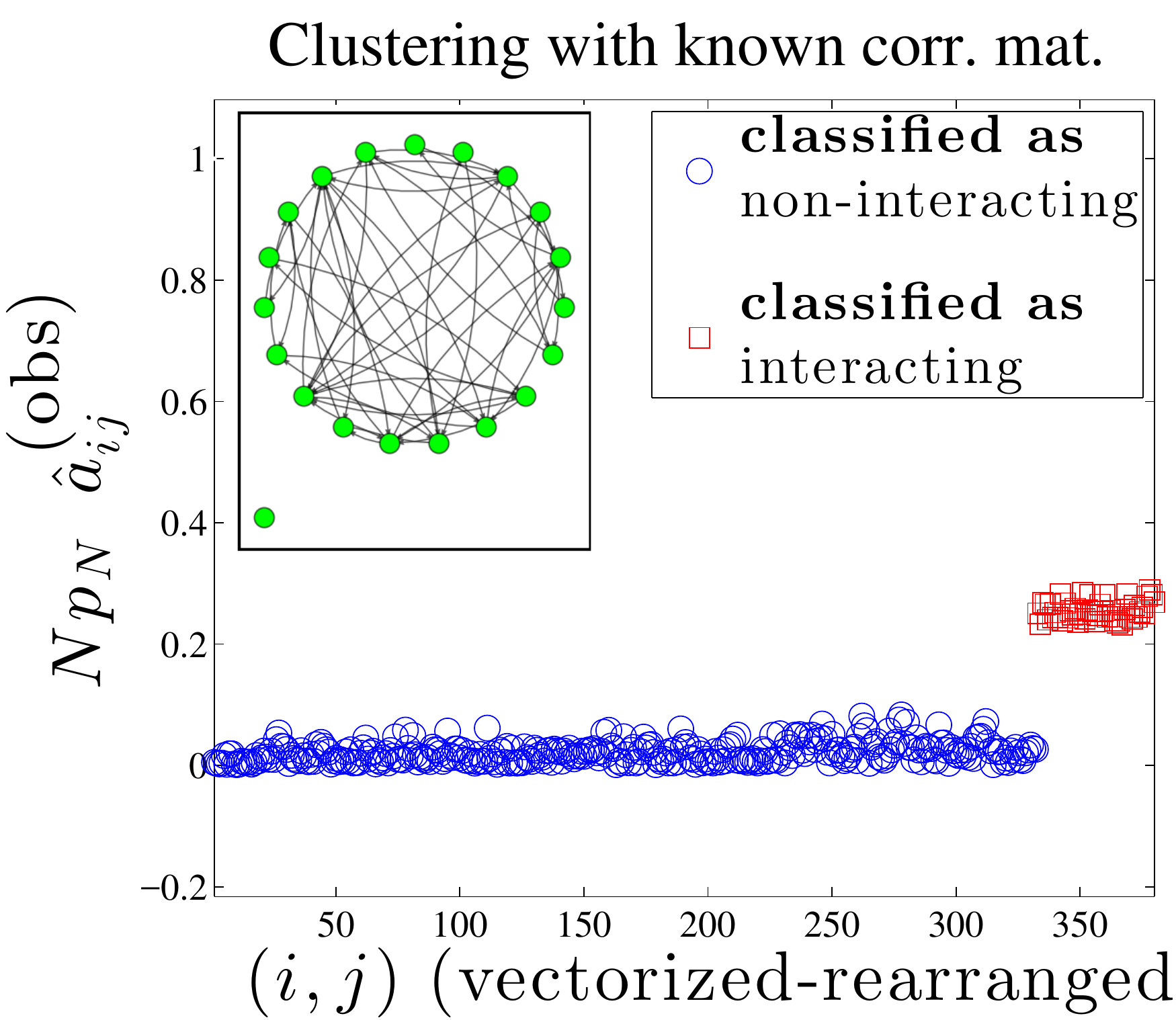}}
\end{minipage}
\begin{minipage}{.33\linewidth}
\centering
{
\includegraphics[scale=.28]{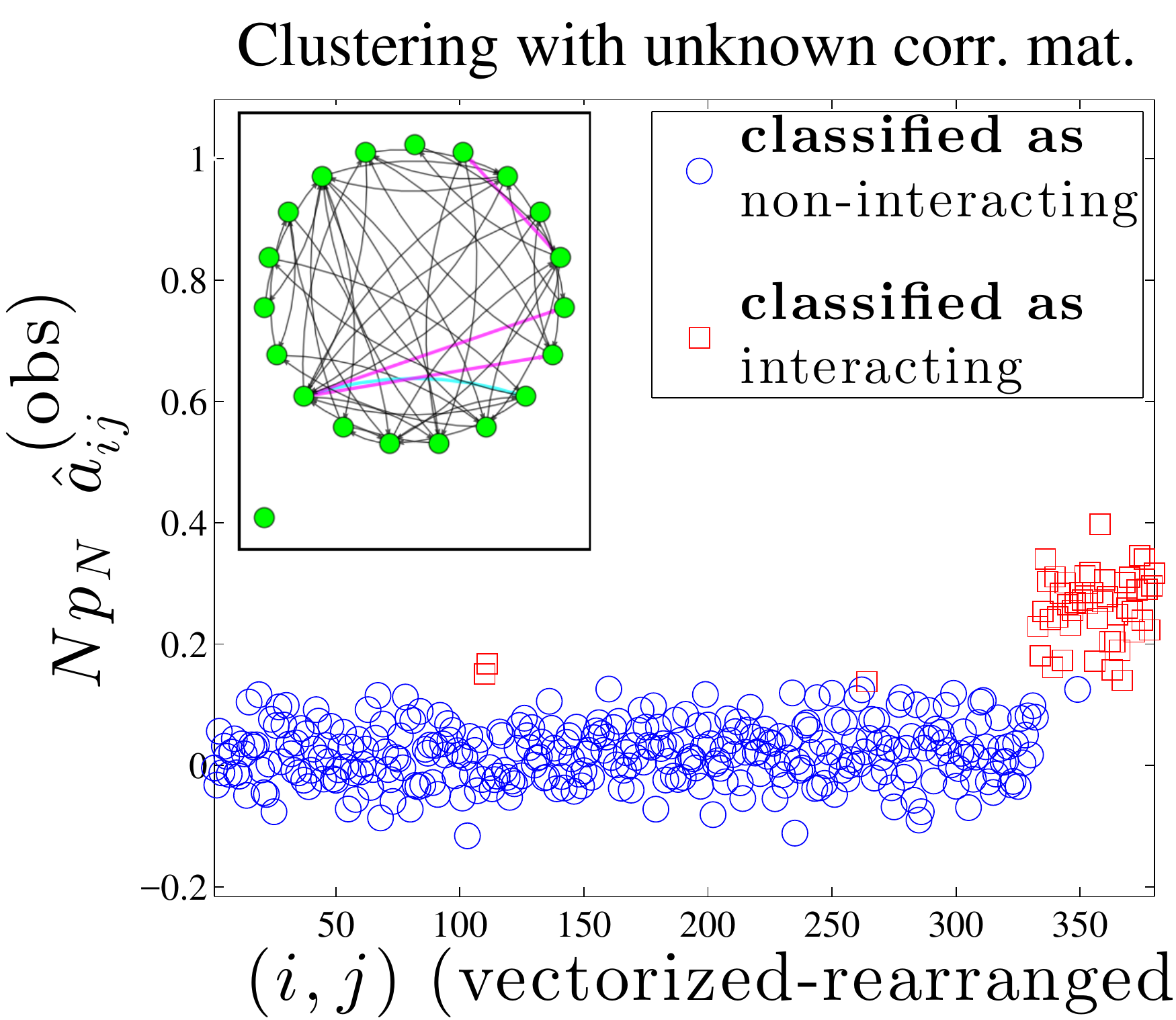}}
\end{minipage}
\caption{Network tomography for the case of a Laplacian combination rule with parameter $\lambda=0.5$ --- see Sec.~\ref{sec:Laplacian}.
The network consists of $N=100$ agents, where only $K=20$ agents are observable ($\xi=0.2$). 
The interaction probability is $p_N = 2\,(\ln N)/N\approx 0.092$, and the value of the step-size is $\mu=0.1$. 
{\bf Leftmost panel}: the {\em true} combination matrix, $A^{\textnormal{(obs)}}$, is vectorized with column-major ordering, and the (vectorized) $(i,j)$ pairs are rearranged in such a way that the zero entries come before the nonzero entries (the same ordering is applied to the estimated matrix, $\hat{A}^{\textnormal{(obs)}}$, in the other two panels). 
The different markers highlight the {\em true} interaction profile of the {\em observable} network portion. 
{\bf Middle panel}: matrix $\hat{A}^{\textnormal{(obs)}}$, computed under perfect knowledge of the steady-state correlation matrices of the {\em observable} diffusion output,  see~(\ref{eq:perfknow}). The different markers highlight the interaction profile as {\em reconstructed} by the $k$-means algorithm. 
{\bf Rightmost panel}: same as middle panel, with matrix $\hat{A}^{\textnormal{(obs)}}$ computed using the correlation matrices {\em estimated empirically} with $n=2 \times 10^4$ samples, see~(\ref{eq:Rest1}),~(\ref{eq:Rest2}), and~(\ref{eq:imperfknow}).
{\bf Inset plots}: interaction profiles represented through the corresponding network graphs. 
In the inset plots of the middle and of the rightmost panels, erroneously detected edges (the edge is not present but it is ``seen'' by the tomography algorithm) are marked in magenta, while missed edged (the edge is present but the tomography algorithm misses it) are marked in cyan.
}
\label{fig:Laplacian}
\end{figure*}

In Fig.~\ref{fig:Laplacian}, leftmost panel, we display the off-diagonal entries of the (scaled) {\em true} combination matrix, $N p_N A^{\textnormal{(obs)}}$, corresponding to the {\em observable} network portion.
For clarity of presentation, the matrix has been vectorized by means of column-major ordering, and the (vectorized) $(i,j)$ pairs have been rearranged in such a way that the zero entries appear before the nonzero entries. 
The same ordering used for $A^{\textnormal{(obs)}}$ will be then applied to the matrices displayed in the remaining panels, i.e., the horizontal axis is homogeneous across the different panels, so as to allow a fair comparison.
If agents $i$ and $j$ do not interact, the pertinent matrix entry is marked with a blue circle, whereas if they do interact, a red square is used. The observed step-function behavior comes from the fact that, for the Laplacian combination rule, the nonzero weights are constant across the network. 

In the middle panel we focus on steps S1) and S2): we display the scaled estimated matrix, $N p_N \hat{A}^{\textnormal{(obs)}}$, computed under perfect knowledge of $R^{\textnormal{(obs)}}_0$ and $R^{\textnormal{(obs)}}_1$, and we display the classification performed by the $k$-means algorithm.
In the rightmost panel we focus instead on steps S3) and S4): we display the scaled estimated matrix, $N p_N \hat{A}^{\textnormal{(obs)}}$, computed with the {\em empirical estimates} $\hat{R}^{\textnormal{(obs)}}_0$ and $\hat{R}^{\textnormal{(obs)}}_1$, and we display the classification performed by the $k$-means algorithm.
In the latter two panels, matrix entries are marked with different colors and symbols, depending on the results of the $k$-means clustering algorithm: blue-circle markers if agents $i$ and $j$ have been {\em classified as} non-interacting, and red-square markers if agents $i$ and $j$ have been {\em classified as} interacting.
 
The network considered in Fig.~\ref{fig:Laplacian} consists of $N=100$ agents, where only $K=20$ agents are observable ($\xi=0.2$). According to the connection properties of the Erd\"os-R\'enyi model, see~ (\ref{eq:pconn}), the interaction probability is chosen as $p_N = 2\,(\ln N)/N\approx 0.092$. 
The parameter of the Laplacian combination matrix is $\lambda=0.5$.
As regards the diffusion algorithm, we choose a typical value for the step-size, i.e., $\mu=0.1$.  

Let us begin with examining the output of steps S1) and S2), middle panel. 
As we can see, the experiments confirm {\em in toto} what is expected from the theoretical analysis: the entries of the matrix $\hat A^{\textnormal{(obs)}}$ appear to be well-separable, since: $i)$ the unavoidable error induced by {\em partial} observation of the network, is relatively small, implying that zero entries of $\hat A^{\textnormal{(obs)}}$ are concentrated around zero; $ii)$ the nonzero entries of the {\em true} combination matrix (leftmost panel) are bounded from below, and, since the error $e_{ij}$ is positive, this fact keeps the nonzero entries of $\hat A^{\textnormal{(obs)}}$ (middle panel) sufficiently far away from the origin.

Next we move on to steps S3) and S4), namely, we move on to examining the behavior in the presence of imperfect knowledge of the correlation matrices. In particular, we use $n=2 \times 10^4$ samples to perform estimation of the correlation matrix from the diffusion output. 
This situation is examined in the rightmost panel of Fig.~\ref{fig:Laplacian}. By comparison with the middle panel, we see that the estimated clusters are more ``noisy'', which makes sense since the procedure {\em must} be affected by the error in estimating $R^{\textnormal{(obs)}}_0$ and $R^{\textnormal{(obs)}}_1$. 
This notwithstanding, tomography is still effective, meaning that the number of observations collected to estimating the correlation matrices is sufficiently high. Few classification errors are committed: the red-square markers appearing among the blue-circle markers, and the blue-circle marker appearing among the red-square markers (approximately at position $350$), represent mistakenly classified agent pairs.

The results arising from the above example are collected, perhaps in a more revealing form, in the inset plots of Fig.~\ref{fig:Laplacian}. 
The displayed network graphs (corresponding only to the observable subnet) are drawn with the following rules. 
An edge drawn from $j$ to $i$ means that agent $i$ {\em is} influenced (leftmost panel) or {\em is estimated to be} influenced (middle and rightmost panels) by agent $j$. 
When an edge is erroneously detected (i.e., the edge is in fact not present but the tomography algorithm ``sees'' it), it is marked in magenta. Likewise, when an edge is not detected (i.e., the edge is present but the tomography algorithm misses it), it is marked in cyan.
The impact of imperfect knowledge of the correlation matrices can be noticed in the inset plot, where we see that some errors are committed in reconstructing the profile of the observed subnet.

\begin{figure*}
{\centering{\bf ~~~~~~Metropolis combination rule}\par\medskip}
\begin{minipage}{.33\linewidth}
\centering
{
\includegraphics[scale=.28]{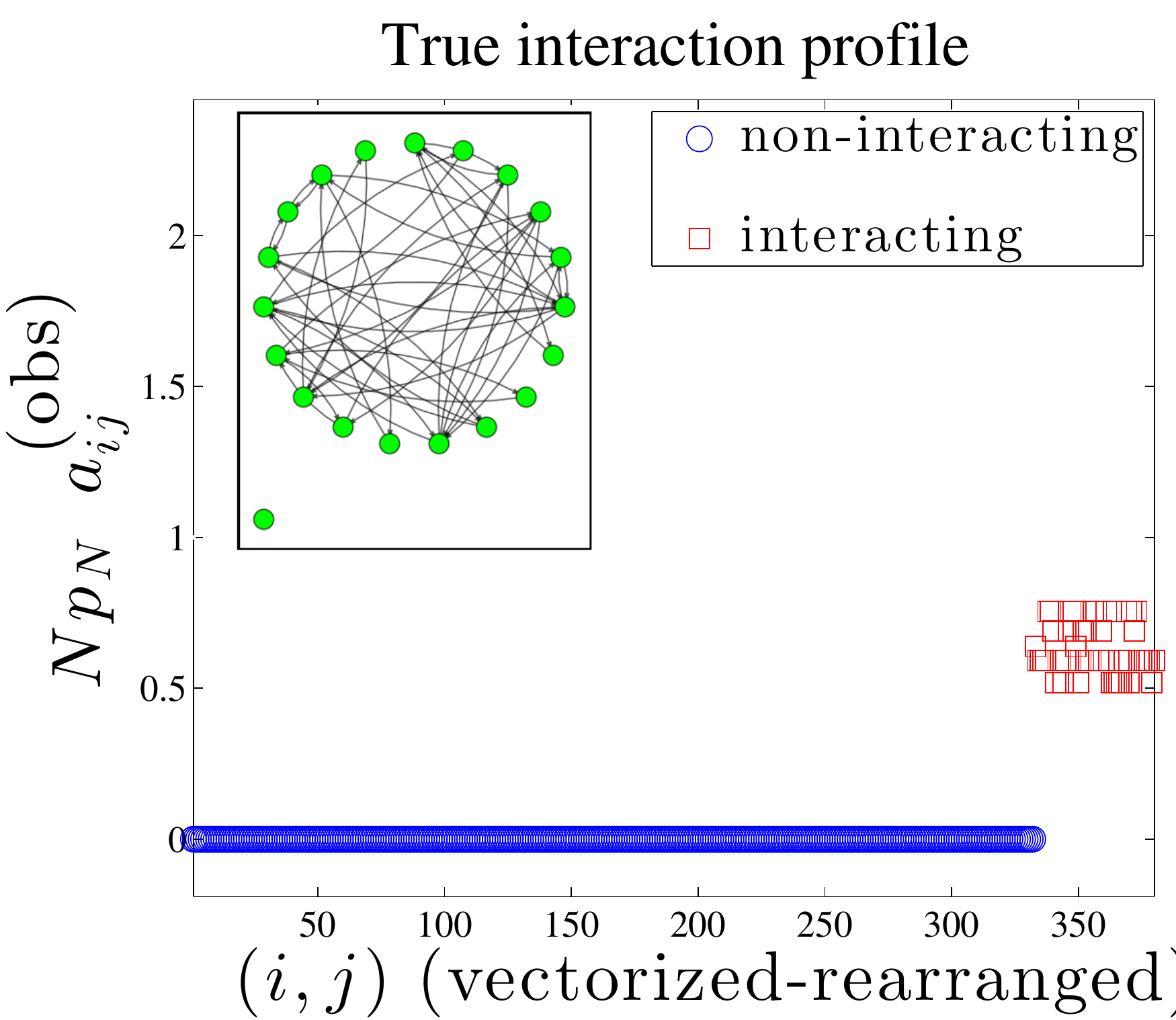}}
\end{minipage}
\begin{minipage}{.33\linewidth}
\centering
{
\includegraphics[scale=.28]{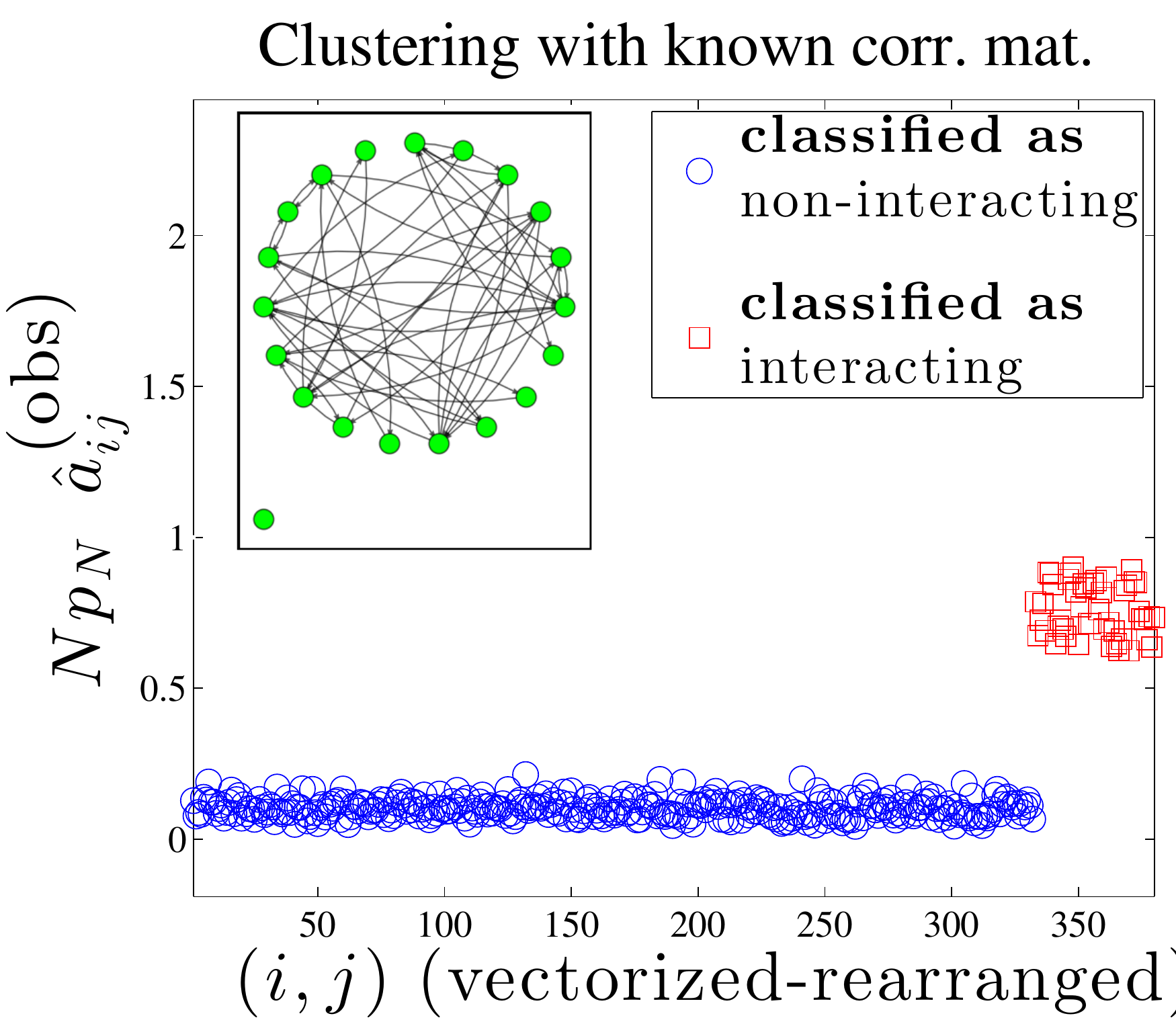}}
\end{minipage}
\begin{minipage}{.33\linewidth}
\centering
{
\includegraphics[scale=.28]{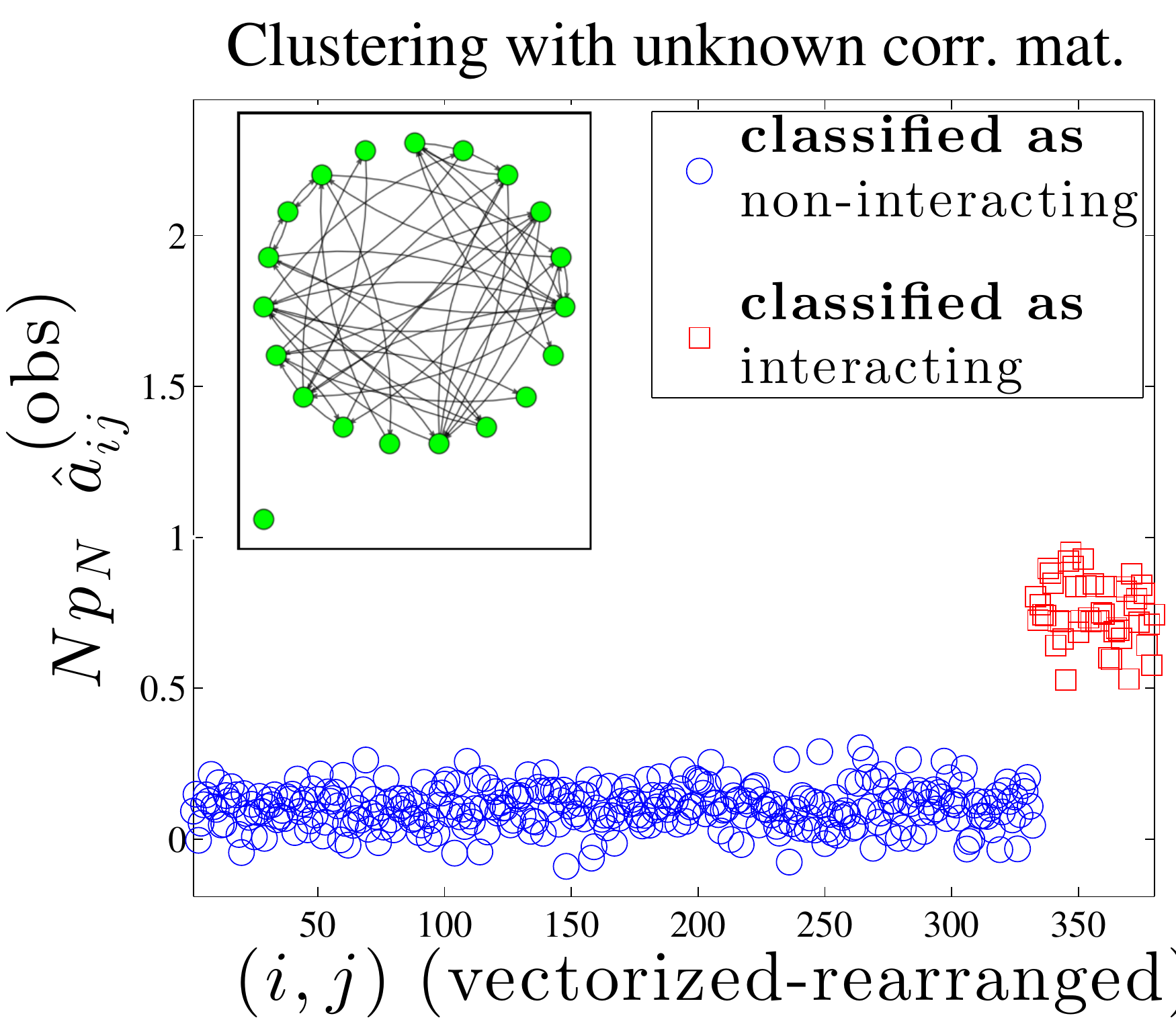}}
\end{minipage}
\caption{
Network tomography for the case of a Metropolis combination rule addressed in Sec.~\ref{sec:Metropolis}.
The network consists of $N=100$ agents, where only $K=20$ agents are observable ($\xi=0.2$). 
The interaction probability is $p_N = 2\,(\ln N)/N\approx 0.092$, and the value of the step-size is $\mu=0.1$. 
{\bf Leftmost panel}: the {\em true} combination matrix, $A^{\textnormal{(obs)}}$, is vectorized with column-major ordering, and the (vectorized) $(i,j)$ pairs are rearranged in such a way that the zero entries come before the nonzero entries (the same ordering is applied to the estimated matrix, $\hat{A}^{\textnormal{(obs)}}$, in the other two panels). 
The different markers highlight the {\em true} interaction profile of the {\em observable} network portion. 
{\bf Middle panel}: matrix $\hat{A}^{\textnormal{(obs)}}$, computed under perfect knowledge of the steady-state correlation matrices of the {\em observable} diffusion output,  see~(\ref{eq:perfknow}). The different markers highlight the interaction profile as {\em reconstructed} by the $k$-means algorithm. 
{\bf Rightmost panel}: same as middle panel, with matrix $\hat{A}^{\textnormal{(obs)}}$ computed using the correlation matrices {\em estimated empirically} with $n=2 \times 10^4$ samples, see~(\ref{eq:Rest1}),~(\ref{eq:Rest2}), and~(\ref{eq:imperfknow}).
{\bf Inset plots}: interaction profiles represented through the corresponding network graphs. 
In the inset plots of the middle and of the rightmost panels, erroneously detected edges (the edge is not present but it is ``seen'' by the tomography algorithm) are marked in magenta, while missed edged (the edge is present but the tomography algorithm misses it) are marked in cyan.
}
\label{fig:Metropolis}
\end{figure*}

\subsection{Metropolis Combination Rule}
\label{sec:Metropolis}
In Fig.~\ref{fig:Metropolis}, the analysis presented in the previous section is applied to a different combination matrix, namely, the Metropolis combination matrix. 
This example is useful because, in a Metropolis rule, the nonzero weights are no longer constant, and exhibit a certain dynamics related to the different neighborhood sizes corresponding to the different agents. 
It is important to examine the impact of such dynamics on the performance of the tomography algorithm. 
From the leftmost panel of Fig.~\ref{fig:Metropolis}, we see that the nonzero entries of the {\em true} combination matrix exhibit a certain variability. This notwithstanding, since the Metropolis rule belongs to class $\mathscr{C}_\tau$, with $\tau=(1-\mu)/e$, the scaled {\em nonzero} entries are still lower bounded in probability. The separability of the clusters is preserved, and considerations similar to those drawn in the previous section as regards the Laplacian rule apply. 

Still, there is an important difference between the two kinds of combination policies. 
By inspecting carefully the rightmost panel in Fig.~\ref{fig:Metropolis} and the related inset, we see that, differently from what happened in the rightmost panel of Fig.~\ref{fig:Laplacian}, the network profile is reconstructed perfectly. This suggests that estimation of the correlation matrices is easier for Metropolis rules than for Laplacian rules. 
The latter effect could be probably ascribed to the improved convergence properties of Metropolis combination matrices (with respect to Laplacian combination matrices), which should allow a more accurate estimation of the correlation matrices for a given number of samples.

\begin{figure*}
{\centering{\bf ~~~~~~Uniform averaging combination rule}\par\medskip}
\begin{minipage}{.33\linewidth}
\centering
{
\includegraphics[scale=.28]{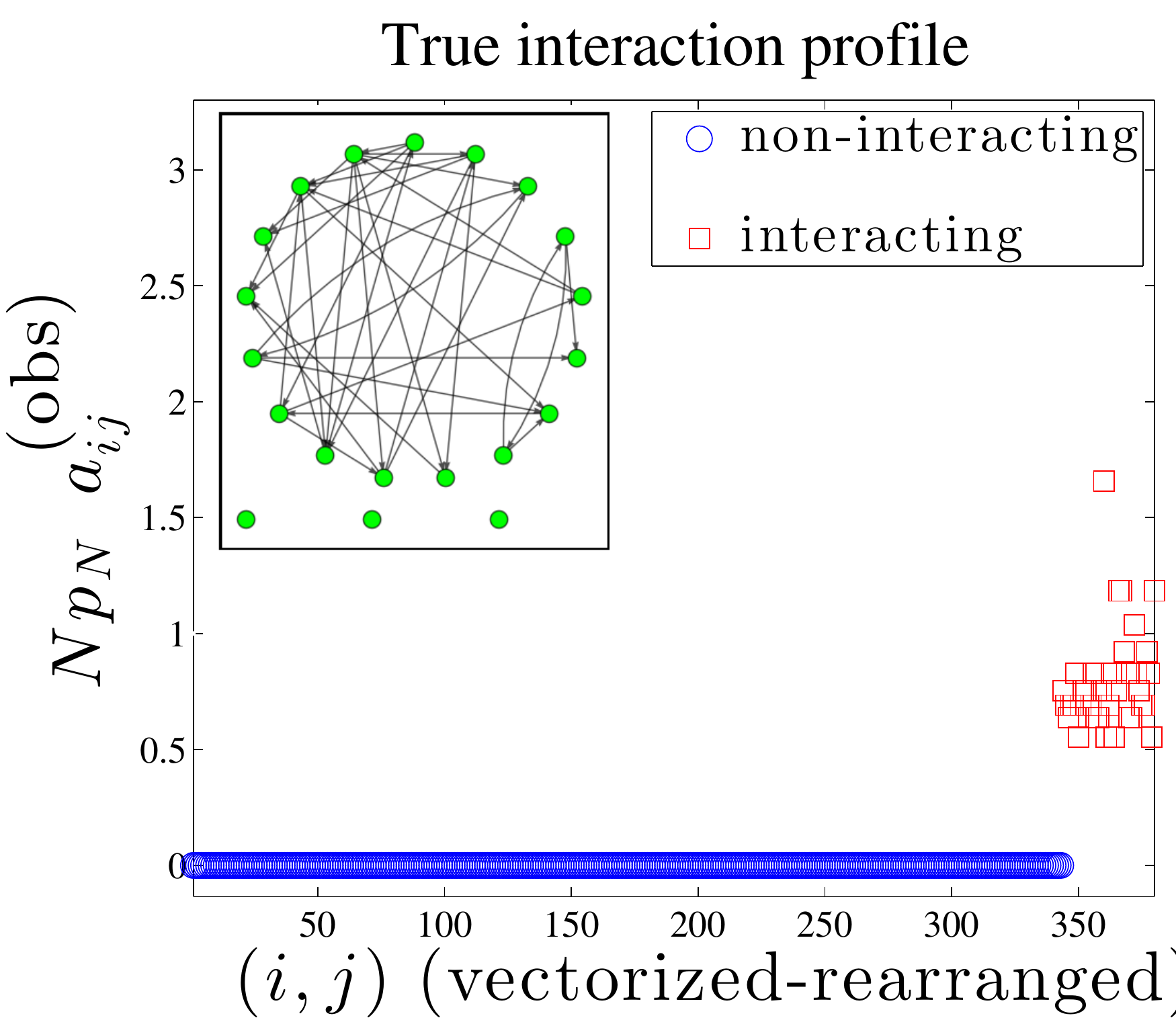}}
\end{minipage}
\begin{minipage}{.33\linewidth}
\centering
{
\includegraphics[scale=.28]{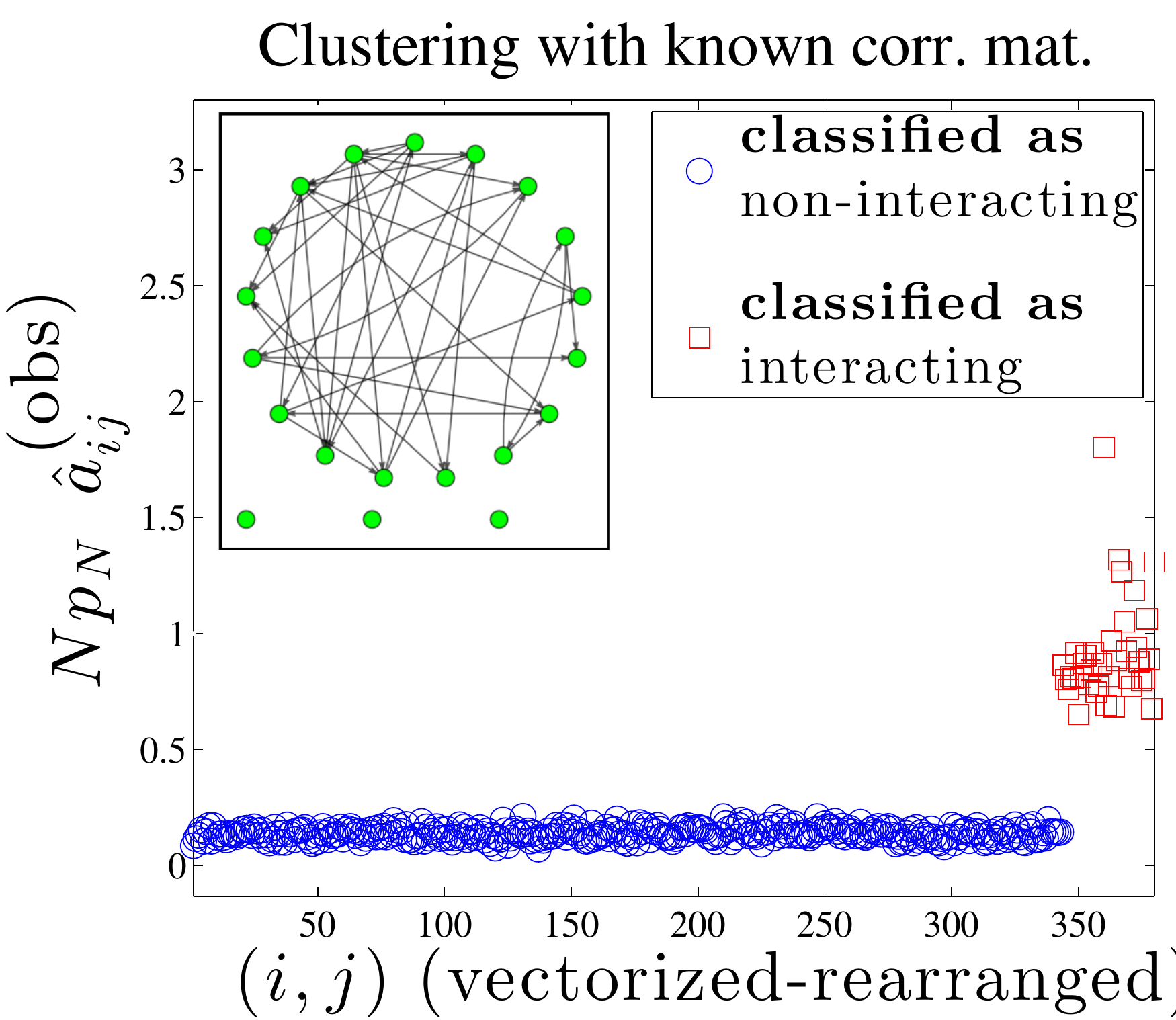}}
\end{minipage}
\begin{minipage}{.33\linewidth}
\centering
{
\includegraphics[scale=.28]{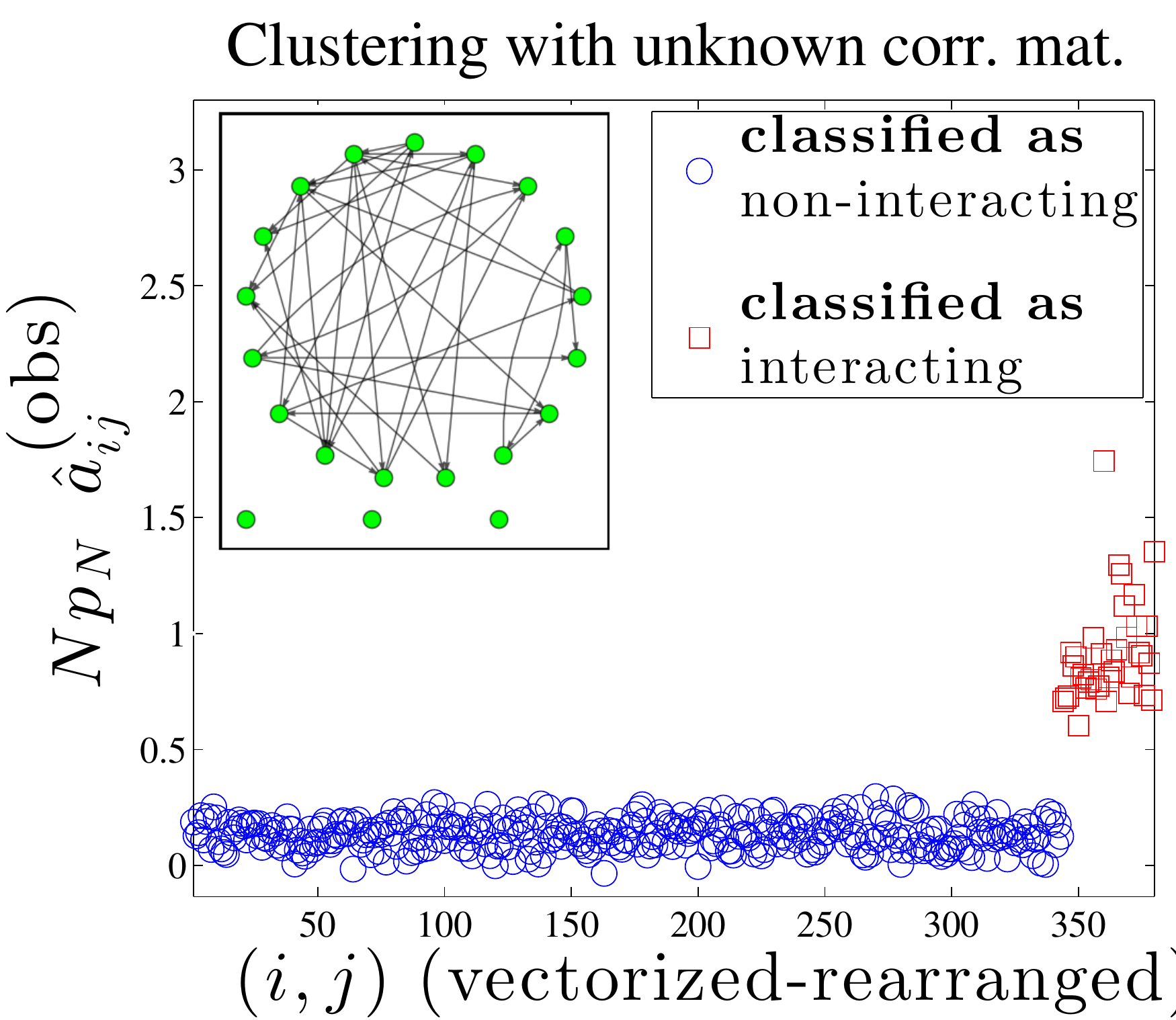}}
\end{minipage}
\caption{
Network tomography for the case of a uniform averaging combination rule addressed in Sec.~\ref{sec:UnifAve}.
The network consists of $N=100$ agents, where only $K=20$ agents are observable ($\xi=0.2$). 
The interaction probability is $p_N = 2\,(\ln N)/N\approx 0.092$, and the value of the step-size is $\mu=0.1$. 
{\bf Leftmost panel}: the {\em true} combination matrix, $A^{\textnormal{(obs)}}$, is vectorized with column-major ordering, and the (vectorized) $(i,j)$ pairs are rearranged in such a way that the zero entries come before the nonzero entries (the same ordering is applied to the estimated matrix, $\hat{A}^{\textnormal{(obs)}}$, in the other two panels). 
The different markers highlight the {\em true} interaction profile of the {\em observable} network portion. 
{\bf Middle panel}: matrix $\hat{A}^{\textnormal{(obs)}}$, computed under perfect knowledge of the steady-state correlation matrices of the {\em observable} diffusion output,  see~(\ref{eq:perfknow}). The different markers highlight the interaction profile as {\em reconstructed} by the $k$-means algorithm. 
{\bf Rightmost panel}: same as middle panel, with matrix $\hat{A}^{\textnormal{(obs)}}$ computed using the correlation matrices {\em estimated empirically} with $n=2 \times 10^4$ samples, see~(\ref{eq:Rest1}),~(\ref{eq:Rest2}), and~(\ref{eq:imperfknow}).
{\bf Inset plots}: interaction profiles represented through the corresponding network graphs. 
In the inset plots of the middle and of the rightmost panels, erroneously detected edges (the edge is not present but it is ``seen'' by the tomography algorithm) are marked in magenta, while missed edged (the edge is present but the tomography algorithm misses it) are marked in cyan.
}
\label{fig:UA}
\end{figure*}

\subsection{Uniform Averaging Rule}
\label{sec:UnifAve}
One of the most important goals of causation is ascertaining whether one agent influences another agent. Such influence is not necessarily symmetrical. 
Although we have obtained an analytical characterization of the error matrix for the case of symmetric combination matrices, it is useful to examine whether the conclusions drawn for the symmetric case find some correspondence in the asymmetric case. 
To this aim, the network interaction profile is now generated according to a {\em binomial random graph} model, $\mathscr{B}(N,p_N)$, where the variables $\bm{g}_{ij}$, for $i=1,2,\dots, N$ and $j\neq i$, are independent Bernoulli random variables with $\P[\bm{g}_{ij}=1]=p_N$ --- see~\cite{binomialgraphs}. Such a procedure leads (in general) to an asymmetric matrix $G$, i.e., to a {\em directed} graph.
As regards the nonzero entries of the combination matrix, we consider the well-known uniform averaging rule (where the weight that an agent assigns to the data of its interacting neighbors is inversely proportional to the size of its neighborhood), which leads in general to an asymmetric combination matrix. 
The results of our experiments are reported in Fig.~\ref{fig:UA}, with reference to the  case of perfect and imperfect knowledge of the correlation matrices, respectively. Remarkably, the tomography algorithm exhibits proper operation even in the asymmetric case. 
From the inset plots, we can appreciate that, differently from the previous examples, it is now possible that the influence of an agent over another agent is in general unidirectional.  

\begin{figure*}
{
\centering{
\bf ~~~~~~The effect of different system parameters} 
\par\medskip
}
%
%
%
\begin{minipage}{.33\linewidth}
{
\centering{
\bf ~~~~~~Parameter $\mu$} 
\par\medskip
}
\centering
{
\includegraphics[scale=.28]{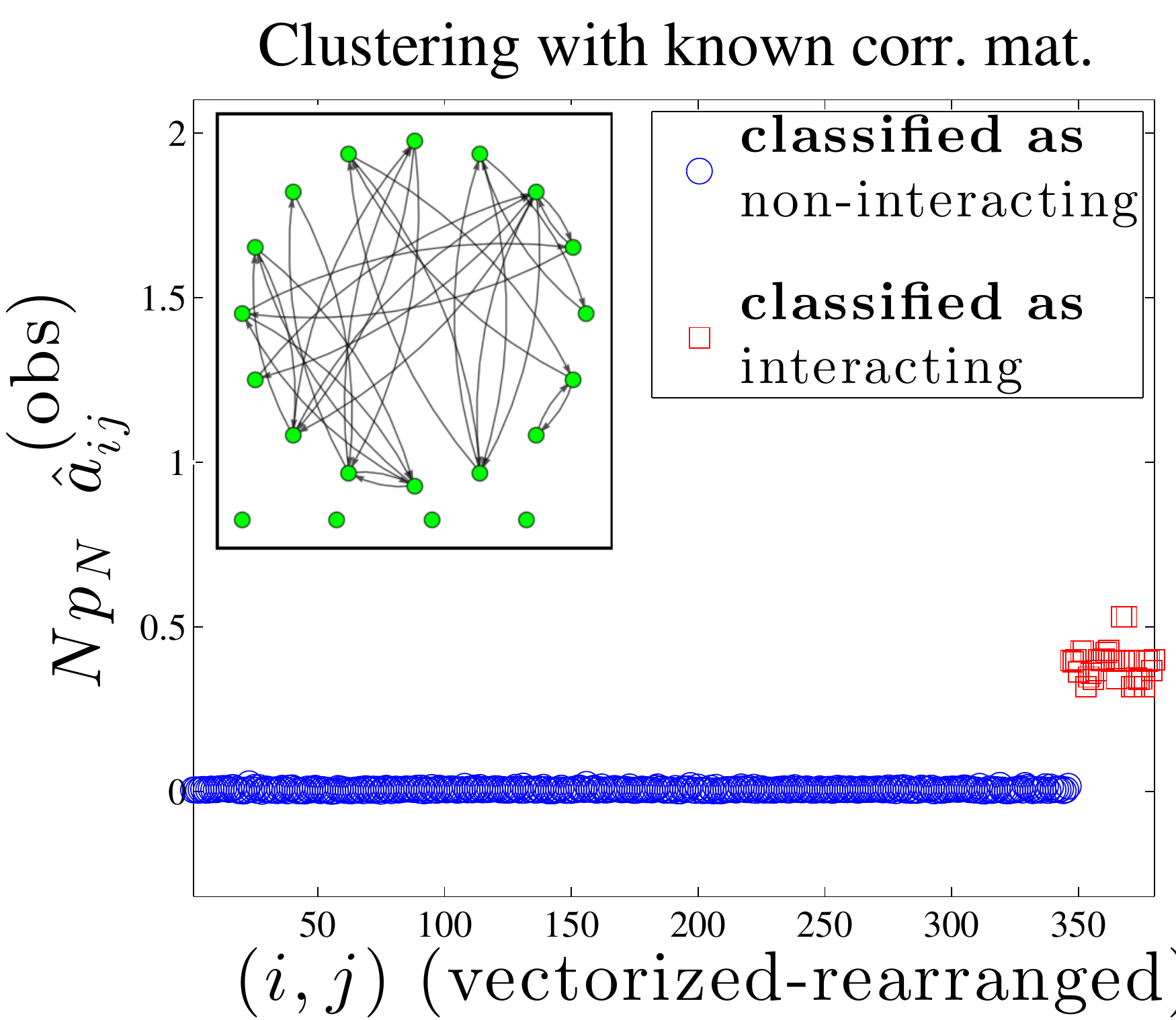}}
\end{minipage}
\begin{minipage}{.33\linewidth}
{
\centering{
\bf ~~~~~~Parameter $\xi$} 
\par\medskip
}
\centering
{
\includegraphics[scale=.28]{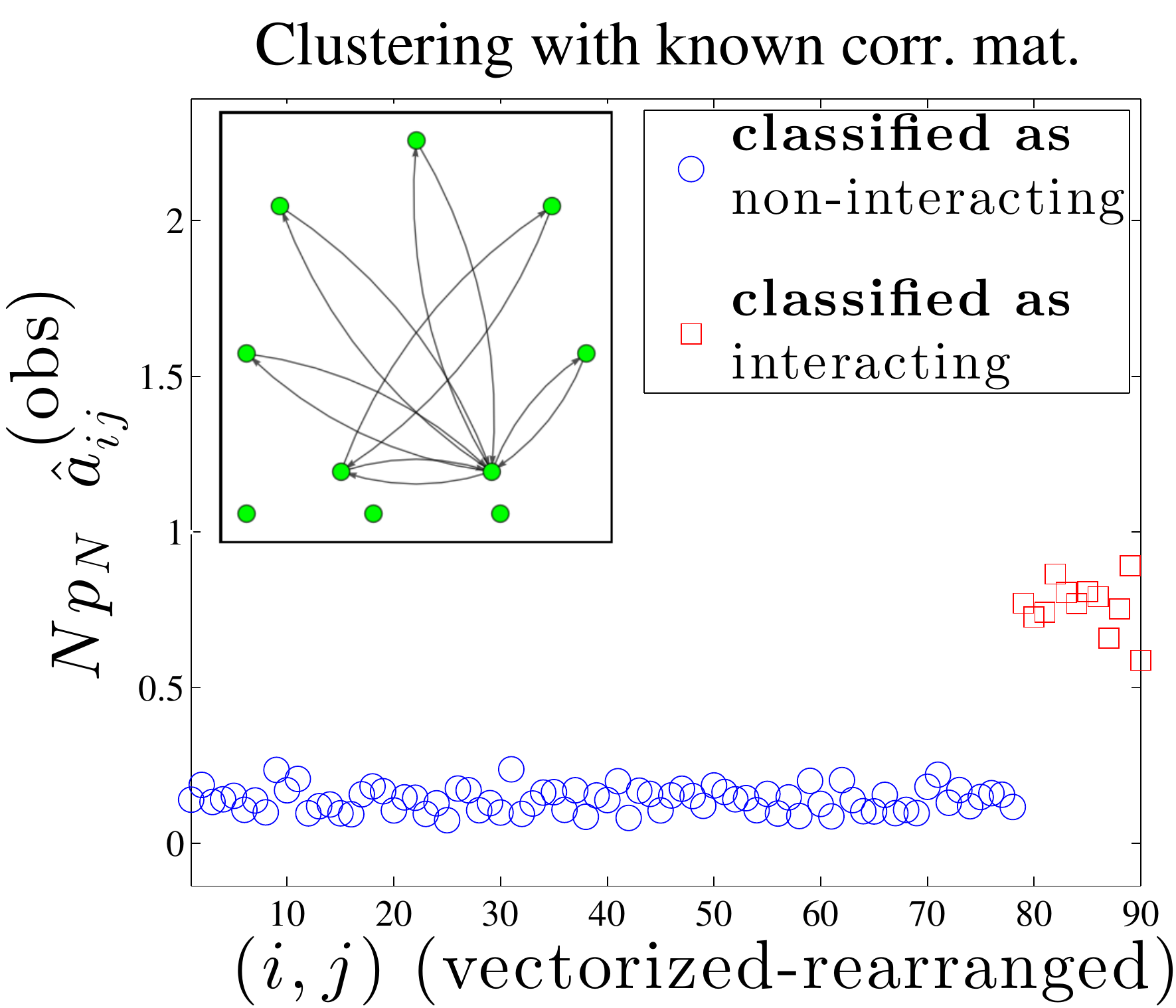}}
\end{minipage}
\begin{minipage}{.33\linewidth}
{
\centering{
\bf ~~~~~~Parameter $p_N$} 
\par\medskip
}
\centering
{
\includegraphics[scale=.28]{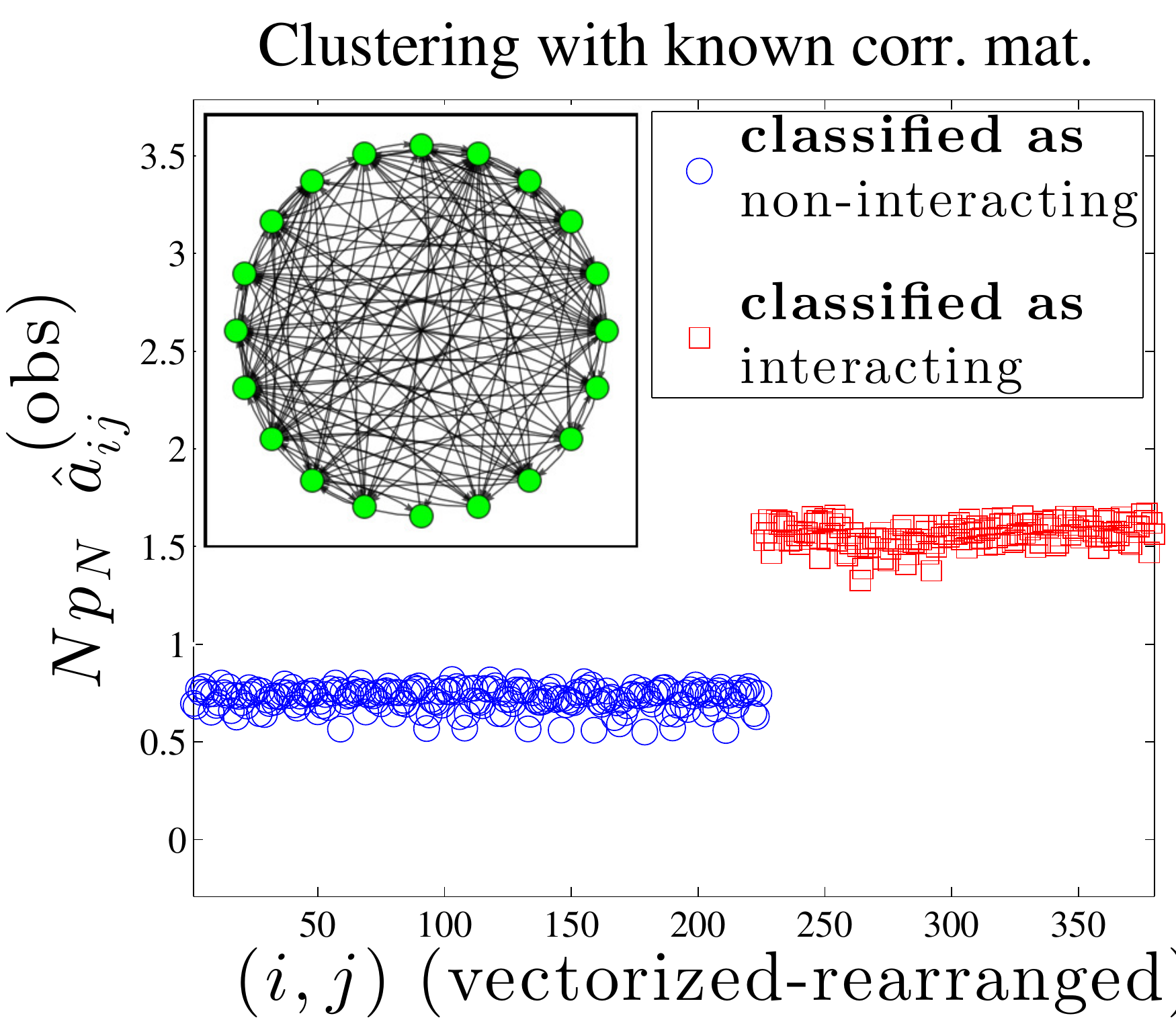}}
\end{minipage}
\caption{
Network tomography for the case of a Metropolis combination rule addressed in Sec.~\ref{sec:Metropolis}, and for a network with $N=100$ agents. 
{\bf Leftmost panel}: the case of ``high'' values for the step-size addressed in Sec.~\ref{sec:mu}. The relevant parameters are: $\mu=0.5$, $\xi=0.2$, and $p_N = 2\,(\ln N)/N\approx 0.092$.
{\bf Middle panel}: the case of ``small'' fractions of observed agents addressed in Sec.~\ref{sec:Obs}. The relevant parameters are: $\xi=0.1$, $p_N\approx 0.092$, and $\mu=0.1$.
{\bf Rightmost panel}: the case of ``high'' values for the interaction probability addressed in Sec.~\ref{sec:Conn}. The relevant parameters are: $p_N\approx 0.46$, $\xi=0.2$, and $\mu=0.1$.
{\bf Lowermost panels}: 
In all panels, the matrix $\hat{A}^{\textnormal{(obs)}}$ (displayed with the same ordering criterion used in the previous figures) is computed under perfect knowledge of the steady-state correlation matrices of the {\em observable} diffusion output,  see~(\ref{eq:perfknow}). The different markers highlight the interaction profile as {\em reconstructed} by the $k$-means algorithm.
{\bf Inset plots}: interaction profiles represented through the corresponding network graphs. 
In the inset plots of the lowermost panels, erroneously detected edges (the edge is not present but it is ``seen'' by the tomography algorithm) are marked in magenta, while missed edged (the edge is present but the tomography algorithm misses it) are marked in cyan.
}
\label{fig:main}
\end{figure*}

\subsection {Role of the Step-Size}
\label{sec:mu}
The step-size, $\mu$, is a fundamental parameter ruling the performance of the adaptive diffusion algorithm implemented by the network agents.
As examined in detail in recent works~\cite{ChenSayedIT2015Part1,ChenSayedIT2015Part2}, the following adaptation/learning trade-off arises: small values of $\mu$ enhance learning precision, while relatively high values of $\mu$ enhance adaptation. 

We now examine the impact of the step-size on the tomography algorithm. 
For the sake of concreteness, we focus on one of the combination rules examined in the previous sections, namely, the Metropolis rule, and consider the noiseless case, namely, steps S1) and S2).
In the previous simulations concerning the Metropolis rule (see Fig.~\ref{fig:Metropolis}), we have used $\mu=0.1$. 
In the new simulation, reported in Fig.~\ref{fig:main} (leftmost panel), we run the diffusion algorithm with $\mu=0.5$, which turns out to be a relatively high value for the step-size in practical applications. 

By joint inspection of Fig.~\ref{fig:Metropolis} (middle panel) and Fig.~\ref{fig:main} (leftmost panel), we can appreciate the presence of different effects.
The first effect relates to the amplitude of the nonzero entries. 
We see that increasing $\mu$ corresponds to diminishing the amplitude of the nonzero entries in the combination matrix. 
This effect is rather obvious, since our definition of $A$ embodies a scaling factor $1-\mu$, see~(\ref{eq:aweightsmatdef}).
The reduction of the distance between zero and nonzero entries could have a negative impact on tomography performance, especially when the correlation matrices must be estimated from the diffusion output, and the additional error associated to such estimation reduces further the separation between the two classes.

The second effect relates to the amplitude of the entries in the error matrix. 
We see that increasing $\mu$ corresponds to diminishing the amplitude of the errors, which could be beneficial for tomography.
This effect can be grasped by Theorem~$1$, since we see that the (columnwise) sum of errors is upper bounded by $1-\mu$.
However, Theorem~$1$ provides only an upper bound, and, hence, the exact behavior of the error as a function of $\mu$ is not known.

As a result, it is not easy to anticipate which of the aforementioned two effects will prevail, and elucidating the trade-offs related to the effect of the step-size on the tomography performance, in connection to the adaptation/learning trade-off, is an interesting issue that deserves further investigation.


\subsection {Degree of Observability}
\label{sec:Obs}
It is not unreasonable to presume that the error induced by {\em partial} observation of the network depends inversely on the observable fraction of agents, $\xi$. 
Accordingly, it is expected that the error grows when the size of the monitored portion of the network decreases, i.e., as $\xi$ decreases. This effect is confirmed by our numerical experiments. 
A sample of such experiments is reported in Fig.~\ref{fig:main} (middle panel).    
In particular, by comparison with Fig.~\ref{fig:Metropolis} (middle panel), it is observed that the averages of the errors shift upward for both the zero and the nonzero entries, and that the error's amplitude is generally increased. 
However, even in the presence of higher errors, separability between the two clusters seems to be preserved.

\subsection{``High'' Interaction Probability}  
\label{sec:Conn}
The theoretical analysis conducted in this work focused on a classic asymptotic regime for random graphs, namely, the regime where the interaction probability goes to zero as the network size increases.          
We were able to conclude that, as $N\rightarrow\infty$, the (scaled) error, $N p_N \bm{e}_{ij}$, is driven to zero in probability. 

On the other hand, in practice one obviously operates with fixed (finite) values of $N$ and nonzero values of $p_N$, and, hence, the error will be never exactly equal to zero. 
In particular, for a fixed value of $N$, a higher value of $p_N$ (which, in the asymptotic setting described by~(\ref{eq:pconn}), would correspond to select a sequence $c_N$ diverging faster), is expected to imply higher values of $N p_N \bm{e}_{ij}$. 
Such features are observed in Fig.~\ref{fig:main}, rightmost panel. 
Comparison with the middle panel of Fig.~\ref{fig:Metropolis} (where the interaction probability was smaller) reveals that the error in Fig.~\ref{fig:main} is, on average, higher.

However, there is an important aspect that emerges here. We see clearly that the average increase of the error does not impair the capability of the system to discriminate between interacting and non-interacting agent pairs. Clustering is still effective, since the two clusters are merely shifted upward. This suggests the following important observation. 
In this case, what matters is the {\em variance} of the error, which should be sufficiently small in order to guarantee separability of the clusters. 

\section{Concluding Remarks and Open Issues}
We considered the problem of tomography over partially observed diffusion networks, where the goal is to infer whether an agent is directly influenced by another agent. 
Under the challenging scenario where only the output of the diffusion algorithm is observable (i.e., there is no direct information about who's talking to whom) {\em and} only a fraction of the agents is accessible, we prove that consistent tomography is possible: as the network size increases, the interacting and non-interacting agent pairs split in two separate clusters, for any given fraction of accessible agents.
The theoretical results are proved under some technical assumptions: $i)$ the network graph is an Erd\"os-R\'enyi graph; $ii)$ the combination matrices belong to a certain symmetric class; $iii)$ perfect knowledge is assumed for the correlation matrices of the {\em observable} diffusion output (steady-state regime). 
As another contribution, we show by means of numerical experiments that the results hold for asymmetric matrices as well (which is useful in the context of {\em causal} inference), and that the theoretical correlation matrices can be replaced by their empirical counterparts, estimated from the diffusion outputs observed for a sufficiently long time.

There are many interesting questions and open issues that deserve future investigation.
First, it would be useful to generalize the theoretical results to broader settings, including: the case of asymmetric combination matrices; graph generative models different from the Erd\"os-R\'enyi model; different classes of combination matrices.

We see from the numerical analysis that the separation between the agents' clusters is very marked, suggesting that vanishing of the error (which, in this work, has been established in terms of the {\em fraction} of erroneously classified agents) might hold under more restrictive criteria, e.g., a maximum error criterion, where we require that the {\em largest} entry in the error matrix vanishes.
It would be also of interest to see whether the characterization of the result in Proposition~$1$ can be obtained without resorting to the independence approximation.
Moreover, we notice that the additive structure of the entries in the error matrix encourages to pursue an asymptotic normality characterization.

From a more practical perspective, an interesting problem is optimizing the algorithms for tomography (exploiting, when possible, further structural properties, such as sparsity), which requires careful managing of the interplay between network size and number of samples collected for the inferential task. 

\appendices
\section{}
\label{app:Theo1}
\begin{IEEEproof}[Proof of Theorem~$1$]
In view of~(\ref{eq:R1nR0n}) and~(\ref{eq:origdef3}), the steady-state zero-lag and one-lag correlation matrices can be written as:
\beq
R_0=\mu^2 Z^{-1}, \quad R_1=\mu^2 A Z^{-1},
\label{eq:R01expl}
\eeq
where we let $Z=I_N - A^2$.
Applying the rules for multiplication between partitioned matrices~\cite[p.~17]{Johnson-Horn}, the observable part of the one-lag correlation matrix is given by:
\beq
[R_1]_{\Omega}= \mu^2 A_{\Omega} [Z^{-1}]_{\Omega} + 
\mu^2 A_{\Omega \Omega^\prime} [Z^{-1}]_{\Omega^\prime\Omega},
\label{eq:R1expl2}
\eeq
where $\Omega^\prime=\{1,2,\dots, N\}\setminus \Omega$ denotes the complement of $\Omega$. 
Therefore, using~(\ref{eq:R01expl}) and~(\ref{eq:R1expl2}), the matrix $\hat A^{\textnormal{(obs)}}$ in~(\ref{eq:A11_first}) can be written as:
\beqa
\hat A^{\textnormal{(obs)}}&=&
(A_{\Omega}[Z^{-1}]_{\Omega} + A_{\Omega\Omega^\prime} [Z^{-1}]_{\Omega^\prime\Omega})
([Z^{-1}]_{\Omega})^{-1}\nonumber\\
&=&
A_{\Omega} + A_{\Omega\Omega^\prime}[Z^{-1}]_{\Omega^\prime\Omega}([Z^{-1}]_{\Omega})^{-1}.
\label{eq:Aobsexp1}
\eeqa
Now, from the matrix inversion lemma we have that~\cite[p.~18]{Johnson-Horn}:
\beq
[Z^{-1}]_{\Omega^\prime\Omega}= - (Z_{\Omega^\prime})^{-1} Z_{\Omega^\prime\Omega} [Z^{-1}]_{\Omega}.
\label{eq:invlem1}
\eeq
Substituting~(\ref{eq:invlem1}) into~(\ref{eq:Aobsexp1}), and making explicit the definition of $Z$, we get:
\beqa
\hat A^{\textnormal{(obs)}}&=&
A_{\Omega} - A_{\Omega\Omega^\prime}(Z_{\Omega^\prime})^{-1} Z_{\Omega^\prime\Omega}
\nonumber\\
&=&
A_{\Omega} - A_{\Omega\Omega^\prime}(\;\underbrace{[I_N - A^2]_{\Omega^\prime}}_{I_{N-K} - [A^2]_{\Omega^\prime}}\;)^{-1}\underbrace{[I_N - A^2]_{\Omega^\prime\Omega}}_{-[A^2]_{\Omega^\prime\Omega}},\nonumber\\
\eeqa
which, in view of~(\ref{eq:errmatdef}), and noting that $A_{\Omega}=A^{\textnormal{(obs)}}$, gives the following expression for the error matrix $E$:
\beq
E=A_{\Omega\Omega^\prime} (I_{N-K} - [A^2]_{\Omega^\prime})^{-1} [A^2]_{\Omega^\prime\Omega}.
\label{eq:errmatinvlem}
\eeq
By introducing the following matrices:
\beq
B\dfz A^2,\quad
H\dfz (I_{N-K} - B_{\Omega^\prime})^{-1},
\label{eq:BCDmat}
\eeq
the error matrix can be represented as:
\beq
E=A_{\Omega\Omega^\prime} H B_{\Omega^\prime\Omega}.
\label{eq:Ematfirstdef}
\eeq
Accordingly, the entries of the error matrix, for $i,j=1,2,\dots,K$, take on the form:
\beq
e_{ij}=\sum_{\ell,m=1}^{N-K} a_{\omega_i\omega^\prime_\ell} \, h_{\ell m}\, b_{\omega^\prime_m\omega_j},
\label{eq:eentries}
\eeq
where indices $\omega_1<\omega_2<\dots<\omega_K$ span the observable set $\Omega$, while indices $\omega^\prime_1<\omega^\prime_2<\dots<\omega^\prime_{N-K}$ span the complement set $\Omega^\prime$.  
We remark that the error matrix $E$ is a nonnegative matrix (i.e., $e_{ij}\geq 0$ for all $i,j=1,2,\dots,K$), because all terms in the summation~(\ref{eq:eentries}) are nonnegative.
In fact, $A_{\Omega\Omega^\prime}$ and $B=A^2$ are nonnegative matrices since $A$ is nonnegative. 
With regard to $H$, from~(\ref{eq:BCDmat}) it can be expressed as:
\beq
H=(I_{N-K} - B_{\Omega^\prime})^{-1}=\sum_{k=0}^\infty (B_{\Omega^\prime})^k,
\label{eq:Hseriesfirstdef}
\eeq
and, hence, $H$ is nonnegative because so is $B_{\Omega^\prime}$.

Let us now consider the (columnwise) sum of the errors, which, in view of~(\ref{eq:eentries}), can be written as:
\beq
\sum_{j=1}^K e_{ij}=\sum_{\ell,m=1}^{N-K} a_{\omega_i\omega^\prime_\ell} \, h_{\ell m}\, \sum_{j=1}^K b_{\omega^\prime_m\omega_j}.
\label{eq:sumoferr}
\eeq
It is convenient to introduce the following auxiliary matrix:
\beq
F\dfz H B_{\Omega^\prime\Omega},
\label{eq:Fmatfirstdef}
\eeq
whose entries, for $\ell=1,2,\dots, N-K$, and $j=1,2,\dots,K$, are accordingly:
\beq
f_{\ell j}=\sum_{m=1}^{N-K} h_{\ell m} b_{\omega^\prime_m\omega_j}.
\label{eq:feljauxil}
\eeq
With such definition, the matrix entry $e_{ij}$ in~(\ref{eq:eentries}) becomes:
\beq
e_{ij}=\sum_{\ell=1}^{N-K} a_{\omega_i\omega^\prime_\ell} f_{\ell j}.
\label{eq:easfuncofef}
\eeq
We start by proving that: 
\beq
\boxed{
\sum_{j=1}^K f_{\ell j}\leq 1
}
\label{eq:fleq1}
\eeq
Recalling the definition of the matrix $B$ in~(\ref{eq:BCDmat}), we have that $\sum_{j=1}^N b_{m j}=(1-\mu)^2$.
Therefore, we can write:
\beqa
\sum_{j=1}^K b_{\omega^\prime_m\omega_j}&=&(1-\mu)^2 - \sum_{j=1}^{N-K} b_{\omega^\prime_m \omega^\prime_j}\nonumber\\
&=& 1 - \mu(2-\mu) - \sum_{j=1}^{N-K} b_{\omega^\prime_m\omega^\prime_j}\nonumber\\
&=& - \mu(2-\mu) + \sum_{j=1}^{N-K} \underbrace{(\delta_{m j} - \{B_{\Omega^\prime}\}_{mj})}_{=\{H^{-1}\}_{mj}},\nonumber\\
\eeqa
and, hence, from~(\ref{eq:feljauxil}) we get:
\beqa
\sum_{j=1}^K f_{\ell j}&=&- \mu(2-\mu) \sum_{m=1}^{N-K} h_{\ell m} + 
\sum_{j=1}^{N-K} \underbrace{\sum_{m=1}^{N-K} h_{\ell m} \{H^{-1}\}_{mj}}_{\delta_{\ell j}}\nonumber\\
&=&
- \mu(2-\mu) \sum_{m=1}^{N-K} h_{\ell m} + 1 \leq 1,
\label{eq:efferrchain}
\eeqa
where the inequality holds since $H$ is a nonnegative matrix, and since $\mu(2-\mu)>0$ because $0<\mu<1$.
Using now~(\ref{eq:easfuncofef}), we can write:
\beq
\sum_{j=1}^K e_{ij}=\sum_{\ell=1}^{N-K} a_{\omega_i\omega^\prime_\ell} 
\sum_{j=1}^{N-K} f_{\ell j}
\leq
\sum_{\ell=1}^{N-K} a_{\omega_i\omega^\prime_\ell}\leq 1-\mu,
\label{eq:errchain}
\eeq
where $i)$ the first inequality comes from~(\ref{eq:fleq1}) (since $A$ is a nonnegative matrix), and $ii)$ the second inequality holds since $\sum_{\ell=1}^N a_{i\ell}=1-\mu$. 
\end{IEEEproof}

\section{}
\label{app:Theo2}
\begin{IEEEproof}[Proof of Lemma~$1$]
We observe that for any nonnegative real number $d$ and $i\neq j$, it holds that:
\beqa
\lefteqn{\P[\bm{d}_{\textnormal{max}}\geq d \, | \, \bm{g}_{ij}=1]\leq
\sum_{\ell=1}^N 
\P[\bm{d}_\ell\geq d \, | \, \bm{g}_{ij}=1]
}\nonumber\\
\nonumber\\
&=&
\sum_{\ell\neq i,j} 
\P[\bm{d}_\ell\geq d \, | \, \bm{g}_{ij}=1] + \nonumber\\
&&
\P[\bm{d}_i\geq d \, | \, \bm{g}_{ij}=1] + 
\P[\bm{d}_j\geq d \, | \, \bm{g}_{ij}=1],
\label{eq:UBbounds}
\eeqa
where the inequality is an application of the union bound.
Now, for $\ell\neq i,j$, we have that $\bm{d}_{\ell}$ is independent of $\bm{g}_{ij}$, and that $\bm{d}_\ell=1+\bm{\beta}(N-1,p_N)$, implying that $\P[\bm{d}_\ell\geq d \, | \, \bm{g}_{ij}=1]=\P\left[1+\bm{\beta}(N-1,p_N)\geq d\right]$.
On the other hand, for $\ell=i$ or $\ell=j$, we have that, conditioned on the event $\bm{g}_{ij}=1$, the degree $\bm{d}_\ell$ is equal to $2$ (because $\bm{g}_{ii}=1$ and $\bm{g}_{ij}=1$) plus the number of neighbors arising from the remaining $N-2$ agents, which yields $\bm{d}_{\ell}=2+\bm{\beta}(N-2,p_N)$.
Therefore, it is legitimate to write the following chain of inequalities, for $\eta\in\mathbb{R}$ and $t>0$:
\beqa
\lefteqn{\P[\bm{d}_{\textnormal{max}}\geq\eta N p_N \, | \, \bm{g}_{ij}=1]}\nonumber\\
&\leq& 
(N-2)\,\P\left[1+\bm{\beta}(N-1,p_N)\geq\eta N p_N\right]
+\nonumber\\
&&
2\,\P\left[2+\bm{\beta}(N-2,p_N)\geq\eta N p_N\right]
\nonumber\\
&\leq& N \, e^{- \eta N p_N t} \,\E[e^{(1+\bm{\beta}(N-1,p_N) )  t}]
+\nonumber\\
&& 2 \, e^{- \eta N p_N t} \,\E[e^{(2+\bm{\beta}(N-2,p_N) )  t}]
\nonumber\\
&=&N e^t\,e^{- \eta N p_N t} [1+p_N(e^t-1)]^{N-1}
+\nonumber\\
&&2 e^{2 t}\,e^{- \eta N p_N t} [1+p_N(e^t-1)]^{N-2}
\nonumber\\
&\leq& (N e^t + 2 e^{2 t})\, e^{-N p_N ( \eta t + 1 - e^t )},
\eeqa
where: $i)$ the first inequality follows from~(\ref{eq:UBbounds}); $ii)$ the second inequality is a classic Chernoff bound\footnote{For a random variable $\bm{x}$, and a positive $t$, the Chernoff bound is given by $\P[\bm{x}\geq\eta]\leq \E[e^{t \bm{x}}]/e^{t \eta}$.}; $iii)$ the equality follows by evaluating explicitly the moment generating function, $\E[e^{\bm{\beta}(N,p) t}]$, of the binomial random variable\footnote{The moment generating function of a Bernoulli random variable with success probability $p$ is given by $p e^t + (1-p)$. Since a binomial random variable $\bm{\beta}(N,p)$ is the sum of $N$ independent Bernoulli variables with success probability $p$, we have $\E[e^{\bm{\beta}(N,p) t}]=[p e^t + (1-p)]^N$.}; $iv)$ the latter inequality follows from $(1+x)^N\leq e^{N x}$ (which holds true for $x>0$), having further replaced the exponents $N-1$ and $N-2$ with the (higher) exponent $N$.
Now, maximizing the exponent $\eta t -e^t$ (w.r.t. $t$) yields the solution $e^t=\eta$, which in turn implies:
\beq
\P[\bm{d}_{\textnormal{max}}\geq\eta N p_N \, | \, \bm{g}_{ij}=1]
\leq 
(\eta N + 2\eta^2) e^{-N p_N ( \eta \ln\eta + 1 -\eta )}.
\eeq
Using now the particular value $\eta=e$ yields: 
\beqa
\P[\bm{d}_{\textnormal{max}}\geq e N p_N \, | \, \bm{g}_{ij}=1]
&\leq& 
(e\, N + 2 e^2) e^{-N p_N}\nonumber\\
&\leq&
\left(e + \frac{2 e^2}{N}\right) e^{-c_N},
\label{eq:convzer}
\eeqa
where, in the latter inequality, we used the fact that $p_N=(\ln N + c_N)/N$.
\end{IEEEproof}

\section{}
\label{app:Theo3}
\begin{IEEEproof}[Proof of Theorem~$2$]
The empirical distributions $\bm{\mathcal{F}}_0(\alpha)$ and $\bm{\mathcal{F}}_1(\alpha)$ in~(\ref{eq:E01indx}) can be written as: 
\beqa
\bm{\mathcal{F}}_0(\alpha)=\frac
{\bm{N}_0(\alpha)}
{\bm{N}_0 + \mathbb{I}\{ \bm{N}_0 = 0\}} 
+ 
\frac 1 2\mathbb{I}\{ \bm{N}_0 = 0\},
\label{eq:E0indx}
\\
\bm{\mathcal{F}}_1(\alpha)=\frac
{\bm{N}_1(\alpha)}
{\bm{N}_1 + \mathbb{I}\{ \bm{N}_1 = 0\}} 
+ 
\frac 1 2\mathbb{I}\{ \bm{N}_1 = 0\},
\label{eq:E1indx}
\eeqa
where we have simply used a formal representation to state that $\bm{\mathcal{F}}_0(\alpha)$) (resp., $\bm{\mathcal{F}}_1(\alpha)$) are set to $1/2$ when $\bm{N}_0=0$ (resp., $\bm{N}_1=0$).

We start by proving that $\bm{\mathcal{F}}_0(\epsilon)\stackrel{\textnormal{p}}{\longrightarrow} 1$ for all $\epsilon>0$.
From~(\ref{eq:E0indx}) we can write:
\beq
1 - \bm{\mathcal{F}}_0(\epsilon)=
\frac
{\bm{N}_0 - \bm{N}_0(\epsilon)}
{\bm{N}_0 + \mathbb{I}\{ \bm{N}_0 = 0\}} 
+
\frac 1 2\mathbb{I}\{ \bm{N}_0 = 0\}.
\label{eq:oneminusF0}
\eeq
Using~(\ref{eq:N0obsemp}), and since $\hat{\bm{a}}_{i j}^{\textnormal{(obs)}}=\bm{a}_{ij}^{\textnormal{(obs)}}+\bm{e}_{ij}$, from~(\ref{eq:oneminusF0}) we can write:
\beqa
\lefteqn{\frac
{\bm{N}_0 - \bm{N}_0(\epsilon)}
{\bm{N}_0 + \mathbb{I}\{ \bm{N}_0 = 0\}}}\nonumber\\
&=&\frac
{\sum_{i=1}^K\sum_{j\neq i} \mathbb{I}\{N p_N \bm{e}_{i j}>\epsilon, \bm{g}_{i j}^{\textnormal{(obs)}}=0\}}
{\bm{N}_0 + \mathbb{I}\{ \bm{N}_0 = 0\}}\nonumber\\
&\leq&
\frac
{\sum_{i=1}^K\sum_{j\neq i} \mathbb{I}\{N p_N \bm{e}_{i j}>\epsilon\}}
{\bm{N}_0 + \mathbb{I}\{ \bm{N}_0 = 0\}}
\nonumber\\
&\leq&
\frac{K N p_N}{\bm{N}_0 + \mathbb{I}\{ \bm{N}_0 = 0\}}\,
\frac{1-\mu}{\epsilon}\nonumber\\
&=&
\left(\frac{K^2}
{\bm{N}_0 + \mathbb{I}\{ \bm{N}_0 = 0\}}
\right)
\frac N K
\frac{1-\mu}{\epsilon}
\,p_N, 
\label{eq:E0bound}
\eeqa
where the latter inequality follows by applying Corollary~$1$.
Now, the asymptotic properties of $\bm{N}_0$ can be deduced from the asymptotic properties of binomial random variables. 
Indeed, since for the Erd\"os-R\'enyi model, $\bm{g}_{ij}=\bm{g}_{ji}$, we have: 
\beq
\bm{N}_0=2\, \sum_{i=1}^K \sum_{j < i} (1-\bm{g}_{ij}^{\textnormal{(obs)}}).
\eeq
Moreover, since the variables $1-\bm{g}_{ij}^{\textnormal{(obs)}}$ form, for $j<i$, a collection of $L=K(K-1)/2$ independent Bernoulli random variables with parameter $1-p_N$, we have that $\bm{N}_0$ is distributed as $2\bm{\beta}(L , 1-p_N)$.
Now, in view of~(\ref{eq:csidef}), $K$ (and, hence, $L$) diverges as $N$ goes to infinity. 
Since by assumption $p_N$ vanishes as $N$ goes to infinity, the product $L(1-p_N)$ diverges as well, and we can invoke~(\ref{eq:binoconv}) to conclude that:
\beq
\frac{\bm{N}_0}{L (1-p_N)}\stackrel{\textnormal{p}}{\longrightarrow} 2\quad\Rightarrow\quad
\frac{\bm{N}_0}{K^2}\stackrel{\textnormal{p}}{\longrightarrow} 1.
\label{eq:binoconv2}
\eeq
Since $\mathbb{I}\{ \bm{N}_0 = 0\}\in\{0,1\}$, using~(\ref{eq:binoconv2}) and~(\ref{eq:csidef}) into~(\ref{eq:E0bound}), and recalling that $p_N$ vanishes by assumption, we have in fact proved that the first term on the RHS in~(\ref{eq:oneminusF0}) goes to zero as $N\rightarrow\infty$. 
With regard to the second term on the RHS in~(\ref{eq:oneminusF0}), it converges to zero in mean (and, hence, in probability~\cite{shao}), since we have: 
\beqa
\E[\mathbb{I}\{ \bm{N}_0 = 0\}]&=&\P[\textnormal{all observed agents are interacting}]\nonumber\\
&=&(p_N)^{L}\stackrel{N\rightarrow\infty}{\longrightarrow} 0.
\label{eq:indic0}
\eeqa
We have in fact proved that $\bm{\mathcal{F}}_0(\epsilon)\stackrel{\textnormal{p}}{\longrightarrow} 1$ for all $\epsilon>0$.

We now show that $\bar{\bm{\mathcal{F}}}_1(\tau)\stackrel{\textnormal{p}}{\longrightarrow} 1$.
Using~(\ref{eq:N1obsemp}), from~(\ref{eq:E1indx}) we can write:
\beqa
\bm{\mathcal{F}}_1(\tau)&=&
\underbrace{\frac{\sum_{i=1}^K \sum_{j\neq i} \mathbb{I}\{N p_N \bm{\hat{a}}_{ij}^{\textnormal{(obs)}}\leq\tau, \bm{g}_{ij}^{\textnormal{(obs)}}=1\}}{K(K-1) p_N}}_{\bm{T}_1}\times\nonumber\\
&& 
\underbrace{\frac{K(K-1) p_N}{\bm{N}_1 + \mathbb{I}\{ \bm{N}_1 = 0\} }}_{\bm{T}_2}
+
\frac 1 2\mathbb{I}\{ \bm{N}_1 = 0\}.
\label{eq:E1eq}
\eeqa
With regard to the term $\bm{T}_1$, we have: 
\beq
\E[\bm{T}_1]=\frac{\sum_{i=1}^K\sum_{j\neq i} \P[N p_N \bm{\hat{a}}_{ij}^{\textnormal{(obs)}}\leq\tau | \bm{g}_{ij}^{\textnormal{(obs)}}=1] }{K(K-1)},
\label{eq:T1}
\eeq
where we used the fact that $\P[\bm{g}_{ij}^{\textnormal{(obs)}}=1]=p_N$.
Moreover, since $\bm{\hat{a}}_{ij}^{\textnormal{(obs)}}=\bm{a}_{ij}^{\textnormal{(obs)}}+\bm{e}_{ij}$, and since $\bm{e}_{ij}\geq 0$, we have that:
\beqa
\lefteqn{\P[N p_N \bm{\hat{a}}_{ij}^{\textnormal{(obs)}}\leq\tau | \bm{g}_{ij}^{\textnormal{(obs)}}=1]}\nonumber\\
&\leq&
\P[N p_N \bm{a}_{ij}^{\textnormal{(obs)}} \leq\tau | \bm{g}_{ij}^{\textnormal{(obs)}}=1].
\label{eq:probchain}
\eeqa
Combining~(\ref{eq:T1}) and~(\ref{eq:probchain}), in view of~(\ref{eq:AssumptionA1}) we can write:
\beq
\E[\bm{T}_1]\leq \epsilon_N \stackrel{N\rightarrow\infty}{\longrightarrow} 0.
\label{eq:T1goestozero}
\eeq
Thus, we have shown that the term $\bm{T}_1$ appearing in~(\ref{eq:E1eq}) converges to zero in mean and, hence, in probability~\cite{shao}. 
Moreover, the term $\bm{T}_2$ in~(\ref{eq:E1eq}) goes to one in probability in view of~(\ref{eq:binoconv}), because: $i)$ for the $\mathscr{G}^\star(N,p_N)$ model, $\bm{N}_1$ is distributed as $2 \bm{\beta}(L , p_N)$, where $L=K(K-1)/2$; $ii)$ the product $L p_N$ diverges as $N$ goes to infinity in view of~(\ref{eq:csidef}), since by assumption the product $N p_N$ diverges; $iii)$ the term $\mathbb{I}\{ \bm{N}_1 = 0\}$ is either $0$ or $1$. 
Finally, as regards the latter term in~(\ref{eq:E1eq}), it is straightforward to see that it converges to zero in mean (and, hence, in probability~\cite{shao}), since we have:
\beqa
\E[\mathbb{I}\{ \bm{N}_1 = 0\}]&=&\P[\textnormal{all observed agents are not interacting}]\nonumber\\
&=&(1-p_N)^{L}\leq e^{- L p_N}\stackrel{N\rightarrow\infty}{\longrightarrow} 0,\nonumber\\
\label{eq:indic1}
\eeqa
where the inequality holds because $p_N\leq 1$.
\end{IEEEproof}

\section{}
\label{app:prop1}
The main result of this appendix is the proof of Proposition~$1$. We start by proving two auxiliary lemmas.
\begin{lemma}
Let $\bm{\beta}(N,p)$ be a binomial random variable of parameters $N$ and $p$. We have, for $m=1,2$:
\beq
\E\left[\frac{1}{(1+\bm{\beta}(N,p))^m}\right]
\leq
\frac{m}{(N p)^m}.
\eeq
\end{lemma}
\begin{IEEEproof}
We prove the claim for the case $m=2$, since the proof for the case $m=1$ follows similar steps.
We have that:
\beqa
\lefteqn{
\E\left[\frac{1}{(1+\bm{\beta}(N,p_N))^2}\right]}\nonumber\\
&=&
\sum_{k=0}^{N} {{N}\choose{k}} p^k (1-p)^{N-k} \frac{1}{(k+1)^2}\nonumber\\
&=&
\sum_{k=0}^{N} {{N}\choose{k}} p^k (1-p)^{N-k} \frac{1}{(k+1)(k+2)}
\underbrace{\frac{k+2}{k+1}}_{\leq 2}\nonumber\\
&\leq&
\sum_{k=0}^{N} {{N}\choose{k}} p^k (1-p)^{N-k} \frac{2}{(k+1)(k+2)}.
\eeqa
Now, since we have
\beq
{{N+2}\choose{k+2}}=\frac{(N+2)!}{(k+2)! (N-k)!}=
{{N}\choose{k}}\frac{(N+1)(N+2)}{(k+1)(k+2)},
\eeq
we can write:
\beqa
\lefteqn{\sum_{k=0}^{N} {{N}\choose{k}} p^k (1-p)^{N-k} \frac{2}{(k+1)(k+2)}}\nonumber\\
&\leq&
\frac{2}{(N+1)(N+2)}\sum_{k=0}^{N} {{N+2}\choose{k+2}}
p^{k} (1-p)^{N-k}
\nonumber\\
&\leq&
\frac{2}{(N p)^2}\sum_{k=0}^{N} {{N+2}\choose{k+2}}
p^{k+2} (1-p)^{(N+2)-(k+2)}
\nonumber\\
&=&
\frac{2}{(N p)^2}\,
\underbrace{\sum_{k=2}^{N+2} {{N+2}\choose{k}} p^{k} (1-p)^{(N+2)-k} }_{\P[\bm{\beta}(N+2,p)\geq 2] < 1} 
<
\frac{2}{(N p)^2},\nonumber\\
\eeqa
and the proof is complete.
\end{IEEEproof}

We recall that $\E$ and $\V$ denote the expectation and variance operators, respectively. 
In the following, the expectation, the variance and the mean-square value of a random variable $\bm{z}$ will be denoted respectively by:
\beq
\bar{z}\dfz \E[\bm{z}],
~
\sigma^2_{z}\dfz\V[\bm{z}]=\E[(\bm{z}-\bar{z})^2],
~
\overline{z^2}\dfz \E[\bm{z}^2]=\sigma^2_z + \bar{z}^2.
\eeq
Moreover, given a sequence of {\em identically distributed} random variables, $\bm{z}_1,\bm{z}_2,\dots,\bm{z}_L$, we shall still write:
\beq
\bar{z}\dfz \E[\bm{z}_\ell],~
\sigma^2_{z}\dfz\V[\bm{z}_\ell]=\E[(\bm{z}_\ell-\bar{z})^2],
~
\overline{z^2}\dfz \E[\bm{z}_\ell^2]=\sigma^2_z + \bar{z}^2,
\eeq
since, owing to identical distribution across $\ell$, both the expectation and the variance do not depend upon $\ell$.
\begin{lemma}
Let $[\bm{u}_1,\bm{u}_2,\dots,\bm{u}_L]$ and $[\bm{z}_1,\bm{z}_2,\dots,\bm{z}_L]$ be mutually independent random vectors, 
with the variables $\bm{z}_{\ell}$, for $\ell=1,2,\dots,L$, being mutually uncorrelated and identically distributed scalar random variables across $\ell$. Let also:
\beq
\bm{s}\dfz\sum_{\ell=1}^L \bm{u}_{\ell} \bm{z}_{\ell}.
\label{eq:sumdef}
\eeq 
Then:
\beq
\boxed{
\sigma^2_s=\sigma^2_z\,\sum_{\ell=1}^L \E[\bm{u}_{\ell}^2] + 
\bar{z}^2
\,\V\left[\sum_{\ell=1}^L \bm{u}_{\ell}\right]
}
\label{eq:lemma2claim}
\eeq
\end{lemma}
\begin{IEEEproof}
We have:
\beqa
\bm{s}^2&=&\sum_{\ell=1}^L\sum_{m=1}^L \bm{u}_{\ell}\bm{u}_{m}\bm{z}_{\ell}\bm{z}_{m}\nonumber\\
&=&\sum_{\ell=1}^L \bm{u}_{\ell}^2 \bm{z}_{\ell}^2 + \sum_{\ell=1}^L\sum_{m\neq \ell} \bm{u}_{\ell}\bm{u}_{m}\bm{z}_{\ell}\bm{z}_{m}.
\eeqa
Since $[\bm{u}_1,\bm{u}_2,\dots,\bm{u}_L]$ and $[\bm{z}_1,\bm{z}_2,\dots,\bm{z}_L]$ are mutually independent random vectors, and since the variables $\bm{z}_\ell$ are mutually uncorrelated, by taking expectations we get:
\beq
\E[\bm{s}^2]=
\sum_{\ell=1}^L \E[\bm{u}_{\ell}^2]\E[\bm{z}_{\ell}^2] + 
\sum_{\ell=1}^L\sum_{m\neq \ell} 
\E[\bm{u}_{\ell} \bm{u}_{m}]
\E[\bm{z}_{\ell}]\E[\bm{z}_{m}].
\label{eq:simpleq1}
\eeq
Moreover, since the variables $\bm{z}_\ell$ are identically distributed, Eq.~(\ref{eq:simpleq1}) yields:
\beq
\E[\bm{s}^2]=
(\sigma^2_z + \bar{z}^2)\,\sum_{\ell=1}^L \E[\bm{u}_{\ell}^2] + 
\bar{z}^2\,\sum_{\ell=1}^L\sum_{m\neq \ell} 
\E[\bm{u}_{\ell} \bm{u}_{m}].
\label{eq:simpleq2}
\eeq
Now, observing that:
\beq
\left(\sum_{\ell=1}^L \bm{u}_{\ell}\right)^2=
\sum_{\ell=1}^L \bm{u}_{\ell}^2 
+ 
\sum_{\ell=1}^L\sum_{m\neq \ell} \bm{u}_{\ell} \bm{u}_{m},
\eeq
we can rearrange~(\ref{eq:simpleq2}) as: 
\beq
\E[\bm{s}^2]=
\sigma^2_z\,\sum_{\ell=1}^L \E[\bm{u}_{\ell}^2]
+ 
\bar{z}^2\,\E\left[\left(\sum_{\ell=1}^L \bm{u}_{\ell}\right)^2\right].
\label{eq:salmostfin}
\eeq
On the other hand, from~(\ref{eq:sumdef})we have:
\beq
\bar{s}=\bar{z}\,\E\left[\sum_{\ell=1}^L \bm{u}_{\ell}\right]
\Rightarrow
\bar{s}^2=\bar{z}^2\,\left(\E\left[\sum_{\ell=1}^L \bm{u}_{\ell}\right]\right)^2,
\eeq
implying, in view of~(\ref{eq:salmostfin}):
\beqa
\sigma^2_s&=&\E[\bm{s}^2] - \bar{s}^2\nonumber\\
&=&
\sigma^2_z\,\sum_{\ell=1}^L \E[\bm{u}_{\ell}^2] +
\nonumber\\ 
&&
\bar{z}^2\,\left\{
\E\left[\left(\sum_{\ell=1}^L \bm{u}_{\ell}\right)^2\right]
-
\left(\E\left[\sum_{\ell=1}^L \bm{u}_{\ell}\right]\right)^2
\right\}\nonumber\\
&=&
\sigma^2_z\,\sum_{\ell=1}^L \E[\bm{u}_{\ell}^2]
+ 
\bar{z}^2\,\V\left[\sum_{\ell=1}^L \bm{u}_{\ell}\right].
\eeqa
\end{IEEEproof}

\begin{IEEEproof}[Proof of Proposition~$1$]
We start by observing that, using~(\ref{eq:oneminusF0}), we can write:
\beq
\frac{1 - \bm{\mathcal{F}}_0(\epsilon)}{p_N}=
\frac
{\bm{N}_0 - \bm{N}_0(\epsilon)}
{\bm{N}_0 + \mathbb{I}\{ \bm{N}_0 = 0\}} \frac{1}{p_N}
+
\frac 1 2\frac{\mathbb{I}\{ \bm{N}_0 = 0\}}{p_N}.
\label{eq:oneminusF0bis}
\eeq
In view of~(\ref{eq:indic0}), the second term appearing on the RHS in~(\ref{eq:oneminusF0bis}) converges to zero in mean, and, hence, in probability~\cite{shao}.
With regard to the first term appearing on the RHS in~(\ref{eq:oneminusF0bis}), from the first inequality in~(\ref{eq:E0bound}) we have:
\beqa
\lefteqn{
\frac
{\bm{N}_0 - \bm{N}_0(\epsilon)}
{\bm{N}_0 + \mathbb{I}\{ \bm{N}_0 = 0\}}
\frac{1}{p_N}
}\nonumber\\
&\leq&
\frac
{\sum_{i=1}^K\sum_{j\neq i} \mathbb{I}\{N p_N \bm{e}_{i j}>\epsilon\}}
{\bm{N}_0 + \mathbb{I}\{ \bm{N}_0 = 0\}}
\frac{1}{p_N}\nonumber\\
&=&
\underbrace{
\frac
{\sum_{i=1}^K\sum_{j\neq i} \mathbb{I}\{N p_N \bm{e}_{i j}>\epsilon\}}
{K(K-1)p_N}
}_{\bm{T}_3}
\times\nonumber\\
&&
\underbrace{
\frac
{K(K-1)}
{\bm{N}_0 + \mathbb{I}\{ \bm{N}_0 = 0\}}}_{\bm{T}_4}.
\label{eq:E0boundbis}
\eeqa
The fact that $\bm{T}_4$ converges to $1$ in probability has been already ascertained in the proof of Theorem~$2$ --- see the rightmost relationship in~(\ref{eq:binoconv2}).
With regard to $\bm{T}_3$, by taking expectations we get:
\beq
\E[\bm{T}_3]=\frac
{\sum_{i=1}^K\sum_{j\neq i} \P[N p_N \bm{e}_{i j}>\epsilon]}
{K(K-1)p_N}
=
\frac{\P[N p_N \bm{e}_{ij}>\epsilon]}{p_N},
\label{eq:expT3}
\eeq
where we used the fact that, in view of~(\ref{eq:Eperminv}), the random variables $\bm{e}_{ij}$, for $i\neq j$, are identically distributed. This is because any renumbering leaves unaltered the statistical distribution of matrix $\bm{E}$. 
In addition, identical distribution across the $K-1$ agents distinct from agent $i$, implies in particular the following identity for expectations:
\beq
\E[\bm{e}_{ij}]=\frac{1}{K-1} \sum_{j\neq i}\E[\bm{e}_{ij}].
\eeq
Multiplying the expected error by the factor $N p_N$, we get:
\beqa
N p_N \E[\bm{e}_{ij}]&=&N p_N \frac{1}{K-1} \sum_{j\neq i}\E[\bm{e}_{ij}]\nonumber\\
&=&
\frac{N p_N}{K-1} \E\left[\sum_{j\neq i} \bm{e}_{ij}\right]
\nonumber\\
&\leq&
(1-\mu) \frac{N}{K-1}  \,  p_N\stackrel{N\rightarrow\infty}{\longrightarrow} 0,
\label{eq:symmexpec}
\eeqa
where the inequality follows from Theorem~$1$, while the convergence follows from~(\ref{eq:csidef}) and from the fact that $p_N$ vanishes. 
Now, we can write, for $i,j=1,2,\dots,K$, and $i\neq j$:
\beqa
\lefteqn{\P[N p_N \bm{e}_{ij}>\epsilon]}
\nonumber\\
&=&
\P[(N p_N \bm{e}_{ij} - N p_N \E[\bm{e}_{ij}])>(\epsilon - N p_N \E[\bm{e}_{ij}])]\nonumber\\
&\leq&
\P[(N p_N)^2 (\bm{e}_{ij} - \E[\bm{e}_{ij}])^2>(\epsilon - N p_N \E[\bm{e}_{ij}])^2]\nonumber\\
&\leq&
\frac{(N p_N)^2 \sigma^2_e}{(\epsilon - N p_N \E[\bm{e}_{ij}])^2}
\leq
\frac{4}{\epsilon^2} (N p_N)^2 \sigma^2_e,
\label{eq:fundvarvanish0}
\eeqa
where we have applied Chebyshev's inequality, and we have assumed that $N$ is sufficiently large to have $N p_N \E[\bm{e}_{ij}]\leq \epsilon/2$ (a condition guaranteed by~(\ref{eq:symmexpec})). 
Therefore, in view of~(\ref{eq:expT3}) and~(\ref{eq:fundvarvanish0}), the claim of the proposition will be proved if we are able to show that
\beq
\boxed{
\frac{(N p_N)^2 \sigma^2_e}{p_N}\stackrel{N\rightarrow\infty}{\longrightarrow} 0
}
\label{eq:fundvarvanish}
\eeq
Accordingly, we now focus on the evaluation of $\sigma^2_e$. 
To this end, we use for $\bm{e}_{ij}$ the representation given in~(\ref{eq:easfuncofef}), and apply Lemma~$3$ to $\bm{e}_{ij}$, for $i\neq j$, with the choices $L=N-K$, and: 
\beqa
&&[\bm{u}_1,\bm{u}_2,\dots,\bm{u}_{L}]
\rightarrow
[\bm{a}_{\omega_i 1},\bm{a}_{\omega_i 2},\dots,\bm{a}_{\omega_i L}],\nonumber\\
&&[\bm{z}_1,\bm{z}_2,\dots,\bm{z}_{L}]
\rightarrow
[\bm{f}_{1j},\bm{f}_{2j},\dots,\bm{f}_{L j}],\nonumber\\
&&\bm{s}\rightarrow \bm{e}_{ij}.
\label{eq:Lemma3choices}
\eeqa
We remark that all the matrices are {\em random}, as a consequence of the randomness implied by the underlying Erd\"os-Renyi graph. 
Preliminarily, it is necessary to verify that the choices in~(\ref{eq:Lemma3choices}) meet the assumptions of Lemma~$3$. 
First, observe that, in view of~(\ref{eq:Fperminv1}), row-permutations leave unaltered the statistical distribution of $\bm{F}$. 
This property implies in particular that the random variables $\bm{f}_{1j},\bm{f}_{2j},\dots,\bm{f}_{L j}$ are identically distributed.
Moreover, under the independence approximation, such entries are approximated as mutually independent (and, hence, uncorrelated). 
Under the same approximation, the random vectors $[\bm{a}_{\omega_i 1},\bm{a}_{\omega_i 2},\dots,\bm{a}_{\omega_i L}]$ and $[\bm{f}_{1j},\bm{f}_{2j},\dots,\bm{f}_{L j}]$ are approximated as mutually independent.
Therefore, we conclude that the hypotheses of Lemma~$3$ are met, and we can write:
\beq
\V[\bm{e}_{ij}]=\sigma^2_e\approx
\sigma^2_f\,
\sum_{\ell=1}^{N-K} \E[\bm{a}^2_{ \omega_i\omega^\prime_\ell}]
+ 
\bar{f}^2\,
\V\left[\sum_{\ell=1}^{N-K} \bm{a}_{\omega_i\omega^\prime_\ell}\right],
\label{eq:VE0}
\eeq
The approximation symbol in~(\ref{eq:VE0}) comes from the fact that the pertinent calculations are done under the independence approximation. 

Now, in view of~(\ref{eq:Aperminv}), any renumbering leaves unaltered the statistical distribution of $\bm{A}$. This property implies that the random variables $\bm{a}_{ij}$, for all $i,j=1,2,\dots, N$, with $i\neq j$, are identically distributed, and, in particular, that they share the common mean-square value $\E[\bm{a}^2_{i j}]=\overline{a^2}$. 
Since $\omega_i\neq \omega^\prime_\ell$ (because $\Omega\cap\Omega^\prime=\emptyset$), Eq.~(\ref{eq:VE0}) can be rewritten as: 
\beq
\sigma^2_e\approx
(N-K)\sigma^2_f\, \overline{a^2} 
+
\bar{f}^2
\,
\V\left[\sum_{\ell=1}^{N-K} \bm{a}_{\omega_i\omega^\prime_\ell}\right].
\label{eq:VE}
\eeq
Likewise, we apply Lemma~$3$ to $\bm{f}_{\ell j}$, with the choices $L=N-K$, and: 
\beqa
&&[\bm{u}_1,\bm{u}_2,\dots,\bm{u}_L]\rightarrow[\bm{h}_{\ell 1},\bm{h}_{\ell 2},\dots,\bm{h}_{\ell L}],\nonumber\\
&&[\bm{z}_1,\bm{z}_2,\dots,\bm{z}_L]\rightarrow[\bm{b}_{1\omega_j},\bm{b}_{2\omega_j},\dots,\bm{b}_{L \omega_j}],\nonumber\\
&&\bm{s}\rightarrow \bm{f}_{\ell j}.
\eeqa
In view of~(\ref{eq:Bpermprop}), any renumbering leaves unaltered the statistical distribution of $\bm{B}$. This property implies in particular that the random variables $\bm{b}_{1\omega_j},\bm{b}_{2\omega_j},\dots,\bm{b}_{L \omega_j}$ are identically distributed.
Moreover, under the independence approximation, such entries are approximated as mutually independent (and, hence, uncorrelated). 
Under the same approximation, the random vectors $[\bm{h}_{\ell 1},\bm{h}_{\ell 2},\dots,\bm{h}_{\ell L}]$ and $[\bm{b}_{1\omega_j},\bm{b}_{2\omega_j},\dots,\bm{b}_{L \omega_j}]$ are approximated as mutually independent.
Therefore, we conclude that also in this case the hypotheses of Lemma~$3$ are met, which allows writing:
\beq
\sigma^2_f
\approx
\sigma^2_b
\,
\sum_{m=1}^{N-K} 
\E[\bm{h}_{\ell m}^2]
+ 
\bar{b}^2
\,
\V\left[\sum_{m=1}^{N-K} \bm{h}_{\ell m}\right].
\label{eq:VF}
\eeq
Substituting~(\ref{eq:VF}) into~(\ref{eq:VE}) gives: 
\beqa
\sigma^2_e&\approx&
(N-K)\,\overline{a^2}\,\sigma^2_b\,
\sum_{m=1}^{N-K} \E[\bm{h}_{\ell m}^2] +
\nonumber\\
&&
(N-K)\,\overline{a^2}\,\bar{b}^2\,
\V\left[\sum_{m=1}^{N-K} \bm{h}_{\ell m}\right] +
\nonumber\\
&&\bar{f}^2\,
\V\left[\sum_{\ell=1}^{N-K} \bm{a}_{\omega_i\omega^\prime_\ell}\right]\nonumber\\
&\dfz&
\tilde\sigma^2_e.
\label{eq:ineqchain3}
\eeqa
A comparison between the true variance, $\sigma^2_e$, and the approximate variance, $\tilde\sigma^2_e$, is given in Fig.~\ref{fig:varver}. 
We see that, for the considered examples, the agreement between the two quantities is satisfying.
Note that, in certain cases, the true curve and the approximate curve (pertaining to the same example) exhibit slightly different slopes, suggesting that the approximation does not always capture the precise asymptotic behavior. 

\begin{figure}[t]
\centerline{\includegraphics[width=.35\textheight]{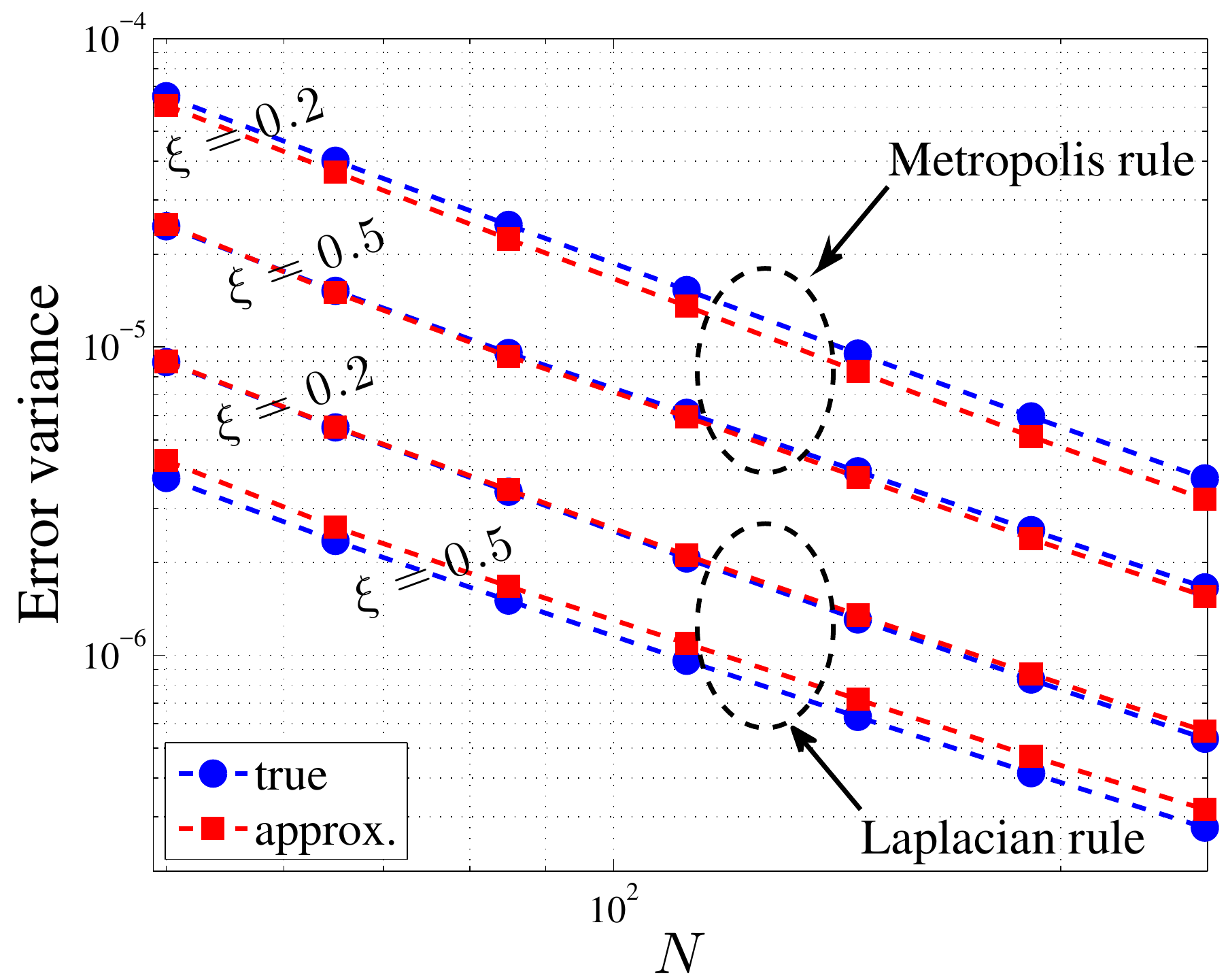}}
\caption{Comparison between the true error variance, $\V[\bm{e}_{ij}]$ (evaluated through $10^3$ Monte Carlo runs), and the error variance arising from the independence approximation, see~(\ref{eq:ineqchain3}). 
The different curves are obtained by varying the type of combination matrix (Laplacian or Metropolis), and the fraction of observable agents ($\xi=0.2$ or $\xi=0.5$).
}
\label{fig:varver}
\end{figure}

We now compute an upper bound for the approximate variance in~(\ref{eq:ineqchain3}). 
First, we find an upper bound for $\bar{f}$. 
In view of~(\ref{eq:Fperminv2}), column-permutations leave unaltered the statistical distribution of $\bm{F}$. This property implies in particular that the random variables $\bm{f}_{\ell 1},\bm{f}_{\ell 2},\dots,\bm{f}_{\ell K}$ are identically distributed, yielding:
\beq
\bar{f}
=
\frac 1 K \sum_{j=1}^K \E[\bm{f}_{\ell j}]
=
\E\left[\frac{\sum_{j=1}^K \bm{f}_{\ell j}}{K}\right]
\leq \frac{1}{K},
\label{eq:fbound}
\eeq
where the inequality comes from~(\ref{eq:fleq1}).

Second, we observe that:
\beqa
\sum_{m=1}^{N-K} \bm{h}_{\ell m}&\stackrel{(a)}{=}&
\sum_{k=0}^\infty 
\sum_{m=1}^{N-K}
[(\bm{B}_{\Omega^\prime})^k]_{\ell m}\nonumber\\
&\stackrel{(b)}{\leq}&
\sum_{k=0}^\infty 
\sum_{m=1}^{N-K}
[[\bm{B}^k]_{\Omega^\prime}]_{\ell m}\nonumber\\
&\stackrel{(c)}{=}&
\sum_{k=0}^\infty 
\sum_{m=1}^{N-K}
[\bm{B}^k]_{\omega^\prime_{\ell}\omega^\prime_m}\nonumber\\
&\stackrel{(d)}{\leq}&
\sum_{k=0}^\infty 
\sum_{j=1}^{N}
[\bm{B}^k]_{\omega^\prime_{\ell} j}\nonumber\\
&\stackrel{(e)}{=}&
\sum_{k=0}^\infty 
(1-\mu)^{2 k}=
\frac{1}{\mu(2-\mu)}\dfz \sqrt{C},\nonumber\\
\label{eq:HHbound}
\eeqa
where $(a)$ follows from~(\ref{eq:Hseriesfirstdef}); $(b)$ holds since for any two nonnegative matrices $P$ and $Q$ we have that $P_{\Omega^\prime}Q_{\Omega^\prime}\leq[P Q]_{\Omega^\prime}$ (with entrywise inequality), and, hence, by iteration we get 
$(\bm{B}_{\Omega^\prime})^k\leq[\bm{B}^k]_{\Omega^\prime}$ because $\bm{B}$ and its powers are nonnegative; $(c)$ holds because indices $\omega^\prime_1<\omega^\prime_2<\dots<\omega^\prime_{N-K}$ span the set $\Omega^\prime$; $(d)$ holds because $\bm{B}^k$ is nonnegative, and, hence, adding more terms can only increase the value of the sum; and $(e)$ holds because $\bm{B}=\bm{A}^2$, yielding $\sum_{j=1}^N \bm{b}_{ij}=(1-\mu)^2 \Rightarrow \sum_{j=1}^N[\bm{B}^k]_{ij}=(1-\mu)^{2 k}$, for all $i=1,2,\dots,N$.
Exploiting now the bounds~(\ref{eq:fbound}) and~(\ref{eq:HHbound}), from~(\ref{eq:ineqchain3}) we get:
\beqa
\tilde\sigma^2_e&\leq&
(N-K)\,\overline{a^2}\,\sigma^2_b\,C
+
(N-K)\,\overline{a^2}\,\bar{b}^2\,C
+
\nonumber\\
&&\frac{1}{K^2}\,
\V\left[\sum_{\ell=1}^{N-K} \bm{a}_{\omega_i\omega^\prime_\ell}\right],
\label{eq:sigmatildeineq}
\eeqa
where we used the following known bounds:
\beqa
\sum_{m=1}^{N-K} \bm{h}_{\ell m}^2&\leq& \left(\sum_{m=1}^{N-K} \bm{h}_{\ell m}\right)^2,\nonumber\\
\V\left[\sum_{m=1}^{N-K} \bm{h}_{\ell m}\right]
&\leq&
\E\left[\left(\sum_{m=1}^{N-K} \bm{h}_{\ell m}\right)^2\right].
\eeqa
Since $\sigma^2_b + \bar{b}^2=\overline{b^2}$, inequality~(\ref{eq:sigmatildeineq}) can be also written as: 
\beq
\tilde\sigma^2_e
\leq
(N-K)\,\overline{a^2}\,\overline{b^2}\,C
+
\frac{1}{K^2}\,
\V\left[\sum_{\ell=1}^{N-K} \bm{a}_{\omega_i\omega^\prime_\ell}\right].
\label{eq:roughupperbound}
\eeq
We must now examine the asymptotic behavior of the terms $\overline{a^2}$ and $\overline{b^2}$.
We start with $\overline{a^2}$. 
We can write:
\beqa
\overline{a^2}&=&
\frac{1}{N-1}\E\left[
\sum_{\ell\neq i}^N\bm{a}_{i\ell}^2
\right]
\leq
\frac{\kappa^2}{(N-1)}
\E\left[
\frac{1}{\bm{d}_i^2}
\sum_{\ell\neq i}^N
\bm{g}_{i\ell}
\right]
\nonumber\\
&\leq&
\frac{\kappa^2}{N-1}\E\left[
\frac{1}{\bm{d}_i}
\right]
=
\frac{\kappa^2}{N-1}
\E\left[\frac{1}{1+\bm{\beta}(N-1,p_N)}\right],\nonumber\\
\label{eq:asqbound}
\eeqa
where the first equality holds because, as already observed, the random variables $\bm{a}_{i\ell}$, for $\ell\neq i$, are identically distributed in view of~(\ref{eq:Aperminv}). The first inequality follows from~(\ref{eq:AssumptionA1bis}), the second inequality follows because $\sum_{\ell\neq i} \bm{g}_{i\ell}=\bm{d}_i-1$, and the last equality follows from the fact that $\bm{d}_i - 1$ is distributed as a binomial random variable of parameters $N-1$ and $p_N$.
In view of Lemma~$2$, Eq.~(\ref{eq:asqbound}) implies the existence of an upper bound, $\epsilon^{(1)}_N$, such that:
\beq
\overline{a^2}\leq \epsilon^{(1)}_N\sim \frac{1}{N^2 p_N},
\label{eq:asqfinal}
\eeq
where the notation $a_N\sim b_N$ means that $a_N$ is on the same order of $b_N$, namely, that the limit $\lim_{N\rightarrow\infty} a_N/b_N$ exists and is finite.  
It remains to examine the asymptotic behavior of the quantity $\overline{b^2}$.
Recalling that $\bm{B}=\bm{A}^2$ and that $\bm{A}$ is symmetric, we have, for $i\neq j$:
\beqa
\bm{b}_{ij}&=&
\sum_{\ell=1}^N \bm{a}_{i\ell} \bm{a}_{\ell j}=(\bm{a}_{ii}+\bm{a}_{jj})\bm{a}_{ij}+
\sum_{\ell\neq i,j}\bm{a}_{i\ell} \bm{a}_{j\ell}\nonumber\\
&\leq&
2 \, \bm{a}_{ij} + \kappa^2 \underbrace{\frac{\sum_{\ell\neq i,j}\bm{g}_{i\ell} \bm{g}_{j\ell}}{\bm{d}_i \bm{d}_j}}_{\dfz \bm{z}_{ij}},
\label{eq:bijnonsquare}
\eeqa
where, in the last step, we used~(\ref{eq:AssumptionA1bis}).
We can thus write:
\beq
\bm{b}_{ij}^2\leq 4 \, \bm{a}_{ij}^2 + \kappa^4 \,\bm{z}_{ij}^2 + 4 \kappa^2\, \bm{a}_{ij}\bm{z}_{ij}.
\label{eq:bijsquareexpress}
\eeq
Let us first focus on the term $\bm{z}_{ij}^2$. We observe that:
\beqa
\bm{z}_{ij}^2&=&\frac{1}{\bm{d}_i^2 \bm{d}_j^2}\left(
\sum_{\ell\neq i,j}\bm{g}_{i\ell} \bm{g}_{j\ell} 
+
\sum_{\substack{\ell\neq i,j \\ m\neq i,j \\ \ell\neq m}} \bm{g}_{i\ell} \bm{g}_{j\ell} \bm{g}_{i m} \bm{g}_{j m}
\right)
\nonumber\\
&\leq&
\underbrace{\sum_{\ell\neq i,j} \frac{\bm{g}_{i\ell} \bm{g}_{j\ell}}{(\bm{d}_i-\bm{g}_{i\ell})^2 (\bm{d}_j-\bm{g}_{j\ell})^2}}_{\bm{Z}_1} + \nonumber\\
&&
\underbrace{\sum_{\substack{\ell\neq i,j \\ m\neq i,j \\ \ell\neq m}}
\frac{\bm{g}_{i\ell} \bm{g}_{j\ell} \bm{g}_{i m} \bm{g}_{j m}}
{(\bm{d}_i-\bm{g}_{i\ell}-\bm{g}_{i m})^2 (\bm{d}_j-\bm{g}_{j\ell} - \bm{g}_{j m})^2}}_{\bm{Z}_2},\nonumber\\
\label{eq:superchain}
\eeqa
where the inequality holds because, in view of~(\ref{eq:degdef}), we have $\bm{d}_i>\bm{d}_i - \bm{g}_{i\ell}>0$ for $\ell\neq i$ and $\bm{d}_i>\bm{d}_i - \bm{g}_{i\ell} - \bm{g}_{i m}>0$ for $\ell\neq i$ and $m\neq i$.
Using again~(\ref{eq:degdef}), the terms $\bm{Z}_1$ and $\bm{Z}_2$ in~(\ref{eq:superchain}) can be rewritten as, respectively:
\beqa
\bm{Z}_1&=&\sum_{\ell\neq i,j} \frac{\bm{g}_{i\ell} \bm{g}_{j\ell}}
{(1+\sum_{k\neq i,\ell}  \bm{g}_{i k})^2 (1+\sum_{k\neq j,\ell}  \bm{g}_{j k})^2}
\nonumber\\
\bm{Z}_2&=&
\sum_{\substack{\ell\neq i,j \\ m\neq i,j \\ \ell\neq m}}
\frac{\bm{g}_{i\ell} \bm{g}_{j\ell} \bm{g}_{i m} \bm{g}_{j m}}
{(1+\sum_{k\neq i,\ell,m}  \bm{g}_{i k})^2 (1+\sum_{k\neq j,\ell,m}  \bm{g}_{j k})^2}.\nonumber\\
\label{eq:superchain2}
\eeqa
We are now interested in evaluating the expectation of $\bm{Z}_1$ and $\bm{Z}_2$. 
To this aim, two observations are useful. 
First, we note that the random variables $\bm{g}_{i\ell}, \bm{g}_{j\ell}, \bm{g}_{i m}, \bm{g}_{j m}$ are, for the indices considered in the pertinent summations, mutually independent Bernoulli variables.
Second, the terms in the denominator are binomial random variables that are independent from the variables present in the numerator.
This is because, for example, the quantity $\sum_{k\neq i,\ell,m}  \bm{g}_{i k}$ appearing at the denominator of the second ratio in~(\ref{eq:superchain2}), does not contain any of the random variables $\bm{g}_{i\ell}, \bm{g}_{j\ell}, \bm{g}_{i m}, \bm{g}_{j m}$ (nor their symmetric counterparts $\bm{g}_{\ell i}, \bm{g}_{\ell j}, \bm{g}_{m i}, \bm{g}_{m j}$) that appear at the numerator. 
Therefore, taking the expectation of $\bm{Z}_1$ in~(\ref{eq:superchain}) yields:
\beqa
\lefteqn{\E[\bm{Z}_1]=\sum_{\ell\neq i,j} \E[\bm{g}_{i\ell}] \E[\bm{g}_{j\ell}] \times}
\nonumber\\
&&
\E\left[\frac{1}{(1+\sum_{k\neq i,\ell}  \bm{g}_{i k})^2}\right]
\E\left[\frac{1}{(1+\sum_{k\neq j,\ell}  \bm{g}_{j k})^2}\right].
\nonumber\\
\label{eq:Z1ineq}
\eeqa
Now, since $\E[\bm{g}_{i\ell}]=\E[\bm{g}_{j\ell}]=p_N$, since $\sum_{k\neq i,\ell}  \bm{g}_{i k}$ is a binomial random variable $\bm{\beta}(N-2,p_N)$, and since the summation in~(\ref{eq:Z1ineq}) contains no more than $N$ terms, we have that:
\beq
\E[\bm{Z}_1]\leq N p^2_N\left(\E\left[\frac{1}{(1+\bm{\beta}(N-2,p_N))^2}\right]\right)^2.
\label{eq:Z1ineq2}
\eeq
Likewise, taking the expectation of $\bm{Z}_2$ in~(\ref{eq:superchain}) yields:
\beqa
\lefteqn{\E[\bm{Z}_2]=
\sum_{\substack{\ell\neq i,j \\ m\neq i,j \\ \ell\neq m}}
\E[\bm{g}_{i\ell}] \E[\bm{g}_{j\ell}] \E[\bm{g}_{i m}] \E[\bm{g}_{j m}] \times}
\nonumber\\
&&
\E\left[\frac{1}{(1+\sum_{k\neq i,\ell,m}  \bm{g}_{i k})^2}\right] 
\E\left[\frac{1}{(1+\sum_{k\neq j,\ell,m}  \bm{g}_{j k})^2}\right].
\nonumber\\
\label{eq:Z2ineq}
\eeqa
Since $\sum_{k\neq i,\ell,m}  \bm{g}_{i k}$ is a binomial random variable $\bm{\beta}(N-3,p_N)$, and since the summation in~(\ref{eq:Z2ineq}) contains no more than $N^2$ terms, we have that:
\beq
\E[\bm{Z}_2]\leq N^2 p_N^4 \left(\E\left[\frac{1}{(1+\bm{\beta}(N-3,p_N))^2}\right]\right)^2.
\label{eq:Z2ineq2}
\eeq
Finally, using~(\ref{eq:Z1ineq2}) and~(\ref{eq:Z2ineq2}) into~(\ref{eq:superchain}), we have:
\beqa
\E[\bm{z}_{ij}^2]&\leq&
N p^2_N\left(\E\left[\frac{1}{(1+\bm{\beta}(N-2,p_N))^2}\right]\right)^2 + \nonumber\\
&&
N^2 p_N^4 \left(\E\left[\frac{1}{(1+\bm{\beta}(N-3,p_N))^2}\right]\right)^2.\nonumber\\
\eeqa
In view of Lemma~$2$, we conclude that:
\beq
\E[\bm{z}_{ij}^2]\leq \epsilon^{(2)}_N \sim \frac{1}{N^3 p_N^2} + \frac{1}{N^2}.
\label{eq:epsi2}
\eeq
Let us move on to examining the term $\bm{a}_{ij} \bm{z}_{ij}$ in~(\ref{eq:bijsquareexpress}).
Using~(\ref{eq:AssumptionA1bis}) and the definition of $\bm{z}_{ij}$ in~(\ref{eq:bijnonsquare}), we can write:
\beq
\bm{a}_{ij} \bm{z}_{ij} \leq \frac{\kappa}{\bm{d}_i} \bm{g}_{ij} \bm{z}_{ij}
=
\kappa\,\bm{g}_{ij} \frac{\sum_{\ell\neq i,j}\bm{g}_{i\ell} \bm{g}_{j\ell}}{\bm{d}_i^2 \bm{d}_j}.
\eeq
Working along the same lines as done to obtain~(\ref{eq:Z2ineq}) and~(\ref{eq:Z2ineq2}), we can write:
\beqa
\lefteqn{\E[\bm{a}_{ij} \bm{z}_{ij}]}
\nonumber\\
&\leq&
\kappa\,
\E\left[
\bm{g}_{ij} \frac{\sum_{\ell\neq i,j}\bm{g}_{i\ell} \bm{g}_{j\ell}}{\bm{d}_i^2 \bm{d}_j}
\right]
\nonumber\\
&\leq&
\kappa\,
\E\left[
\sum_{\ell\neq i,j}
\frac{
\bm{g}_{ij} \bm{g}_{i\ell} \bm{g}_{j\ell}
}
{(\bm{d}_i -\bm{g}_{i\ell}-\bm{g}_{ij})^2 (\bm{d}_j - \bm{g}_{j\ell} - \bm{g}_{ij})}
\right]
\nonumber\\
&=&
\kappa\,
\sum_{\ell\neq i,j}
\E[\bm{g}_{ij}] \E[\bm{g}_{i\ell}] \E[\bm{g}_{j\ell}] \times\nonumber\\
&&
\E\left[\frac{1}{(1+\sum_{k\neq i,j,\ell}  \bm{g}_{i k})^2}\right]
\E\left[\frac{1}{(1+\sum_{k\neq i,j,\ell}  \bm{g}_{j k})}\right]
\nonumber\\
&\leq&
\kappa\,
N p_N^3\times\nonumber\\
&&
\E\left[
\frac{1}{(1+\bm{\beta}(N-3,p_N))^2}
\right] 
\E\left[
\frac{1}{1+\bm{\beta}(N-3,p_N)}
\right],\nonumber\\
\eeqa
which, using again Lemma~$2$, yields:
\beq
\E[\bm{a}_{ij} \bm{z}_{ij}] \leq \epsilon^{(3)}_N \sim \frac{1}{N^2}.
\label{eq:epsi3}
\eeq
We can now apply to~(\ref{eq:bijsquareexpress}) the upper bounds obtained in~(\ref{eq:asqfinal}),~(\ref{eq:epsi2}) and~(\ref{eq:epsi3}), getting:
\beqa
\E[\bm{b}_{ij}^2] &\leq& 
4 \, \E[\bm{a}_{ij}^2] + \kappa^4 \, \E[\bm{z}_{ij}^2] + 4 \kappa^2\, \E[\bm{a}_{ij}\bm{z}_{ij}]
\nonumber\\
&\leq&
4 \, \epsilon^{(1)}_N + \kappa^4 \, \epsilon^{(2)}_N + 4 \kappa^2\, \epsilon^{(3)}_N\nonumber\\
&\sim&
\frac{1}{N^2 p_N} + \frac{1}{N^3 p_N^2} + \frac{1}{N^2}\nonumber\\
&=&
\frac{1}{N^2 p_N}
\left(
1 + \frac{1}{N p_N} + p_N
\right)\sim \frac{1}{N^2 p_N}, \nonumber\\
\label{eq:bsqfinal}
\eeqa
where the last estimate follows because, under the $\mathscr{G}^\star(N,p_N)$ model, $N p_N=\ln N +c_N\rightarrow\infty$, and $p_N\rightarrow 0$, as $N\rightarrow\infty$.

It remains to use the results obtained as regards the decaying rate of $\overline{a^2}$ and $\overline{b^2}$ into~(\ref{eq:roughupperbound}). 
To this end, it is useful to recast~(\ref{eq:roughupperbound}) in the following form:
\beq
\tilde\sigma^2_e \leq
C (1- K/N)\, N \, \overline{a^2}\,\overline{b^2}
+
\frac{1/N^2}{(K/N)^2} (1-\mu)^2,
\label{eq:roughupperbound2}
\eeq
where we used the fact that $\sum_{\ell=1}^{N-K} \bm{a}_{\omega_i\omega^\prime_\ell}\leq 1-\mu$, and that the variance is upper bounded by the mean-square value. 
Using now~(\ref{eq:asqfinal}) and~(\ref{eq:bsqfinal}) into~(\ref{eq:roughupperbound2}), and noticing further that $K/N\rightarrow \xi$ in view of~(\ref{eq:csidef}), we conclude that: 
\beq
\tilde\sigma^2_e\leq \epsilon_N^{(4)}
\sim 
\frac{1}{N^3 p^2_N} + \frac{1}{N^2},
\eeq
implying:
\beq
\frac{(N p_N)^2 \tilde\sigma^2_e}{p_N}\leq N^2 p_N \,  \epsilon_N^{(4)}
\sim 
\frac{1}{N p_N} + p_N\stackrel{N\rightarrow\infty}{\longrightarrow} 0,
\eeq
which, in view of~(\ref{eq:fundvarvanish}), establishes the proposition.
\end{IEEEproof}

\section{}
\label{app:permprop}
We start by introducing the operations of matrix permutation and renumbering.
A matrix $P$ is a permutation matrix if exactly one entry in each row and column is equal to $1$, and all other entries are $0$ --- see~\cite{Johnson-Horn}.
Left multiplication of a matrix $Z$ by a permutation matrix $P$ permutes the rows of $Z$, whereas right multiplication permutes the columns. For example, if:
\beq
P=\left[\begin{array}{cccc}0 & 0 & 0 & 1 \\0 & 1 & 0 & 0 \\1 & 0 & 0 & 0 \\0 & 0 & 1 & 0\end{array}\right]
\eeq
then $\tilde{Z}=P Z$ sends the first row of $Z$ to the third row of $\tilde{Z}$, leaves the second row of $Z$ in the second row of $\tilde{Z}$, sends the third row of $Z$ to the fourth row of $\tilde{Z}$, and sends the fourth row of $Z$ to the first row of $\tilde{Z}$. 

Permutation matrices have the following property~\cite{Johnson-Horn}:
\beq
P^T=P^{-1}.
\label{eq:permatinv}
\eeq
Since $P^T$ permutes columns in the same way that $P$ permutes rows, the transformation $Z\rightarrow P Z P^T$ permutes the rows and columns of $Z$.
In our network setting, where the $(i,j)$-th matrix entry is associated to a property of agents $i$ and $j$, the latter transformation corresponds to {\em renumbering} the agents. 

A random matrix $\bm{Z}$ will be said to be statistically invariant to renumbering if, for any permutation matrix $P$:
\beq
P\bm{Z}P^T \stackrel{(\textnormal{d})}{=} \bm{Z},
\eeq 
where the symbol $\stackrel{(\textnormal{d})}{=}$ denotes equality in distribution.
Likewise, $\bm{Z}$ is invariant to row (resp., to column) permutation, if we have, for any permutation matrix $P$:
\beq
P\bm{Z} \stackrel{(\textnormal{d})}{=} \bm{Z} \qquad \textnormal{(resp., $\bm{Z}P^T \stackrel{(\textnormal{d})}{=} \bm{Z}$)}.
\eeq 
\begin{lemma}[Renumbered combination matrix]
Under the Erd\"os-Renyi model, if the combination policy possesses property P2, the $N\times N$ combination matrix $\bm{A}$ and the $K\times K$ error matrix $\bm{E}$ in~(\ref{eq:Ematfirstdef}) are statistically invariant to renumbering. 
Likewise, the $(N-K)\times K$ matrix $\bm{F}$ in~(\ref{eq:Fmatfirstdef}) is invariant to row and to column permutation. 
Formally, letting $\mathcal{P}_M$ be the ensemble of all $M\times M$ permutation matrices, we have that:
\beqa
&&P\bm{A}P^T \stackrel{(\textnormal{d})}{=} \bm{A},\quad \forall P\in\mathcal{P}_N,
\label{eq:Aperminv}
\\
&&P\bm{E}P^T \stackrel{(\textnormal{d})}{=} \bm{E},\quad \forall P\in\mathcal{P}_K,
\label{eq:Eperminv}
\\
&&P\bm{F} \stackrel{(\textnormal{d})}{=} \bm{F}, \quad \forall P\in\mathcal{P}_{N - K},
\label{eq:Fperminv1}
\\
&&\bm{F}P^T \stackrel{(\textnormal{d})}{=} \bm{F}, \quad \forall P\in\mathcal{P}_{K}.
\label{eq:Fperminv2}
\eeqa
\end{lemma}
\begin{IEEEproof}
Under the Erd\"os-Renyi model, the variables $\bm{g}_{ij}$, for $i=1,2,\dots, N$ and $j>i$, are independent Bernoulli random variables with $\P[\bm{g}_{ij}=1]=p_N$, and the matrix $\bm{G}$ is a symmetric matrix.
Therefore, exchanging the agents does not alter the statistical properties of $\bm{G}$, namely, any renumbered version of $\bm{G}$ has the same probability of occurrence:
\beq
P\bm{G}P^T\stackrel{(d)}{=}\bm{G}\Rightarrow
\gamma(P\bm{G}P^T)\stackrel{(d)}{=}\gamma(\bm{G}),
\label{eq:ERinv}
\eeq 
where the latter equality in distribution holds because $\gamma(\cdot)$ is a deterministic function.
Moreover, by property P$2$ we have that:
\beq
\gamma(P\bm{G}P^T)=P\gamma(\bm{G})P^T.
\label{eq:a2perm}
\eeq
Since $\bm{A}=\gamma(\bm{G})$, Eqs.~(\ref{eq:ERinv}) and~(\ref{eq:a2perm}) immediately imply~(\ref{eq:Aperminv}).

Next we prove~(\ref{eq:Eperminv}). 
Using~(\ref{eq:Ematfirstdef}) and~(\ref{eq:BCDmat}), the matrix $\bm{E}$ can be formally written as a function of $\bm{A}$ as follows:
\beqa
\bm{E}&=&
\bm{A}_{\Omega\Omega^\prime}\bm{H}\bm{B}_{\Omega^\prime\Omega}
\nonumber\\
&=&\bm{A}_{\Omega\Omega^\prime}(I_{N-K} - [\bm{A}^2]_{\Omega^\prime})^{-1} [\bm{A}^2]_{\Omega^\prime\Omega}
\nonumber\\
&\dfz& \psi(\bm{A}).
\label{eq:Efunc}
\eeqa
Let now $\tilde{\bm{A}}$ and $\tilde{\bm{E}}$ denote arbitrarily renumbered versions of $\bm{A}$ and $\bm{E}$, respectively.
Since we know that $\bm{A}$ and $\tilde{\bm{A}}$ share the same distribution, and since $\psi(\cdot)$ is a deterministic function, also $\psi(\bm{A})$ and $\psi(\tilde{\bm{A}})$ will have the same distribution. 
Therefore, in view of~(\ref{eq:Efunc}), claim~(\ref{eq:Eperminv}) will be proved if we show that any renumbered error matrix, $\tilde{\bm{E}}$, can be always written as $\tilde{\bm{E}}=\psi(\tilde{\bm{A}})$, for a certain renumbered combination matrix, $\tilde{\bm{A}}$.
To this end, let us introduce an $N\times N$ permutation matrix $P$ that permutes rows belonging to the index set $\Omega$ only with rows belonging to $\Omega$, and rows belonging to the complement set $\Omega^\prime$ only with rows belonging to $\Omega^\prime$.
Formally, this assumption implies that the principal sub-matrix $P_\Omega$ is a $K\times K$ permutation matrix, that $P_{\Omega^\prime}$ is an $(N-K)\times(N-K)$ permutation matrix, and that $P_{\Omega\Omega^\prime}$ and $P_{\Omega^\prime\Omega}$ are matrices containing only zero entries.
Therefore, since any renumbering of matrix $\bm{E}$ is accomplished by using a certain $K\times K$ permutation matrix, we can always write, without losing generality: 
\beq
\tilde{\bm{E}}=P_{\Omega} \bm{E} P_{\Omega}^T=
P_{\Omega}\bm{A}_{\Omega\Omega^\prime}P_{\Omega^\prime}^T
P_{\Omega^\prime} \bm{H} P_{\Omega^\prime}^T
P_{\Omega^\prime}\bm{B}_{\Omega^\prime\Omega}P_{\Omega}^T,
\label{eq:Etildechain}
\eeq
where we used~(\ref{eq:permatinv}).
From the rules for multiplication of partitioned matrices (and since $P_{\Omega\Omega^\prime}$ and $P_{\Omega^\prime\Omega}$ are matrices with all zeros), we have, for a generic $N\times N$ matrix $Z$:
\beqa
\,[P Z P^T]_{\Omega}&=&P_\Omega Z_{\Omega} P^T_{\Omega},
\label{eq:pomeg}\\
\,[P Z P^T]_{\Omega^\prime}&=&P_{\Omega^\prime} Z_{\Omega^\prime} P^T_{\Omega^\prime},
\label{eq:pomegprim}\\
\,[P Z P^T]_{\Omega\Omega^\prime}&=&P_\Omega Z_{\Omega\Omega^\prime} P^T_{\Omega^\prime}.
\label{eq:pomegomegprim}\\
\,[P Z P^T]_{\Omega^\prime\Omega}&=&P_{\Omega^\prime} Z_{\Omega^\prime\Omega} P^T_{\Omega},
\label{eq:pomegprimomeg}
\eeqa
Therefore, applying~(\ref{eq:pomegomegprim}) and~(\ref{eq:pomegprimomeg}) to~(\ref{eq:Etildechain}), we get: 
\beq
\tilde{\bm{E}}=P_{\Omega} \bm{E} P_{\Omega}^T=
[P\bm{A}P^T]_{\Omega\Omega^\prime}\,
P_{\Omega^\prime}\bm{H} P_{\Omega^\prime}^T\,
[P\bm{B}P^T]_{\Omega^\prime\Omega}.
\label{eq:Etildechain2}
\eeq
Moreover, we observe that:
\beqa
P_{\Omega^\prime} \bm{H} P_{\Omega^\prime}^T&=&
P_{\Omega^\prime}(I_{N-K} - \bm{B}_{\Omega^\prime})^{-1}P_{\Omega^\prime}^T\nonumber\\
&=&(P_{\Omega^\prime}(I_{N-K} - \bm{B}_{\Omega^\prime})P_{\Omega^\prime}^T)^{-1}\nonumber\\
&=&
(I_{N-K} - P_{\Omega^\prime}\bm{B}_{\Omega^\prime}P_{\Omega^\prime}^T)^{-1}\nonumber\\
&=&
(I_{N-K} - [P\bm{B}P^T]_{\Omega^\prime})^{-1}
\label{eq:Hpermprop}
\eeqa
where we used again~(\ref{eq:permatinv}).
On the other hand, we have that:
\beq
P\bm{B}P^T=P \bm{A}^2 P^T=(P \bm{A} P^T) (P \bm{A} P^T)=
(P\bm{A}P^T)^2.
\label{eq:Bpermprop}
\eeq
Finally, letting $\tilde{\bm{A}}=P\bm{A}P^T$, and using~(\ref{eq:Hpermprop}) and~(\ref{eq:Bpermprop})  into~(\ref{eq:Etildechain2}), we get:
\beq
\tilde{\bm{E}}=\tilde{\bm{A}}_{\Omega\Omega^\prime}
(I_{N-K} - [\tilde{\bm{A}}^2]_{\Omega^\prime})^{-1}
[\tilde{\bm{A}}^2]_{\Omega^\prime\Omega}=\psi(\tilde{\bm{A}}),
\eeq
which completes the proof of~(\ref{eq:Eperminv}). 

Let us now focus on proving~(\ref{eq:Fperminv1}).
Using~(\ref{eq:Fmatfirstdef}) and~(\ref{eq:BCDmat}), the matrix $\bm{F}$ can be formally written as a function of $\bm{A}$ as follows:
\beq
\bm{F}=
\bm{H}\bm{B}_{\Omega^\prime\Omega}
=(I_{N-K} - [\bm{A}^2]_{\Omega^\prime})^{-1} [\bm{A}^2]_{\Omega^\prime\Omega}
\dfz \varphi(\bm{A}).
\label{eq:Ffunc}
\eeq
Let now $\tilde{\bm{F}}$ denote an arbitrarily row-permuted version of $\bm{F}$.
Claim~(\ref{eq:Fperminv1}) will be proved if we show that any row-permuted matrix $\tilde{\bm{F}}$ can be always written as $\tilde{\bm{F}}=\varphi(\tilde{\bm{A}})$, for a certain renumbered combination matrix, $\tilde{\bm{A}}$.
To this end, let us introduce an $N\times N$ permutation matrix $P$ that permutes rows belonging to the complement set $\Omega^\prime$ only with rows belonging to $\Omega^\prime$, while leaving unaltered the rows belonging to the index set $\Omega$.
Such assumption implies that the principal sub-matrix $P_{\Omega^\prime}$ is an $(N-K)\times(N-K)$ permutation matrix, that $P_\Omega=I_K$, and that $P_{\Omega\Omega^\prime}$ and $P_{\Omega^\prime\Omega}$ are matrices containing only zero entries.
Thus, an arbitrary row-permutation of matrix $\bm{F}$ can be represented as:
\beqa
\tilde{\bm{F}}=P_{\Omega^\prime} \bm{F}&=&
P_{\Omega^\prime} \bm{H} P_{\Omega^\prime}^T
P_{\Omega^\prime}\bm{B}_{\Omega^\prime\Omega}P_{\Omega}^T
\nonumber\\
&=&
P_{\Omega^\prime}\bm{H} P_{\Omega^\prime}^T\,
[P\bm{B}P^T]_{\Omega^\prime\Omega},
\label{eq:Ftildechain}
\eeqa
where we used~(\ref{eq:permatinv}), the equality $P_\Omega=I_K$, and~(\ref{eq:pomegprimomeg}).
Letting now $\tilde{\bm{A}}=P\bm{A}P^T$, and using~(\ref{eq:Hpermprop}) and~(\ref{eq:Bpermprop})  into~(\ref{eq:Ftildechain}), we get:
\beq
\tilde{\bm{F}}=(I_{N-K} - [\tilde{\bm{A}}^2]_{\Omega^\prime})^{-1}
[\tilde{\bm{A}}^2]_{\Omega^\prime\Omega}=\varphi(\tilde{\bm{A}}),
\eeq
which completes the proof of~(\ref{eq:Fperminv1}).
The proof of~(\ref{eq:Fperminv2}) is similar to the proof of~(\ref{eq:Fperminv1}), and is accordingly omitted. 
\end{IEEEproof}


\end{document}